\documentclass[aps,prl,twocolumn,superscriptaddress]{revtex4-2}
\usepackage{graphicx,amssymb,amsmath,booktabs,multirow,bm,dsfont}
\usepackage[dvipsnames]{xcolor}
\usepackage{CJKutf8}
\usepackage{hyperref}
\usepackage{tikz}
\usetikzlibrary{arrows.meta}
\hypersetup{
    colorlinks=true,
    urlcolor=blue,
    linkcolor = blue,
    urlcolor  = blue,
    citecolor = blue,
    anchorcolor = blue,
    pdftitle={pi1300},
    pdfauthor={AGEMOS},
    pdfsubject={QCD}}

\renewcommand{\Re}{{\rm Re}}

\renewcommand{\Im}{{\rm Im}}
\newcommand{\MeV}{{\rm MeV}}

\usepackage{cleveref} 
    \crefformat{figure}{Fig.~#2#1#3}
    \crefformat{table}{Table~#2#1#3}
    \crefformat{equation}{Eq.~(#2#1#3)}
    \crefformat{section}{Sect.~#2#1#3}
    \crefformat{appendix}{App.~#2#1#3}
\usepackage[normalem]{ulem}
\allowdisplaybreaks
\begin{document}

\title{Emergence of the $\pi(1300)$ Resonance from Lattice QCD}

\author{Haobo~Yan (\begin{CJK*}{UTF8}{gbsn}燕浩波\end{CJK*})}
\email{haobo@stu.pku.edu.cn}
\affiliation{School of Physics, Peking University, Beijing 100871, China}
\affiliation{Helmholtz-Institut f\"ur Strahlen- und Kernphysik (Theorie) and Bethe Center for Theoretical Physics,  Universit\"at Bonn, 53115 Bonn, Germany}
\author{Maxim~Mai}
\email{maxim.mai@faculty.unibe.ch}
\affiliation{Albert Einstein Center for Fundamental Physics, Institute for Theoretical Physics, University of Bern, Sidlerstrasse 5, 3012 Bern, Switzerland}
\affiliation{The George Washington University, Washington, DC 20052, USA}
\author{Marco~Garofalo}
\email{garofalo@hiskp.uni-bonn.de}
\affiliation{Helmholtz-Institut f\"ur Strahlen- und Kernphysik (Theorie) and Bethe Center for Theoretical Physics,  Universit\"at Bonn, 53115 Bonn, Germany}
\author{Yuchuan~Feng}
\email{fengyuchuan@gwmail.gwu.edu}
\affiliation{The George Washington University, Washington, DC 20052, USA}
\author{Michael Döring}
\email{doring@gwu.edu}
\affiliation{The George Washington University, Washington, DC 20052, USA}
\author{Chuan~Liu (\begin{CJK*}{UTF8}{gbsn}刘川\end{CJK*})}
\email{liuchuan@pku.edu.cn}
\affiliation{School of Physics, Peking University, Beijing 100871, China}
\affiliation{Center for High Energy Physics, Peking University, Beijing 100871, China}
\affiliation{Collaborative Innovation Center of Quantum Matter, Beijing 100871, China}
\author{Liuming~Liu (\begin{CJK*}{UTF8}{gbsn}刘柳明\end{CJK*})}
\email{liuming@impcas.ac.cn}
\affiliation{Institute of Modern Physics, Chinese Academy of Sciences, Lanzhou 730000, China}
\affiliation{University of Chinese Academy of Sciences, Beijing 100049, China}
\author{Ulf-G.~Mei{\ss}ner}
\email{meissner@hiskp.uni-bonn.de}
\affiliation{Helmholtz-Institut f\"ur Strahlen- und Kernphysik (Theorie) and Bethe Center for Theoretical Physics,  Universit\"at Bonn, 53115 Bonn, Germany}
\affiliation{Institute for Advanced Simulation (IAS-4), Forschungszentrum J\"ulich, 52425 J\"ulich, Germany}
\affiliation{Peng Huanwu Collaborative Center for Research and Education, International Institute for Interdisciplinary and Frontiers, Beihang University, Beijing 100191, China}
\author{Carsten~Urbach}
\email{urbach@hiskp.uni-bonn.de}
\affiliation{Helmholtz-Institut f\"ur Strahlen- und Kernphysik (Theorie) and Bethe Center for Theoretical Physics,  Universit\"at Bonn, 53115 Bonn, Germany}

\begin{abstract}
The mass of the lightest hadron in nature, the pion, is one seventh of that of the nucleon and one tenth of the mass of its first excited state, the $\pi(1300)$. This enormous energy difference opens an interesting window into the confinement of quarks and the structure of the lightest hadrons. In this Letter, we provide the first calculation of resonance parameters of the $\pi(1300)$ from lattice quantum chromodynamics (QCD). For this purpose, recently derived state-of-the-art tools are adapted and applied both in the construction of three-hadron operators and for mapping finite-volume spectra to infinite-volume amplitudes, subsequently analytically continuing these to complex energies. For our heavy pion mass ensembles, we find a clear signal of the resonance. Making a simple assumption of vanishing pion mass dependence for the three-body force, but incorporating constraints from Chiral Perturbation Theory for all the two-body channels, enables a robust extrapolation to the physical point. Applying model averaging, we extract a pole position of $M_{\pi(1300)}=(1169\pm46)-i(62_{-62}^{+168})\,\MeV$ supporting values from phenomenology.
\end{abstract}
\maketitle

\emph{Introduction}---%
The strong interaction of quarks and gluons as fundamental particles of nature is fascinating and contradicts our daily experience: they behave as individual particles at large energies but only confined into so-called hadrons at low energies. The theory describing this interaction is called quantum chromodynamics (QCD).
Unraveling the spectrum of QCD is at the core of many theoretical endeavors and is also driving large experimental efforts; for recent reviews, see Refs.~\cite{Hyodo:2020czb, Chen:2022asf, Mai:2022eur, Pelaez:2025wma}. For example, investigations of states with strange or heavy charm and bottom quarks have been investigated, for instance, with the LHCb experiment at the Large Hadron Collider (LHC) or with the Beijing Spectrometer III (BES III) at the Beijing Electron–Positron Collider II (BEPC II), see, e.g., Ref.~\cite{Ketzer:2019wmd} or the upcoming/proposed experiments such as AMBER at LHC~\cite{Friedrich:2024ylw}, SIS100 at GSI/FAIR~\cite{Messchendorp:2025men}, Klong at JLAB~\cite{KLF:2020gai}, STCF~\cite{Achasov:2023gey, Ai:2025xop}, or CEPC~\cite{CEPCStudyGroup:2025kmw}.

In the hadron spectrum, the pion, the lightest hadron with a mass of one seventh of the proton, is identified as a Goldstone boson of the spontaneously broken chiral symmetry and, therefore, plays a pivotal role in our understanding of quark-gluon interactions. However, it turns out that its first excited state, listed as $\pi(1300)$ and observed by Ananeva and collaborators in 1981~\cite{Ananeva:1981sb} in diffractive dissociation on nuclei, is much less understood. It is currently assumed to have a mass of $1300\pm100\ \mathrm{MeV}$ and a width of $200 - 600\ \mathrm{MeV}$~\cite{ParticleDataGroup:2024cfk}. However, even the existence of the $\pi(1300)$ as a hadronic resonance might be debatable. While strong evidence for it was found by CLEO~\cite{dArgent:2017gzv}, an ``elusive $\pi(1300)$'' bump disappeared at low $t'$ in COMPASS~\cite{COMPASS:2015gxz} resulting in no entry for it in the particle data group (PDG)~\cite{ParticleDataGroup:2024cfk}. If it exists, it is about ten times heavier than the pion, almost mass degenerate with the other members of the first excited pseudoscalar octet, $\eta(1295)$ and $K(1460)$~\cite{Doring:2025phq}, and as heavy as the excited $N(1440)$ state of the nucleon, which is only $1.5$ times heavier than the proton. Unraveling this pattern is, therefore, fundamental to understanding the emergence of the hadron spectrum. In addition, the $\pi(1300)$ is special because its $J^{PC}=0^{-+}$ quantum numbers allow for a decay to three pions without any centrifugal barrier. Indeed, all pions are in relative S-wave enhancing three-body effects, which distinguishes this resonance from the previously studied $\omega$~\cite{Yan:2024gwp} and $a_1$~\cite{Mai:2021nul} resonances. The $S$-wave system contains the emblematic $f_0(500)$ resonance~\cite{Pelaez:2015qba} in a subchannel and enhances the elusive three-body effects through the strong $\pi\pi$ decays of this broad isobar to recombine with the $S$-wave spectator pion, making the $\pi(1300)$ a unique testing ground for few-body studies.

The $\pi(1300)$ is important for several fields. First, while it is primarily a $q\bar{q}$ state, it may contain significant $q\bar{q}g$ hybrid components~\cite{Kalashnikova:1993xb, Donnachie:1999re}. Second, its poorly constrained properties contribute to dominant uncertainties in QCD sum-rule determinations of light-quark masses~\cite{Narison:2014vka, Boyle:2012jad}. Third, in $\tau \to 3\pi \nu_\tau$ decays, the large width of $\pi(1300)$ can enhance charge-parity (CP)-violating effects in extensions of the standard model~\cite{Choi:1994ch}.

Theoretically, there is little known about the $\pi(1300)$ from first principles, which makes such a computation highly desirable and important. The theoretical method of choice is lattice QCD (LQCD), which allows for a nonperturbative computation of the properties of such states from first principles. LQCD is formulated in Euclidean space-time such that unstable states can be accessed only after a mapping of LQCD spectra to scattering amplitudes, possible for two- or three-hadron systems (see reviews~\cite{Hansen:2019nir, Mai:2021lwb, Briceno:2017max}, recent status updates~\cite{Romero-Lopez:2022usb,Green:2026jgv,Sharpe:2026mtt}, and related works~\cite{Polejaeva:2012ut, Briceno:2012rv,Briceno:2018aml,Briceno:2024ehy,Alexandru:2020xqf,Alotaibi:2025pxz,Meng:2017jgx,Mai:2018djl,Blanton:2019vdk,Culver:2019vvu,Hansen:2020otl,Fischer:2020jzp,Alexandru:2020xqf,Draper:2023boj,Blanton:2021llb,Dawid:2025doq,Dawid:2025zxc,Hansen:2025oag,Briceno:2025yuq,Feng:2026ixm,Severt:2022jtg}). The finite-volume framework for three-body decays has been formulated~\cite{Muller:2020wjo, Hansen:2021ofl, Pang:2023jri}, but so far only the $a_1$~\cite{Mai:2021nul} and the $\omega$~\cite{Yan:2024gwp} mesons have been investigated directly from the lattice. For the first time, we go beyond the postdiction of well-established resonances and carry out exploratory spectroscopy on the lattice. The $\pi(1300)$ has so far only been investigated without treating it as a resonant state~\cite{Dudek:2010wm, McNeile:2006qy}; for a review, see Ref.~\cite{Kronfeld:2012uk}. In this Letter, we go beyond this by performing the first investigation of the $\pi(1300)$ using the full three-body finite-volume formalism. We also study it at two values of the pion mass, which allows us, using effective field theories, to obtain a first theoretical determination of its pole position at the physical point. For the schematic overview of the methodology, see~\cref{fig:Workflow} and the subsequent section; for further theoretical works, see Refs.~\cite{Steele:1997gh, Holl:2004fr, Wang:2007zzb, MartinezTorres:2011vh, Williams:2015cvx}.


\begin{figure}[t]
\centering
\resizebox{\linewidth}{!}{
\begin{tikzpicture}[
    >=Latex,
    font=\large,
    box/.style={draw,
        align=center,
        minimum height=1.cm}
]

\node[box, fill=green!40, rounded corners=15pt,text width=2cm] (chpt) at (0,6) {CHPT};
\node[box, fill=green!40, rounded corners=15pt,text width=5cm] (lqcd) at (10,6) {Lattice QCD};
\node[box, fill=Cerulean!20, text width=2cm] (tau) at (0,4) {$\tau(\vec{\ell}_r)$};
\node[box, fill=Cerulean!20, text width=4.cm] (cparam) at (3.5,4) {\scriptsize $\displaystyle C = c_c + \frac{c_p}{s - m_0^2} + \dots$};
\node[box, fill=Cerulean!20, text width=5cm] (spectra) at (10,4) {$\pi\pi$ \ \& \ \ $\pi\pi\pi$ spectra};
\node[box, fill=Goldenrod, text width=13.5cm, rounded corners=35pt,minimum height=2.5cm] (middle) at (5.7,1.3){
\begin{align*}
    &\det\!\left[B + C - \tau^{-1}E_L\right] = 0&&\text{\footnotesize \color{black}3-body quantization condition}\\
    &T_3 = (B + C) + \int_l (B + C)\frac{\tau}{2E_l} T_3&&\text{\footnotesize \color{black} 3-body scattering amplitude}
\end{align*}
};

\node[box, fill=CarnationPink!20, text width=13.5cm] (bottom) at (5.7,-1.5) {$\rho(770)$, $f_0(500)$, $\pi(1300)$ pole positions $(s^*$ $\in$ $\mathds{C})$};


\node[align=center,text width=1.5cm] at (13.5,6) {\small theory};
\node[align=center,text width=1.5cm] at (13.5,4) {\small input};
\node[align=center,text width=1.5cm] at (13.5,1.3) {\small amplitudes};
\node[align=center,text width=1.5cm] at (13.5,-1.5) {\small output};

\draw[->, thick,gray] (chpt) -- (tau);
\draw[->, thick,gray] (lqcd) -- (spectra);
\draw[->, thick,gray] (tau) -- (0,2.55);
\draw[->, thick,gray] (cparam) -- (3.5,2.55);
\draw[<->, thick,gray] (spectra) -- (10,2.55);
\draw[->, thick,gray] (middle) -- (bottom);
\end{tikzpicture}
}
\caption{Workflow of the pole position determination for two- and three-body systems from lattice QCD. Finite-volume spectra are determined from the lattice, while information about short range two- or three-body forces is augmented by chiral perturbation theory (CHPT).}
\label{fig:Workflow}
\end{figure}

\bigskip
\emph{Lattice QCD spectrum}---%
LQCD is formulated in a finite volume $L^3\times T$ in space-time discretized with a lattice spacing, denoted $a$. 
Monte Carlo simulations are performed to generate so-called gauge ensembles. In this Letter, we use four such ensembles with $N_{\mathrm{f}}=2+1$ dynamical quark flavors at a single value of the lattice spacing $a=0.07746(18)\,\mathrm{fm}$ generated by the Chinese LQCD (CLQCD) collaboration; see Ref.~\cite{CLQCD:2023sdb} for details.
The four ensembles (denoted as F32P21, F48P21, F32P30, and F48P30) feature two values of the pion mass $M_\pi \approx 305\ \mathrm{MeV}$ (P30) and $M_\pi \approx 208\ \mathrm{MeV}$ (P21), as well as two spatial volumes with $L=32$ (F32) and $L=48$ (F48), respectively.
In addition, we have $T=2L$ for the temporal extent, apart from ensemble F32P30, where $T=3L$.

\begin{figure}[b]
\centering
\begin{tabular}{ccccc}
    \raisebox{-0.5\height}{\includegraphics[width=0.18\linewidth]{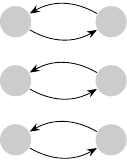}} & 
    \raisebox{-0.5\height}{\includegraphics[width=0.18\linewidth]{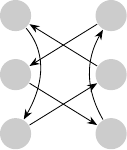}} & 
    \raisebox{-0.5\height}{\includegraphics[width=0.18\linewidth]{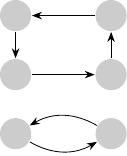}} & 
    \raisebox{-0.5\height}{\includegraphics[width=0.18\linewidth]{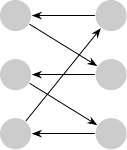}} & 
    \raisebox{-0.5\height}{\includegraphics[width=0.18\linewidth]{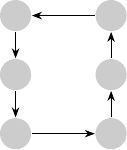}} \\
    TD & TS & TR & TZ & TB \\
    \raisebox{-0.5\height}{\includegraphics[width=0.18\linewidth]{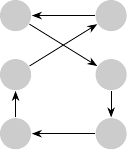}} & 
    \raisebox{-0.5\height}{\includegraphics[width=0.18\linewidth]{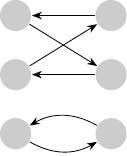}} & 
    \raisebox{-0.5\height}{\includegraphics[width=0.18\linewidth]{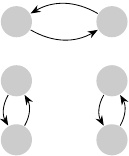}} & 
    \raisebox{-0.5\height}{\includegraphics[width=0.18\linewidth]{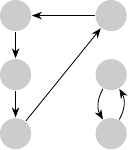}} & 
    \raisebox{-0.5\height}{\includegraphics[width=0.18\linewidth]{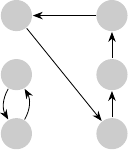}} \\
    TW & TE & TC & TYL & TYR 
\end{tabular}
\caption{Diagram topologies for $I=1$ $\pi\pi\pi \to \pi\pi\pi$. All diagrams with the same source and sink are permutations and recombinations of these topologies. The other topologies can be found in Supplemental Material~\cite{supp}.}
\label{fig:topologies}
\end{figure}

The discrete finite-volume spectra are extracted by diagonalizing a correlator matrix $C_{ij}(t) = \langle O_i(t) O_j(0) \rangle - \langle O_i\rangle\langle O_j\rangle$. 
Its elements are Euclidean two-point correlators of operators $O_i$.
By solving a generalized eigenvalue problem (GEVP)~\cite{Michael:1982gb,Luscher:1990ck,Blossier:2009kd,Fischer:2020bgv}, so-called principal correlators $\lambda_n(t)$ are determined, which decay as $\lambda_n(t)\sim \exp(-E_n t)$ for large enough $t$ values, such that $aE_n$ can be estimated. Details of the applied analysis procedure, including a discussion of the removal of so-called thermal pollutions, can be found in Supplemental Material~\cite{supp}. Because of the breaking of the rotational symmetry, the operators $O_i$ are constructed in irreducible representations (irreps) of the cubic group. It has been observed in the literature (for a review, see, e.g., Ref.~\cite{Mai:2022eur}) that the inclusion of multiparticle operators in $C(t)$ is essential for a reliable determination of the relevant lattice energy levels. Thus, we use the publicly available \verb|OpTion| package~\cite{Yan:2025jlq} to construct all relevant single- and multiparticle operators in the two-pion ($I=0,1,2$) and three-pion ($I=1$) channels. In the three-$\pi$ channel, for instance, we work in the $A_1^-$ irrep with isospin $I=1$ and appropriate relative momenta including three-$\pi$, $\rho\pi$, $\sigma\pi$ and single-$\pi$ operators.
The list of all operators is compiled in Supplemental Material~\cite{supp}.

For the contractions of the correlator matrix, we apply distillation~\cite{HadronSpectrum:2009krc}, with example contraction topologies shown in \cref{fig:topologies}. The other topologies and implementation details can be found in Supplemental Material~\cite{supp}. We note in passing that there are several hundred contractions needed in the three-pion sector, factorially more than in the two-particle channels.

\begin{figure}[t]
\centering
\includegraphics[width=\linewidth,trim=0.9cm 0.9cm 1.7cm 1.0cm,clip]{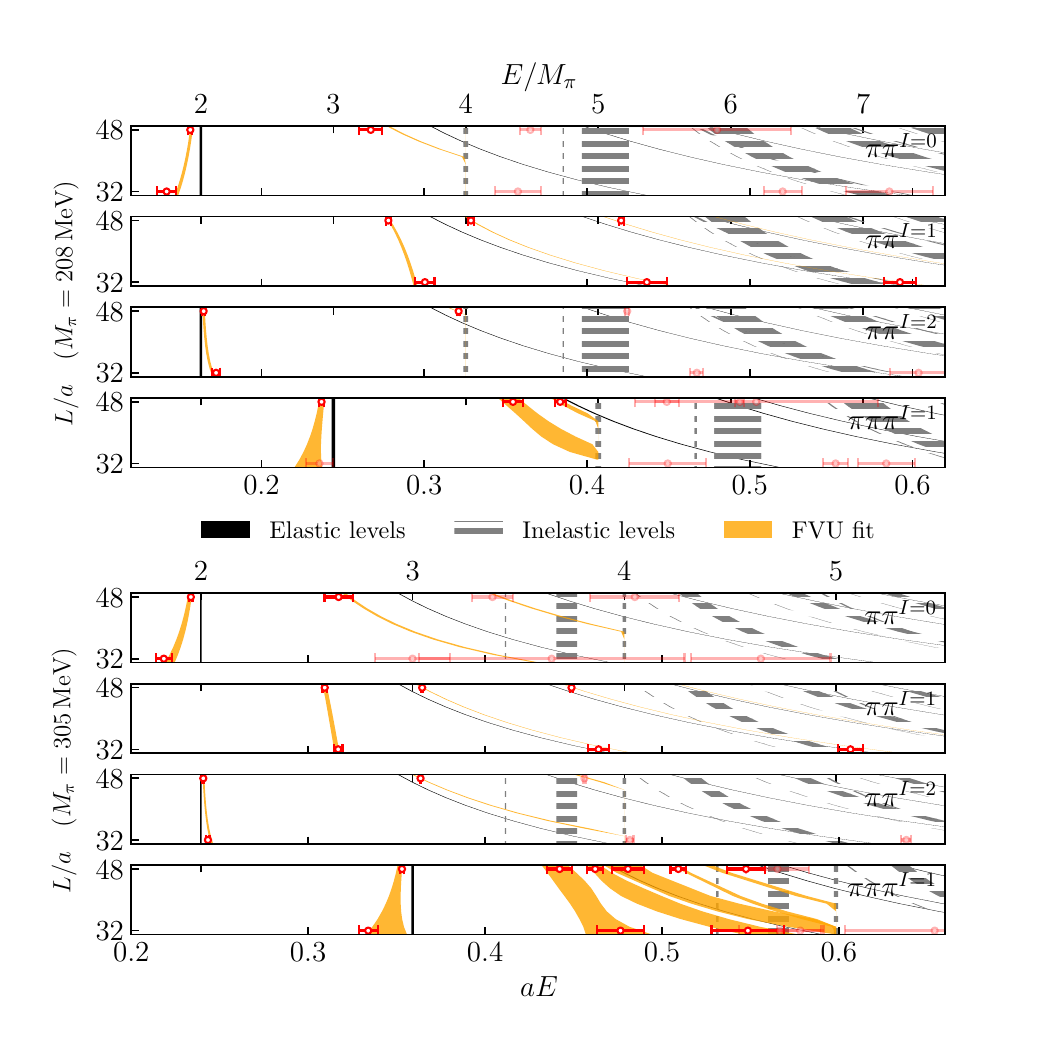}
\caption{
Finite-volume spectra for $I=0,1,2$ $\pi\pi$ and $I=1$ $\pi\pi\pi$ for heavy and light pion mass. Red points represent the interacting lattice energy levels, the faded ones of which are not included in the analysis. Solid bands are the noninteracting elastic levels, and dashed bands depict the inelastic levels. The orange bands are the solutions from the overall best fit.
}
\label{fig:spectra}
\end{figure}

The extracted lattice energy levels are shown in Fig.~\ref{fig:spectra} in lattice units (lower $x$ axis) and in units of $M_\pi$ (upper $x$ axis) as the red circles for the different ensembles and channels. 
In the two-pion sector, for $I=0,1$ the lowest $aE_n$ is below the $2M_\pi$ threshold, indicating attractive interaction, while $aE_n$ above this threshold is consistent with a small repulsive interaction in the $I = 2$ channel, as expected.
In the three-pion sector, a volume-independent ground-state level consistent with the pion mass is observed for both $M_{\pi}$ values, but omitted in the figure. 
Above the $3\pi$ threshold, simple identification of levels with states is not possible. We note, however, that there is a visible aggregation of states for $\pi\pi\pi(I=1)$ with $M_\pi\approx 305\,\MeV$ around and above the $4M_\pi$, indicating strong attraction possibly due to the presence of a resonance. This hypothesis will be tested below.

\bigskip
\emph{From LQCD spectra to scattering amplitudes}---%
The lattice spectra provide the first-principle information on the two- and three-body interactions. 
Practically, this information can be accessed through the so-called quantization condition, for which we employ the 
state-of-the-art finite-volume unitarity (FVU) three-body quantization conditions~\cite{Mai:2017bge}; see Refs.~\cite{Mai:2018djl, Mai:2019fba, Alexandru:2020xqf, Brett:2021wyd, Mai:2021nul, Garofalo:2022pux, Yan:2024gwp} for applications. As such it determines the finite-volume spectrum (fixed irrep, $A_1^-$) given two- and three-body short range interaction terms $K$ and $C$ as roots of
%
\begin{align}
    \det\left[(B+C)-\tau^{-1} E_L\right]=0\,,
    ~~
    \tau=(\tilde K^{-1}-\Sigma^{\rm FV})^{-1}
    \label{eq:QC-tilde}
\end{align}
%
with matrices of dimension corresponding to  ${\rm isospin}(I_2)\times{\rm helicity}(\lambda)\times {\rm spectator-momentum}(\bm{p}\in (2\pi)/L\,\mathds{Z}^3)$ space. For the quantum number $I^G(J^{PC})=1^-(0^{-+})$ in the center of mass system, this yields the following basis 
$\{\sigma(-\bm{p})\pi(\bm{p})$,
$\rho(\lambda=-1,-\bm{p})\pi(\bm{p})$, 
$\rho(\lambda= 0,-\bm{p})\pi(\bm{p})$, 
$\rho(\lambda=+1,-\bm{p})\pi(\bm{p})$,  
$G(-\bm{p})\pi(\bm{p})\}$. Here, $\sigma$, $\rho$, and $G$ serve as placeholders for two-pion subsystems of isospin $I_2=0,1,2$, respectively. Matrices $\Sigma^{\rm FV}$, $B$, and $E_L$ only depend on momenta and masses, representing all possible on-shell configurations of three pions, while the two- or three-body forces ($\tilde K^{-1}/C$) contain the physical information about a particular channel~\cite{Mai:2021nul, Yan:2024gwp, Feng:2024wyg}. We note that the roots of $\tau^{-1}$ determine the two-body energy eigenvalues in the pertinent isospin channel.

In the two-body sector, each finite-volume energy eigenvalue determines one value of $\tilde K^{-1}$ (neglecting higher partial waves, below inelastic channels). However, in a three-particle system, one particle (spectator) can take certain momentum away, necessitating a detailed understanding of the functional behavior of $C$ and $K^{-1}$. Little is known about the form of the three-body force for this system, forcing us to pick a generic form for a transition $\alpha \overset{C}{\to} \beta$ in the orbital angular momentum basis~\cite{Chung:1971ri}
\begin{align}
    c^{\alpha\beta}=c_c^{\alpha\beta}+\frac{c_p^{\alpha\beta}}{s-m_0^2}\,,
    \quad
    {\alpha=\beta}\in \{\sigma\pi,\rho\pi\} \,,
    \label{eq:C-term}
\end{align}
where $s$ is the total three-body energy squared, and only the phenomenologically most relevant channels are turned on. In contrast to this, for the relevant two-body subsystem, we can take advantage of many existing theoretical studies and relate its dynamics directly to phase shifts as 
\begin{align}
    &\tilde K^{-1}_{I_2}(\sigma)=K^{-1}_{I_2}(\sigma)+\Re\,\Sigma^{\rm IV}_{I_2}(\sigma),\nonumber\\
    &K^{-1}_{I_2}(\sigma)=
    \begin{cases}
        -\frac{p(\sigma)\cot\delta^{I_20}(\sigma)}{16\pi\sqrt{\sigma}}\,,
        \quad \text{for } I_2=0,2,\\
        -\frac{p^3(\sigma)\cot\delta^{11}(\sigma)}{12\pi\sqrt{\sigma}}\,, \quad \text{for } I_2=1.
    \end{cases}
    \label{eq:Ktilde+SigmaIV}
\end{align}
Here, $\sigma$, $p$, and $\Sigma^{\rm IV}$ denote two-body invariant mass squared, two-body center-of-mass momentum and two-body self-energy integral, respectively. For explicit (now standard) formulas, see Refs.~\cite{Feng:2024wyg,Yan:2024gwp}. To parametrize the phase shifts, we adopt a well-tested~\cite{Mai:2019pqr,Doring:2016bdr,Acharya:2015pya} modified inverse amplitude method (mIAM) approach~\cite{Hanhart:2008mx} that synthesizes S-matrix theory and chiral perturbation theory (CHPT). The amplitude behaves correctly (including Adler zero) deep below the two-body threshold, which is needed for the three-body formalism. However, at some very low $\sigma$, it is unrealistic, mitigated here by $K^{-1}_{I_2}(\sigma)\mapsto K^{-1}_{I_2}(\sigma_{\rm MP})e^{-\sigma+\sigma_{\rm MP}}$, effectively turning the interaction smoothly to zero below the matching point $\sigma_{\rm MP}$. The latter, together with the largest spectator momentum $|l_{\rm max}|$ and the largest self-energy $|k_{\rm max}|$, represents all necessary cutoffs of the three-body formalism. The procedure outlined above is depicted graphically in \cref{fig:Workflow}.

\begin{figure}[b]
    \centering
    \includegraphics[width=\linewidth]{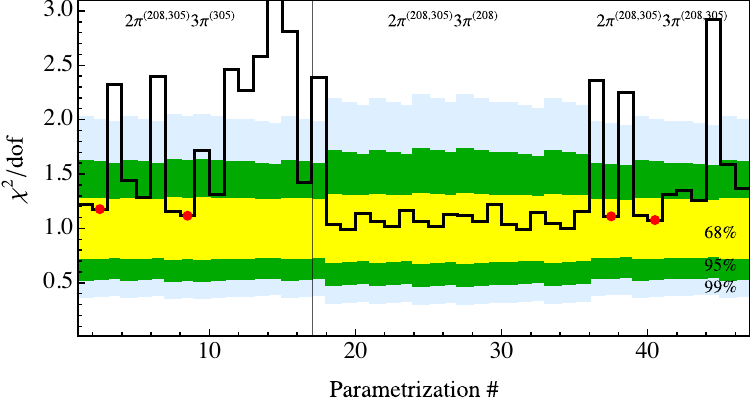}
    \caption{Quality of the 47 central fit results. The used LQCD input is marked as $2\pi^{(M_\pi)}$ and $3\pi^{(M_\pi)}$, respectively. The $68\%$, $95\%$, and $99\%$ confidence intervals around $\mathbb{E}[\chi^2]$ of the $\chi^2$ distribution are shown by the shaded bands; dots mark our best fits, used to determine the pole positions in \cref{fig/main:poles}.}
    \label{fig/main:fit-quality-all-fits}
\end{figure}

\bigskip
\emph{Analysis and results}---%
When applying the three-body quantization to finite-volume spectra, certain choices [e.g., parametrization in \cref{eq:C-term}, or cutoffs] need to be made. To assess the systematics associated with this, we have tested $\sim 2000$ different scenarios, all of which are described in Supplemental Material~\cite{supp}.
We found that varying cutoffs from $\sigma_{\rm MP}=1M_\pi^2$, $|l_{max}|=\sqrt{3}\frac{2\pi}{aL}$, $|k_{max}|=\sqrt{3\cdot5^2} \frac{2\pi}{aL}\}$ does not have any significant impact on our fit/final results. This leaves us with 47 main fits with their quality shown in \cref{fig/main:fit-quality-all-fits}. The quality of the overall best fit ($\chi^2/{\rm dof}=1.08$, see parameter table in Supplemental Material~\cite{supp}) is contrasted with the available finite-volume spectra in \cref{fig:spectra}.

We observe that, when only light pion mass three-body input (F32/48P21) is considered, nearly all types of parametrizations of $C$ lead to excellent fit quality within the $68\%$ confidence interval of the $\chi^2$ distribution. The likely reason for this is that there are only very few energy levels below the relevant threshold (see \cref{fig:spectra}).
Besides this, we consistently observe that a slightly better $\chi^2/{\rm dof}$ is obtained for mIAM3. We, therefore, only use mIAM3 fits in the following to avoid overcounting similar models. 

The parameters obtained in the fit to the finite-volume spectra are used in an infinite-volume three-body formalism (IVU)~\cite{Sadasivan:2021emk, Garofalo:2022pux, Dawid:2023jrj, Yan:2024gwp, Feng:2024wyg}, allowing us to extract universal parameters of hadron resonances; see \cref{fig:Workflow} for the general workflow. This requires solving an integral equation, accomplished efficiently using the complex contour method. For details and comparisons between methods, see Refs.~\cite{Feng:2024wyg, Sakthivasan:2024uwd, Doring:2025sgb}. We note that integrating over the momentum of the spectator up to a certain cutoff defines the maximal energy range up to which the amplitude fulfills unitarity, i.e., $\sqrt{s_{\rm max}}=\sqrt{\sigma+l_{\rm max}^2}+\sqrt{M_\pi^2+l_{\rm max}^2}|_{\sigma=4M_\pi^2}$. On the other hand, for a fixed $s$, larger cutoff values can lead to $\sigma\ll\sigma_{\rm MP}$, rendering the two-body input unrealistic. We observe that our cutoff choice maximizes the range of applicability of the methodology.

For the fits to light pion mass data only, we observe that only a few of the fits lead to a $\pi(1300)$ pole. This, but also a very sparse finite-volume spectrum in \cref{fig:spectra}, indicates that $\pi(1300)$ is weakly pion mass dependent with a pole expected at $M_{\pi(1300)}\approx 1300/208\gtrsim 6 M_\pi$---well outside of the energy range that can be tackled by current lattice and finite-volume methods. Luckily, this is not the case for the heavy pion mass $M_{\pi(1300)}\approx 1300/305\approx 4.2 M_\pi$, such that all fits in $68\%$, $95\%$, and $99\%$ confidence intervals around the observed $\chi^2_{\rm obs}$ indeed lead to a $\pi(1300)$ pole. Interestingly, these fits include also those with no explicit pole term in the parametrization of the three-body force, thus generating a $\pi(1300)$ pole dynamically.

\begin{figure}
    \centering
    \includegraphics[width=\linewidth]{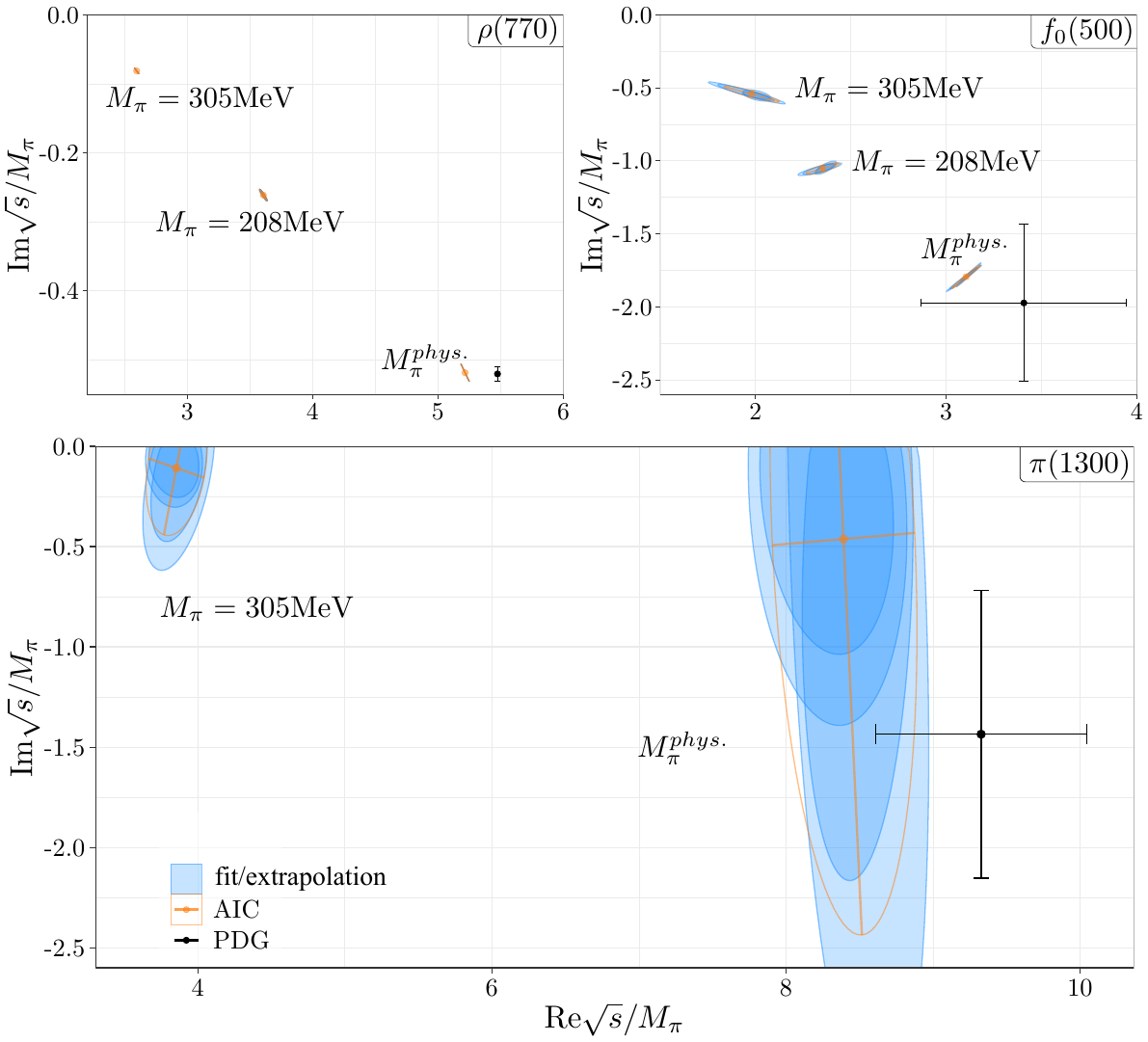}
    \caption{Pole positions on the second Riemann sheet of the two- and three-hadron systems with total energy $E\in\mathds{C}$, respectively, to $\rho(770),f_0(500),\pi(1300)$ states. Shaded areas represent results from individual fits in the $68\%$ confidence interval around the observed $\chi^2_{\rm obs}$. Orange crosses depict the AIC model-averaged results. PDG results are provided by the black crosses for comparison.
    }
    \label{fig/main:poles}
\end{figure}

Taking into account fits in the $68\%$ confidence interval around $\mathbb{E}[\chi^2]$ of the $\chi^2$ distribution, which include heavy three-body input (red dots in \cref{fig/main:fit-quality-all-fits}), we extract pole positions depicted in \cref{fig/main:poles}, including the pole positions of the two-pion resonances in the isovector and isoscalar subsystems. Extrapolating to the physical point in the two-body sector is achieved by setting $M_\pi$ to its physical value as dictated by CHPT. In the three-body case, the simplest ansatz (see above) is to use parameters obtained from fits to the heavy pion mass data, assuming negligible pion mass dependence of all terms in \cref{eq:C-term} evaluated in physical/lattice units. Results of individual fits and a model average of those using the Akaike information criterion~\cite{Akaike} are depicted in \cref{fig/main:poles}. The model average result constitutes our principal finding
\begin{align}
    M_{\pi(1300)}&=(1169\pm46)-i(62\pm169)\,\MeV\,,\nonumber\\
    M_{\rho(770)}&=(727\pm3)-i(72\pm1)\,\MeV\,,\\
    M_{f_{0}(500)}&=(433\pm 7)-i(250\pm 7)\,\MeV\,.\nonumber
\end{align}
Our results for the two-body resonances agree quite well with the PDG values~\cite{ParticleDataGroup:2024cfk}, except for $\Re M_{\rho(770)}$. 
Since our result for $\Re M_\rho(770)$ is consistent with previous lattice studies on these but also other ensembles~\cite{Mai:2019pqr, Yan:2024gwp}, we attribute this discrepancy to finite lattice spacing effects, discussed in Ref.~\cite{Wang:2025hew} using the same lattice configurations. The agreement in all other results indicates (a similar observation was made recently in Ref.~\cite{Dawid:2025zxc}) that the three-body data actually constrain the two-body sector, for which larger uncertainties would be expected otherwise.

Regarding the main goal and novel result of this Letter---the prediction of the parameters of the excited state of the pion from QCD---we observe agreement with the PDG values within $1\sigma$ uncertainties. We emphasize that, for all extracted pole positions $\Im\, M_{\pi(1300)}<0$, while the extracted errors are unconstrained due to the simplistic Gaussian statistics employed here.

\bigskip
\emph{Conclusion}---%
We have determined the resonance parameter of the $\pi(1300)$ from QCD through a lattice calculation of finite-volume spectra including one-, two-, and three-hadron operators, and by utilizing state-of-the-art three-particle quantization conditions. Our light pion three-body spectra do not provide sufficient input for a reliable extraction of resonance parameters. At the heavier pion mass, all fits lead to a scattering amplitude featuring a resonance, effectively ruling out the no-resonance scenario. With these results and the simplest possible ansatz for the quark-mass dependence of the three-body force, the resonance parameters at the physical point are extracted and found to be comparable with PDG values.

Various sources of systematics have been tested, such as different parametrizations of the two-/three-body input or cutoff dependencies, which only mildly affect our main conclusions. Still, in the future, it would be interesting to map out the quark-mass dependence in more detail, for example, at even heavier pion mass values. Additionally, providing line shapes and Dalitz plots would be of great interest for comparing them to phenomenological studies of multihadron.

\noindent Software \verb|QUDA|~\cite{Clark:2009wm, Babich:2011np, Clark:2016rdz} is used to solve the perambulators, as performed in Refs.~\cite{Yan:2024yuq, Yan:2023gvq}.

\bigskip
\noindent\emph{Acknowledgments}---
H.~Y. is grateful to Y.~Chen and Z.~Zhang for helpful discussions. H.~Y. and M.~M. thank C.~Culver for helpful discussions. H.~Y. acknowledges support from NSFC under Grant No.~124B2096.
H.~Y., C.~L., and L.~L. acknowledge support from NSFC under Grants No.~12293060, No.~12293061, No.~12293063, No.~12175279, and No.~11935017.
The work of M.~M. was further funded through the Heisenberg Programme by the Deutsche Forschungsgemeinschaft (DFG, German Research Foundation)---No.~532635001.
This project was funded by the Deutsche Forschungsgemeinschaft (DFG, German Research Foundation) as part of the CRC 1639 NuMeriQS---Project No.~511713970 and by the MKW NRW under the funding code No.~NW21-024-A. 
The work of U.-G.~M. was supported in part by the CAS President's International Fellowship Initiative (PIFI) (Grant No.~2025PD0022). The work of M.~D. and Y.~F. is supported by the National Science Foundation under Grant No.~PHY-2310036.
The gauge configurations are generated on the HPC Cluster of ITP-CAS, the Southern Nuclear Science Computing Center (SNSC), the Siyuan-1 cluster supported by the Center for High Performance Computing at Shanghai Jiao Tong University, and the Dongjiang Yuan Intelligent Computing Center.
Part of the simulations were performed on the High-performance Computing Platform of Peking University and the Southern Nuclear Science Computing Center (SNSC).

\bigskip
\noindent\emph{Data availability}---
The data that support the findings of this article are openly available~\cite{data}.

\bibliography{BIB}

\clearpage
\appendix
\onecolumngrid
\setcounter{equation}{0}
\setcounter{figure}{0}
\setcounter{table}{0}
\makeatletter
\renewcommand{\theequation}{S\arabic{equation}}
\renewcommand{\thefigure}{S\arabic{figure}}
\renewcommand{\thetable}{S\arabic{table}}
\setcounter{secnumdepth}{2}

\section*{Supplemental Material}
\subsection{Operator constructions}
\label{SUPP/SEC:operators}
The operators used to interpolate the single-hadron, two-body, and three-body systems are constructed from linear combinations of quark bilinears. We employ only local, light-flavor operators of the form $O = \bar{l}^{\prime} \Gamma l$, where $l, l' \in {u, d}$ and $\Gamma$ determines the Dirac quantum number. We define the elementary building blocks $\sigma^{l} = \bar{l} \mathds{1} l$, $\rho_i^{l} = \bar{l} \gamma_i l$, $\pi^{l} = \bar{l} \gamma_5 l$, $\rho^+_i/\pi^+ = \bar{d} \gamma_{i/5} u$ and $\rho^-_i/\pi^- = \bar{u} \gamma_{i/5} d$.

The operator basis is constructed to capture all relevant non-interacting levels below the inelastic thresholds: $\bar{K}K$ and $\eta\eta$ for the $\pi\pi$ channel, and $\bar{K}K\pi$, $\eta\eta\pi$, and $5\pi$ for the $\pi\pi\pi$ channel. In the $\pi\pi$ channel, two-pion operators are constructed for all isospins, along with $\sigma$ for the $I = 0$ channel and $\rho$ for the $I = 1$ channel. For the $\pi\pi\pi$ channel, we focus on the $A_1^-$ irrep and isospin $I = 1$, which includes the usual pion and the $\pi(1300)$. We construct single-pion operators, two-body $\rho\pi$ (dominated by $P$-wave) and $\sigma\pi$ (dominated by $S$-wave) operators, as well as operators that resemble three-pion states with the appropriate relative momenta.

\subsubsection{Isospin construction}

Operators are first constructed in the isospin space. The one-meson operators are straightforward, e.g., the $\sigma$-meson for $I=0$ $\pi\pi$ system is related to $\frac{1}{\sqrt{2}} \left[ \sigma^u + \sigma^d \right]$. For two-body systems like $\pi\pi, \sigma\pi$ and $\rho\pi$, the isospin decomposition is $3 \otimes 3 = 1 \oplus 3 \oplus 5$. The corresponding operators, which resemble two particles, are
\begin{equation}
\begin{cases}
    O_{\rho\pi}^{I=2, I_z=2} &= \rho^+ \pi^+, \\
    O_{\rho\pi}^{I=1, I_z=1} &= \frac{1}{2} \left[ -\rho^+ \pi^u + \rho^+ \pi^d + \rho^u \pi^+ - \rho^d \pi^+ \right], \\
    O_{\rho\pi}^{I=0, I_z=0} &= -\frac{1}{\sqrt{3}} \left[ \rho^+ \pi^- + \rho^- \pi^+ + \frac{1}{2} [\rho^u \pi^u - \rho^u \pi^d - \rho^d \pi^u + \rho^d \pi^d] \right], \\
    O_{\sigma\pi}^{I=1, I_z=1} &= -\frac{1}{\sqrt{2}} \left[ \sigma^u \pi^+ + \sigma^d \pi^+ \right], \\
    O_{\sigma\pi}^{I=0, I_z=0} &= \frac{1}{2} \left[ \sigma^u \pi^u - \sigma^u \pi^d + \sigma^d \pi^u - \sigma^d \pi^d \right].
\end{cases}
\end{equation}
The $\pi\pi$ operators share the same flavor structure as the $\rho\pi$ operators listed above and are, therefore, not repeated.

For the three-pion system, the isospin decomposition is
\begin{equation}
    3 \otimes 3 \otimes 3 = (1 \oplus 3 \oplus 5) \otimes 3 = 1 \oplus 3^3 \oplus 5^2 \oplus 7.
\end{equation}
The multiplicity of three in the $I=1$ channel implies that up to three independent flavor structures can be constructed, distinguished by the intermediate isospin $I_{12}$ of the first two pions. Corresponding operators are
\begin{equation}
\begin{cases}
    O_{\pi\pi\pi}^{I=1, I_z=1, I_{12}=0} &= \frac{1}{\sqrt{12}} \left[ 2 \pi^+\pi^-\pi^+ + \pi^u\pi^u\pi^+ - \pi^d\pi^u\pi^+ - \pi^u\pi^d\pi^+ + \pi^d\pi^d\pi^+ + 2 \pi^-\pi^+\pi^+ \right], \\
    O_{\pi\pi\pi}^{I=1, I_z=1, I_{12}=1} &= \frac{1}{4} \left[ - \pi^+\pi^u\pi^u + \pi^+\pi^d\pi^u + \pi^+\pi^u\pi^d - \pi^+\pi^d\pi^d + \pi^u\pi^+\pi^u \right. \\
    & \qquad \left. - \pi^d\pi^+\pi^u - \pi^u\pi^+\pi^d + \pi^d\pi^+\pi^d - 2 \pi^+\pi^-\pi^+ + 2 \pi^-\pi^+\pi^+ \right], \\
    O_{\pi\pi\pi}^{I=1, I_z=1, I_{12}=2} &= \frac{1}{\sqrt{240}} \left[ 12 \pi^+\pi^+\pi^- + 3 \pi^+\pi^u\pi^u - 3 \pi^+\pi^d\pi^u - 3 \pi^+\pi^u\pi^d + 3 \pi^+\pi^d\pi^d \right. \\
    & \qquad \left. + 3 \pi^u\pi^+\pi^u - 3 \pi^d\pi^+\pi^u - 3 \pi^u\pi^+\pi^d + 3 \pi^d\pi^+\pi^d + 2 \pi^+\pi^-\pi^+ \right. \\
    & \qquad \left. - 2 \pi^u\pi^u\pi^+ + 2 \pi^d\pi^u\pi^+ + 2 \pi^u\pi^d\pi^+ - 2 \pi^d\pi^d\pi^+ + 2 \pi^-\pi^+\pi^+ \right]. \\
\end{cases}
\end{equation}
The physical multiplicity of independent operators depends on the momentum configuration. For all pions at rest ($p_1^2 = p_2^2 = p_3^2 = 0$), $O_{\pi\pi\pi}^{I=1, I_z=1, I_{12}=0} = O_{\pi\pi\pi}^{I=1, I_z=1, I_{12}=2}$ and $O_{\pi\pi\pi}^{I=1, I_z=1, I_{12}=1} = 0$, leaving one independent operator. For the momentum configuration $p_1^2 = p_2^2 = 1, p_3^2 = 0$, $O_{\pi\pi\pi}^{I=1, I_z=1, I_{12}=0} \neq O_{\pi\pi\pi}^{I=1, I_z=1, I_{12}=2} \neq 0$ and $O_{\pi\pi\pi}^{I=1, I_z=1, I_{12}=1} = 0$, resulting in two independent operators.

In this procedure, the charge parity and $G$ parity of the operators are automatically projected onto the corresponding quantum numbers of the system.

\subsubsection{Momentum projection}
The operators are projected onto irreducible representations (irreps) of the cubic group using a publicly available package, \verb|OpTion|~\cite{Yan:2025jlq}. The explicit forms of the operators in the $A_1^-$ irrep for the $I=1$ $\pi\pi\pi$ channel are
\begin{equation}
\begin{cases}
    O_{\pi_1} &= \pi(0), \\
    O_{(\rho\pi)_1} &= \rho_x(-e_x)\pi(e_x) - \rho_x(e_x)\pi(-e_x) + \rho_y(-e_y)\pi(e_y) - \rho_y(e_y)\pi(-e_y) + \rho_z(-e_z)\pi(e_z) - \rho_z(e_z)\pi(-e_z), \\
    O_{(\sigma\pi)_1} &= \sigma(0)\pi(0), \\
    O_{(\sigma\pi)_2} &= \sigma(-e_x)\pi(e_x) + \sigma(e_x)\pi(-e_x) + \sigma(-e_y)\pi(e_y) + \sigma(e_y)\pi(-e_y) + \sigma(-e_z)\pi(e_z) + \sigma(e_z)\pi(-e_z), \\
    O_{(\sigma\pi)_3} &= \sigma(e_{-x,-y})\pi(e_{x,y}) + \sigma(e_{-x,y})\pi(e_{x,-y})+ \sigma(e_{-x,-z})\pi(e_{x,z}) + \sigma(e_{-x,z})\pi(e_{x,-z}) \\
    & \quad + \sigma(e_{x,-y})\pi(e_{-x,y}) + \sigma(e_{x,y})\pi(e_{-x,-y}) + \sigma(e_{x,-z})\pi(e_{-x,z}) + \sigma(e_{x,z})\pi(e_{-x,-z}) \\
    & \quad + \sigma(e_{-y,-z})\pi(e_{y,z}) + \sigma(e_{-y,z})\pi(e_{y,-z}) + \sigma(e_{y,-z})\pi(e_{-y,z}) + \sigma(e_{y,z})\pi(e_{-y,-z}), \\
    O_{(\pi\pi\pi)_1} &= \pi(0)\pi(0)\pi(0), \\
    O_{(\pi\pi\pi)_2} &= \pi(e_x)\pi(-e_x)\pi(0) + \pi(-e_x)\pi(e_x)\pi(0) + \pi(e_y)\pi(-e_y)\pi(0) \\
    & \quad + \pi(-e_y)\pi(e_y)\pi(0) + \pi(e_z)\pi(-e_z)\pi(0) + \pi(-e_z)\pi(e_z)\pi(0).
\end{cases}
\end{equation}

The operators in the $A_1^+$ irrep for the $I=0$ $\pi\pi$ channel are
\begin{equation}
\begin{cases}
    O_{\sigma_1} &= \sigma(0), \\
    O_{(\pi\pi)_1} &= \pi(0)\pi(0), \\
    O_{(\pi\pi)_2} &= \pi(-e_x)\pi(e_x) + \pi(e_x)\pi(-e_x) + \pi(-e_y)\pi(e_y) + \pi(e_y)\pi(-e_y) + \pi(-e_z)\pi(e_z) + \pi(e_z)\pi(-e_z), \\
    O_{(\pi\pi)_3} &= \pi(e_{-x,-y})\pi(e_{x,y}) + \pi(e_{-x,y})\pi(e_{x,-y})+ \pi(e_{-x,-z})\pi(e_{x,z}) + \pi(e_{-x,z})\pi(e_{x,-z}), \\
\end{cases}
\end{equation}
The operators for the $A_1^+$ irrep in the $I=2$ $\pi\pi$ channel share the same momentum structure as $O_{(\pi\pi){1,2,3}}$, but exclude the single-body $O{\sigma_1}$.

The operators in the $T_1^-$ irrep for the $I=1$ $\pi\pi$ channel are
\begin{equation}
\begin{cases}
    O_{\rho_1} &= \rho_z(0), \\
    O_{(\pi\pi)_1} &= \pi(e_x)\pi(-e_x) - \pi(-e_x)\pi(e_x), \\
    O_{(\pi\pi)_2} &= \pi(e_{x,z})\pi(e_{-x,-z}) + \pi(e_{-x,z})\pi(e_{x,-z}) + \pi(e_{y,z})\pi(e_{-y,-z}) + \pi(e_{-y,z})\pi(e_{y,-z}) \\
    & \quad - \pi(e_{x,-z})\pi(e_{-x,z}) - \pi(e_{-x,-z})\pi(e_{x,z}) - \pi(e_{y,-z})\pi(e_{-y,z}) - \pi(e_{-y,-z})\pi(e_{y,z}).
\end{cases}
\end{equation}

The operators used in this work are cataloged in Tab.~\ref{tab:operators} for easy reference. They can be universally represented as
\begin{equation}
\begin{cases}
    O_{\text{one}} = \sum_{i} \eta_i \Gamma_{\mu_i}(0), \\
    O_{\text{two}} = \sum_{i} \eta_i \Gamma_{\mu_i}(\vec{p}_i) \pi(-\vec{p}_i), \\
    O_{\text{three}} = \sum_{i} \eta_i \Gamma_{\mu_i}(\vec{p}_{i1}) \pi(\vec{p}_{i2}) \pi(-\vec{p}_{i1}-\vec{p}_{i2}),
\end{cases}
\end{equation}
where $\Gamma_{\mu_i}$ is a quark bilinear with $\mu_i \in { 0, 5, x, y, z }$ corresponding to the Dirac matrices $\gamma_0, \gamma_5, \gamma_x, \gamma_y, \gamma_z$. Each operator is uniquely identified by its coefficients $\eta_i$, Dirac indices $\mu_i$, and momenta $\vec{p}_i$.

For compactness, the parameters are denoted as $\eta_{\mu_i}^{\alpha_{i1}(;\alpha_{i2})}$, where the superscript $\alpha$ encodes the momentum vector. A single direction (e.g., $x$) denotes one unit of momentum in that direction ($\vec{p} = [100]$). Combined directions (e.g., $yz$) denote $\vec{p} = [011]$, and a prefactor (e.g., $-2x$) denotes $\vec{p} = [-200]$.

\begin{table*}[htbp]
\centering
\caption{Catalog of interpolating operators. The operator basis for each channel (defined by isospin $I$ and cubic group irrep) includes single-meson ($\pi$, $\rho$, $\sigma$), two-meson ($\pi\pi$, $\rho\pi$, $\sigma\pi$), and three-meson ($\pi\pi\pi$) operators. Parameters $\eta_{\mu_i}^{\alpha_{i1}(;\alpha_{i2})}$ encode the Dirac structure and momenta as defined in the text; overall constants are suppressed.}
\addtolength{\tabcolsep}{6pt}
\begin{tabular}{cccc}
\toprule
channel & isospin & type & operator \\
\midrule
\multirow{13}{*}{$\pi\pi$} & \multirow{5}{*}{$I=0$} & \multirow{1}{*}{one} & $(+1)^{0}_{0}$ \\
\cmidrule(lr){3-4}
& & \multirow{4}{*}{two} & $(+1)^{0}_{0}$ \\
\cmidrule(lr){4-4}
& & & $(+1)^{x}_{5}, (+1)^{-x}_{5}, (+1)^{y}_{5}, (+1)^{-y}_{5}, (+1)^{z}_{5}, (+1)^{-z}_{5}$ \\
\cmidrule(lr){4-4}
& & & $(+1)^{yz}_{5}, (+1)^{xz}_{5}, (+1)^{xy}_{5}, (+1)^{-y,z}_{5}, (+1)^{-x,z}_{5}, (+1)^{-x,y}_{5},$ \\
& & & $ (+1)^{y,-z}_{5}, (+1)^{x,-z}_{5}, (+1)^{x,-y}_{5}, (+1)^{-y,-z}_{5}, (+1)^{-x,-z}_{5}, (+1)^{-x,-y}_{5}$ \\
\cmidrule(lr){2-4}
& \multirow{4}{*}{$I=1$} & \multirow{1}{*}{one} & $(+1)^{0}_{z}$ \\
\cmidrule(lr){3-4}
& & \multirow{3}{*}{two} & $(+1)^{z}_{5}, (-1)^{-z}_{5}$ \\
\cmidrule(lr){4-4}
& & & $(+1)^{xz}_{5}, (+1)^{-x,z}_{5}, (+1)^{yz}_{5}, (+1)^{-y,z}_{5},$ \\
& & & $ (-1)^{x,-z}_{5}, (-1)^{-x,-z}_{5}, (-1)^{y,-z}_{5}, (-1)^{-y,-z}_{5}$ \\
\cmidrule(lr){2-4}
& \multirow{4}{*}{$I=2$} & \multirow{4}{*}{two} & $(+1)^{0}_{0}$ \\
\cmidrule(lr){4-4}
& & & $(+1)^{x}_{5}, (+1)^{-x}_{5}, (+1)^{y}_{5}, (+1)^{-y}_{5}, (+1)^{z}_{5}, (+1)^{-z}_{5}$ \\
\cmidrule(lr){4-4}
& & & $(+1)^{yz}_{5}, (+1)^{xz}_{5}, (+1)^{xy}_{5}, (+1)^{-y,z}_{5}, (+1)^{-x,z}_{5}, (+1)^{-x,y}_{5},$ \\
& & & $ (+1)^{y,-z}_{5}, (+1)^{x,-z}_{5}, (+1)^{x,-y}_{5}, (+1)^{-y,-z}_{5}, (+1)^{-x,-z}_{5}, (+1)^{-x,-y}_{5}$ \\
\midrule
\multirow{9}{*}{$\pi\pi\pi$} & \multirow{9}{*}{$I=1$} & \multirow{1}{*}{one} & $(+1)^{0}_{5}$ \\
\cmidrule(lr){3-4}
& & \multirow{5}{*}{two} & $(+1)^{-x}_{x}, (-1)^{x}_{x}, (+1)^{-y}_{y}, (-1)^{y}_{y}, (+1)^{-z}_{z}, (-1)^{z}_{z}$ \\
\cmidrule(lr){4-4}
& & & $(+1)^{0}_{0}$ \\
\cmidrule(lr){4-4}
& & & $(+1)^{-x}_{0}, (+1)^{x}_{0}, (+1)^{-y}_{0}, (+1)^{y}_{0}, (+1)^{-z}_{0}, (+1)^{z}_{0}$ \\
\cmidrule(lr){4-4}
& & & $(+1)^{yz}_{0}, (+1)^{xz}_{0}, (+1)^{xy}_{0}, (+1)^{-y,z}_{0}, (+1)^{-x,z}_{0}, (+1)^{-x,y}_{0},$ \\
& & & $ (+1)^{y,-z}_{0}, (+1)^{x,-z}_{0}, (+1)^{x,-y}_{0}, (+1)^{-y,-z}_{0}, (+1)^{-x,-z}_{0}, (+1)^{-x,-y}_{0}$ \\
\cmidrule(lr){3-4}
& & \multirow{3}{*}{three} & $(+1)^{0;0}_{5}$ \\
\cmidrule(lr){4-4}
& & & $(+1)^{x;-x}_{5}, (+1)^{-x;x}_{5}, (+1)^{y;-y}_{5}, (+1)^{-y;y}_{5}, (+1)^{z;-z}_{5}, (+1)^{-z;z}_{5}$ \\
\cmidrule(lr){4-4}
& & & $(+1)^{x;-x}_{5}, (+1)^{-x;x}_{5}, (+1)^{y;-y}_{5}, (+1)^{-y;y}_{5}, (+1)^{z;-z}_{5}, (+1)^{-z;z}_{5}$ \\
\bottomrule
\end{tabular}
\addtolength{\tabcolsep}{-6pt}
\label{tab:operators}
\end{table*}

\subsection{Contraction details and topologies of the diagrams}
\label{SUPP/SEC:topologies}
In this work, we employ the distillation method~\cite{HadronSpectrum:2009krc} to enable efficient all-to-all evaluation of quark propagators. The distillation method is constructed using the three-dimensional lattice Laplacian operator:
\begin{equation}
    \sum_{i=1}^3 U_i(\vec{x}, t) \delta_{\vec{x}+\hat{i}, \vec{y}} + U_i^{\dagger}(\vec{x}-a\hat{i}, t) \delta_{\vec{x}-\hat{i}, \vec{y}} - 6 \delta_{\vec{x}, \vec{y}}.
\end{equation}
By retaining only the lowest $N_v$ eigenmodes $V(t)$, one could define the distillation operator $\square(t) = V(t) V^{\dagger}(t)$. Applying this operator to a quark field, $\square_{\vec{x} \vec{y}}(t) \psi(\vec{y},t)$, effectively smears the field and suppresses high-momentum modes. 

Within this framework, the correlation function contractions factorize into combinations of momentum-projected eigenmode products $\sum_{\vec{x}} \mathrm{e}^{-i \vec{p} \cdot \vec{x}} V^{\dagger}(t) V(t)$ and perambulators:
\begin{equation}
    \tau_{\alpha \beta}(t_1, t_2) = V^{\dagger}(t_1) M^{-1}_{\alpha \beta}(t_1, t_2) V(t_2).
\end{equation}
The key computational advantage is that the required matrix inversions scale only with the number of eigenmodes $N_v$, which is significantly smaller than the full spatial volume needed for exact all-to-all propagators. Throughout this study, we consistently employ $N_v = 200$ eigenmodes for the larger volume ($L=48$) and $N_v = 100$ for the smaller volume ($L=32$). To enhance statistical precision, we perform calculations using all available time sources across all gauge configurations.

\begin{figure}[t]
\centering
\begin{tabular}{c}
    \raisebox{-0.5\height}{\includegraphics[width=0.12\linewidth]{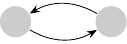}} \\
    D \\
\end{tabular}
\caption{Topologies of the diagrams for one-one-type contractions.}
\label{fig:topologies-1-1}
\end{figure}

\begin{figure}[t]
\centering
\begin{tabular}{cccccc}
    \raisebox{-0.5\height}{\includegraphics[width=0.12\linewidth]{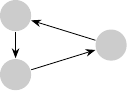}} & \quad
    \raisebox{-0.5\height}{\includegraphics[width=0.12\linewidth]{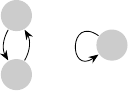}} & \quad
    \raisebox{-0.5\height}{\includegraphics[width=0.12\linewidth]{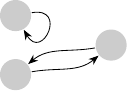}} & \quad
    \raisebox{-0.5\height}{\includegraphics[width=0.12\linewidth]{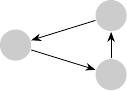}} & \quad
    \raisebox{-0.5\height}{\includegraphics[width=0.12\linewidth]{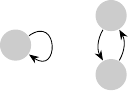}} & \quad
    \raisebox{-0.5\height}{\includegraphics[width=0.12\linewidth]{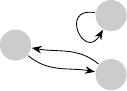}} \\
    T & A & S & T & A & S \\
\end{tabular}
\caption{Topologies of the diagrams for one-two-type contractions.}
\label{fig:topologies-1-2}
\end{figure}

\begin{figure}[t]
\centering
\begin{tabular}{cccccc}
    \raisebox{-0.5\height}{\includegraphics[width=0.12\linewidth]{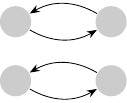}} & \quad
    \raisebox{-0.5\height}{\includegraphics[width=0.12\linewidth]{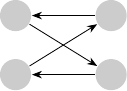}} & \quad
    \raisebox{-0.5\height}{\includegraphics[width=0.12\linewidth]{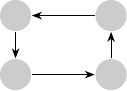}} & \quad
    \raisebox{-0.5\height}{\includegraphics[width=0.12\linewidth]{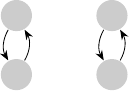}} & \quad
    \raisebox{-0.5\height}{\includegraphics[width=0.12\linewidth]{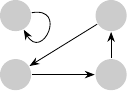}} & \quad
    \raisebox{-0.5\height}{\includegraphics[width=0.12\linewidth]{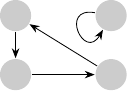}} \\
    D & E & B & A & ML & MR \\
\end{tabular}
\caption{Topologies of the diagrams for two-two-type contractions.}
\label{fig:topologies-2-2}
\end{figure}

\begin{figure}[t]
\centering
\begin{tabular}{cccc}
    \raisebox{-0.5\height}{\includegraphics[width=0.12\linewidth]{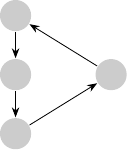}} & \quad
    \raisebox{-0.5\height}{\includegraphics[width=0.12\linewidth]{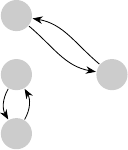}} & \quad
    \raisebox{-0.5\height}{\includegraphics[width=0.12\linewidth]{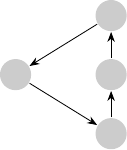}} & \quad
    \raisebox{-0.5\height}{\includegraphics[width=0.12\linewidth]{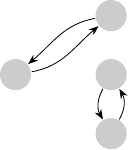}} \\
    TT & TU & TT & TU \\
\end{tabular}
\caption{Topologies of the diagrams for one-three-type contractions.}
\label{fig:topologies-1-3}
\end{figure}

\begin{figure}[t]
\centering
\begin{tabular}{cccccc}
    \raisebox{-0.5\height}{\includegraphics[width=0.12\linewidth]{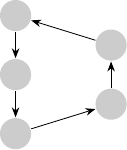}} & \quad
    \raisebox{-0.5\height}{\includegraphics[width=0.12\linewidth]{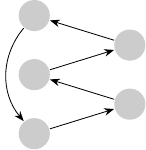}} & \quad
    \raisebox{-0.5\height}{\includegraphics[width=0.12\linewidth]{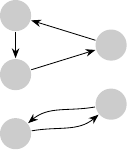}} & \quad
    \raisebox{-0.5\height}{\includegraphics[width=0.12\linewidth]{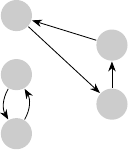}} & \quad
    \raisebox{-0.5\height}{\includegraphics[width=0.12\linewidth]{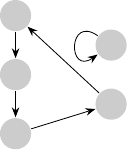}} & \quad
    \raisebox{-0.5\height}{\includegraphics[width=0.12\linewidth]{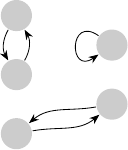}} \\
    TT & TM & TD & TU & TO & TG \\
    \raisebox{-0.5\height}{\includegraphics[width=0.12\linewidth]{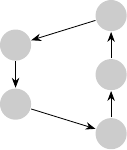}} & \quad
    \raisebox{-0.5\height}{\includegraphics[width=0.12\linewidth]{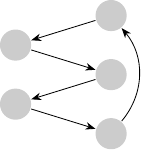}} & \quad
    \raisebox{-0.5\height}{\includegraphics[width=0.12\linewidth]{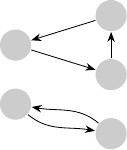}} & \quad
    \raisebox{-0.5\height}{\includegraphics[width=0.12\linewidth]{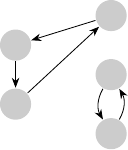}} & \quad
    \raisebox{-0.5\height}{\includegraphics[width=0.12\linewidth]{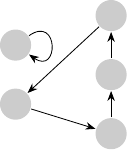}} & \quad
    \raisebox{-0.5\height}{\includegraphics[width=0.12\linewidth]{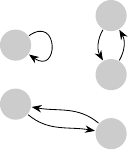}} \\
    TT & TM & TD & TU & TO & TG \\
\end{tabular}
\caption{Topologies of the diagrams for two-three-type contractions.}
\label{fig:topologies-2-3}
\end{figure}

\begin{figure}[t]
\centering
\begin{tabular}{ccccc}
    \raisebox{-0.5\height}{\includegraphics[width=0.12\linewidth]{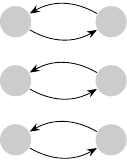}} & \quad
    \raisebox{-0.5\height}{\includegraphics[width=0.12\linewidth]{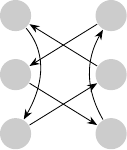}} & \quad
    \raisebox{-0.5\height}{\includegraphics[width=0.12\linewidth]{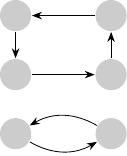}} & \quad
    \raisebox{-0.5\height}{\includegraphics[width=0.12\linewidth]{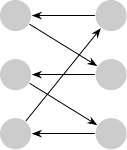}} & \quad
    \raisebox{-0.5\height}{\includegraphics[width=0.12\linewidth]{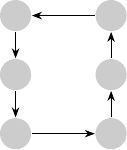}} \\
    TD & TS & TR & TZ & TB \\
    \raisebox{-0.5\height}{\includegraphics[width=0.12\linewidth]{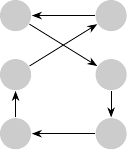}} & \quad
    \raisebox{-0.5\height}{\includegraphics[width=0.12\linewidth]{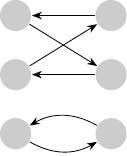}} & \quad
    \raisebox{-0.5\height}{\includegraphics[width=0.12\linewidth]{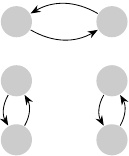}} & \quad
    \raisebox{-0.5\height}{\includegraphics[width=0.12\linewidth]{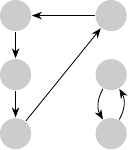}} & \quad
    \raisebox{-0.5\height}{\includegraphics[width=0.12\linewidth]{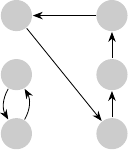}} \\
    TW & TE & TC & TYL & TYR \\
\end{tabular}
\caption{Topologies of the diagrams for three-three-type contractions.}
\label{fig:topologies-3-3}
\end{figure}

Given that the total number of contraction diagrams reaches several hundred, presenting explicit expressions for all correlation functions would be impractical. Instead, we systematically catalog the topological structures of all contraction diagrams. For the $I=1$ $\pi\pi\pi$ channel, we display the complete set of topologies organized by operator type in Figs.~\ref{fig:topologies-1-1}, \ref{fig:topologies-1-2}, \ref{fig:topologies-2-2}, \ref{fig:topologies-1-3}, \ref{fig:topologies-2-3}, and \ref{fig:topologies-3-3}. All computational diagrams represent specific permutations of these fundamental topological structures.

\subsection{Finite-volume spectra}
\label{SUPP/SEC:spectra}

We provide technical details and fitting plots for the extraction of finite-volume energies. The generalized eigenvalue problem (GEVP) of the correlation matrix $C_{ij}$ reads
\begin{equation}
C(t) v_n(t, t_0) = \lambda_n(t, t_0) C(t_0) v_n(t, t_0),
\end{equation}
where $\lambda_n(t, t_0)$ and $v_n(t, t_0)$ are the eigenvalues and eigenvectors, respectively. The reference time $t_0$ is chosen as late as possible while maintaining an acceptable signal-to-noise ratio for the resulting effective mass.

In the $I=0$ $\pi\pi$ channel, the vacuum expectation value (VEV) must be subtracted from the correlator. We evaluate the loops of the sigma operator and the $\pi\pi$ operators. We observed that subtracting a time-dependent VEV leads to a better signal quality than a constant VEV subtraction.

We observe nonzero thermal pollution in the spectra, which is removed at leading order by weighting and shifting $C_{ij}$~\cite{Dudek:2012gj}:
\begin{equation}
\tilde{C}(t) = e^{-\mathcal{E} t} \big[ e^{\mathcal{E} t} C(t) - e^{\mathcal{E} (t+1)} C(t+1) \big],
\label{eq:thermal}
\end{equation}
where $\mathcal{E} = 0$ in our case. The resulting $\tilde{C}(t)$ is then processed through the conventional GEVP procedure. This procedure is applied to $\pi\pi$ with $I=0$ for F32P30 and F32P21, and to $\pi\pi$ with $I=0,2$ for all ensembles. Eigenvalues from the GEVP are sorted according to their relative magnitudes for $\pi\pi$ with $I=0,1$ and $\pi\pi\pi$ with $I=1$, while for $\pi\pi$ with $I=2$, they are sorted based on overlap with a reference time $t_0 + 1$. This selection of strategy yields the cleanest plateaus.

In the $I=1$ $\pi\pi\pi$ channel, since the ground-state single-pion energy is known exactly, it is sometimes beneficial to subtract a pion state before solving the GEVP. This corresponds to using \cref{eq:thermal} with $\mathcal{E} = M_{\pi}$. We find that for smaller volumes (F32P30 and F32P21), this subtraction significantly improves the signal of excited states. The trick does not work for higher excited states because the signal deteriorates very fast.

Energy levels of the $n^{\rm th}$ excited states are extracted via a two-state fit of $\lambda_n(t, t_0)$:
\begin{equation}
\lambda_n(t, t_0) = (1-A_n) e^{-E_n (t-t_0)} + A_n e^{-E_n^{\prime} (t-t_0)},
\end{equation}
where $E_n$ denotes the $n^{\rm th}$ energy level.

\begin{figure}[htbp]
\centering
\includegraphics[width=0.4\columnwidth]{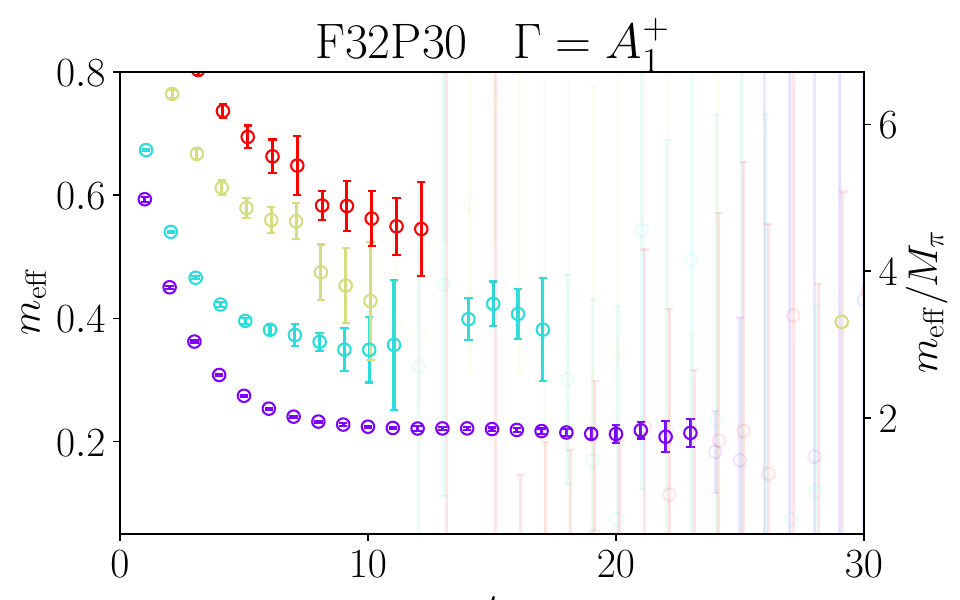}
\includegraphics[width=0.4\columnwidth]{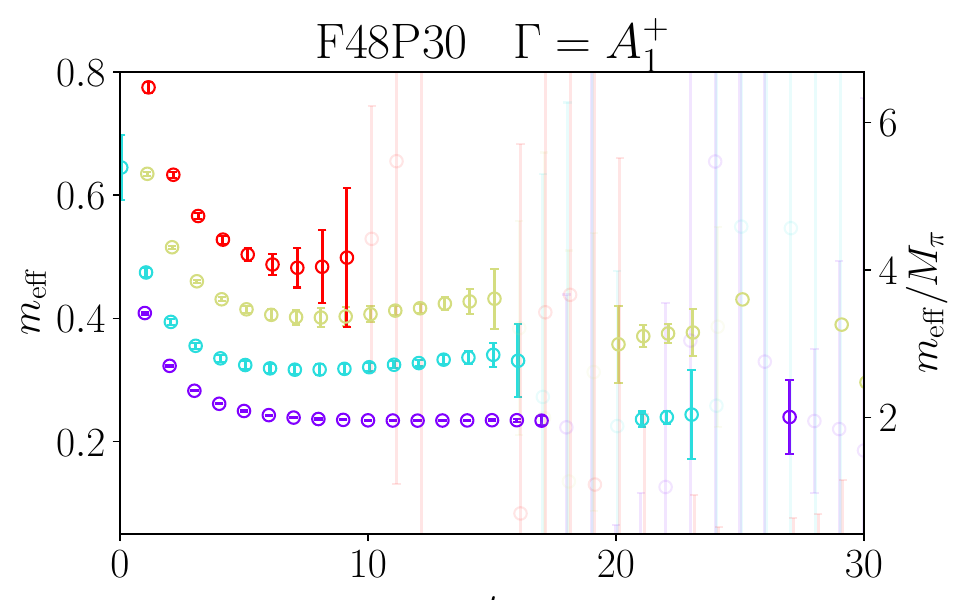}
\\
\includegraphics[width=0.4\columnwidth]{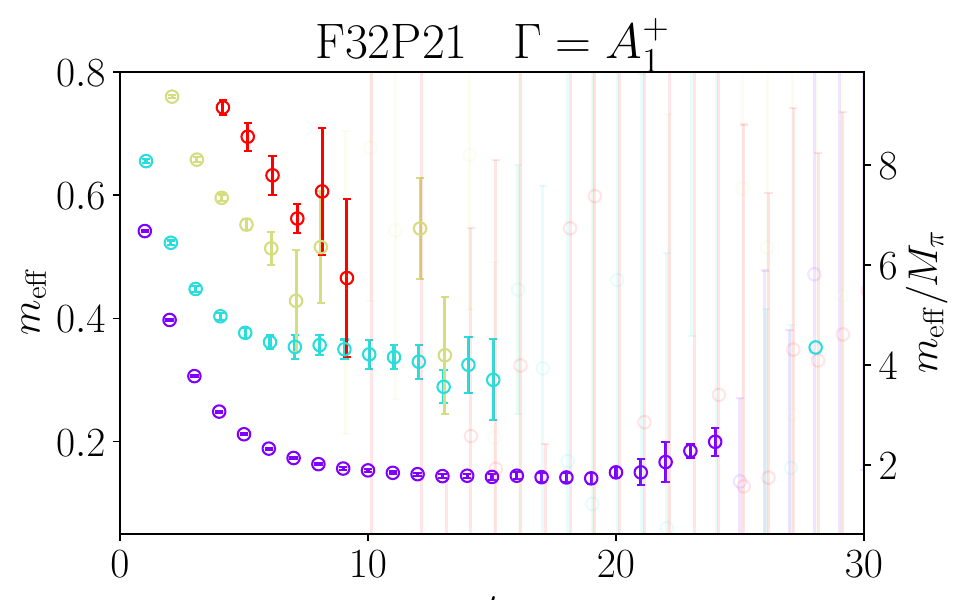}
\includegraphics[width=0.4\columnwidth]{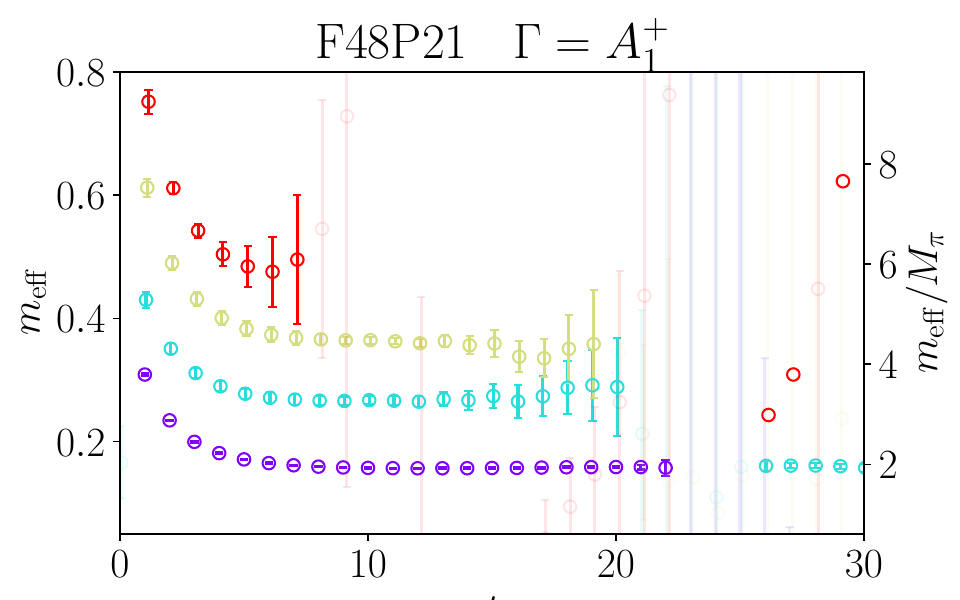}
\caption{Effective mass plots of the eigenvalues $\lambda_n(t)$ for the $I=0$ $\pi\pi$ channel. Each color represents a different state. The left $y$-axis is in lattice units, while the right $y$-axis is in pion mass units. Data points with particularly large statistical uncertainties are shown with reduced opacity to improve readability.}
\label{fig:pipi-I=0-meff}
\end{figure}

\begin{figure}[htbp]
\centering
\includegraphics[width=0.4\columnwidth]{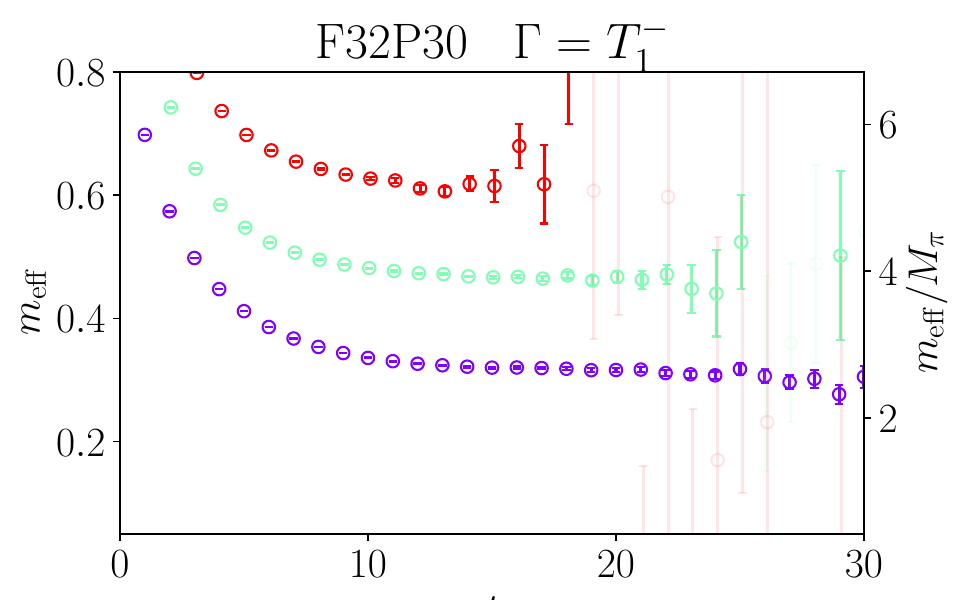}
\includegraphics[width=0.4\columnwidth]{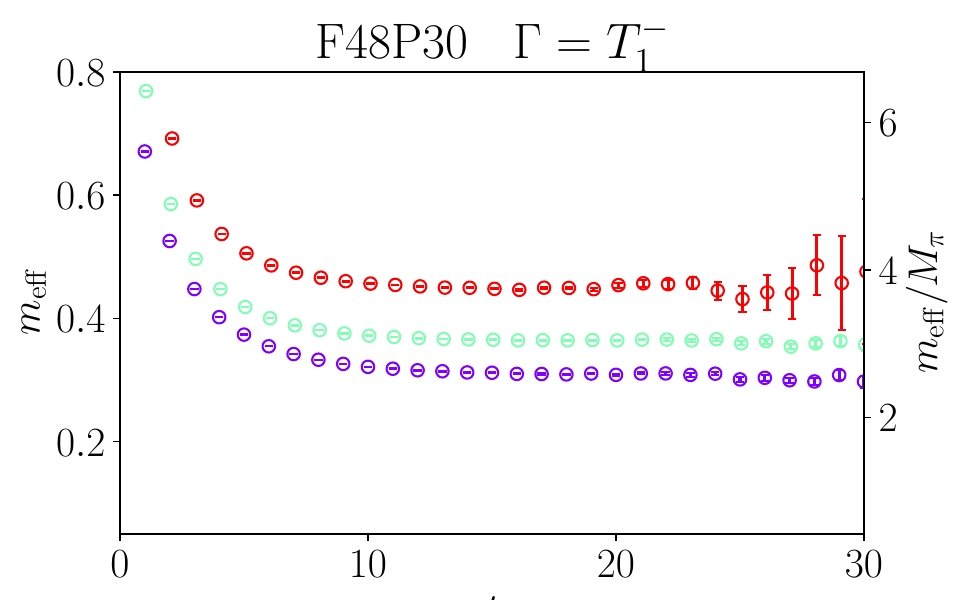}
\\
\includegraphics[width=0.4\columnwidth]{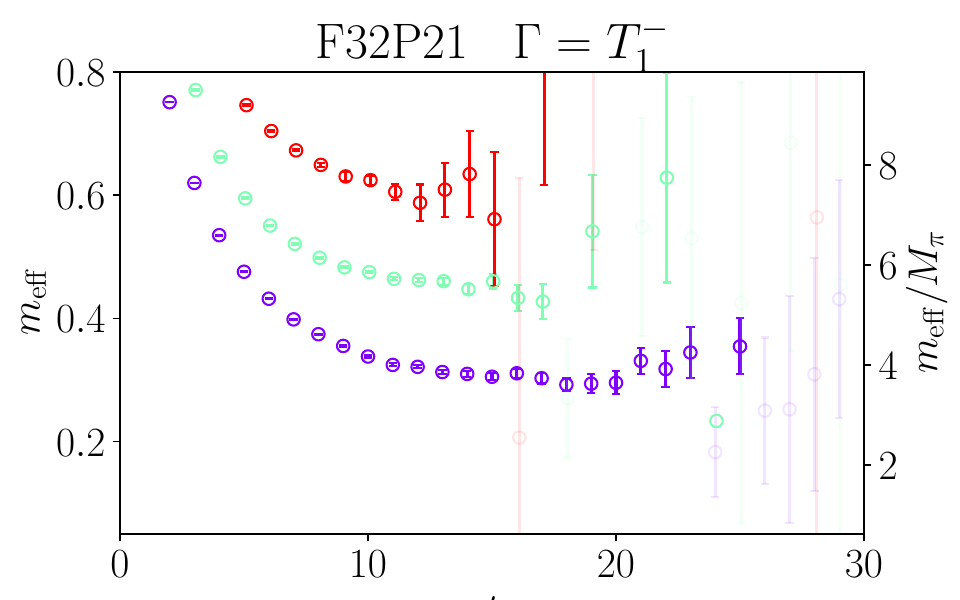}
\includegraphics[width=0.4\columnwidth]{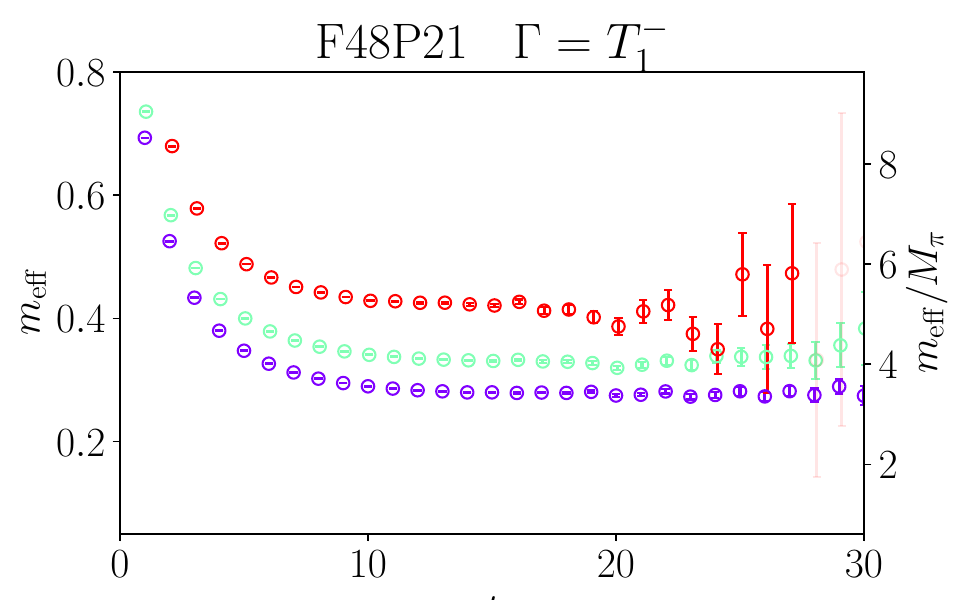}
\caption{Effective mass plots for the $I=1$ $\pi\pi$ channel across all ensembles used.}
\label{fig:pipi-I=1-meff}
\end{figure}

\begin{figure}[htbp]
\centering
\includegraphics[width=0.4\columnwidth]{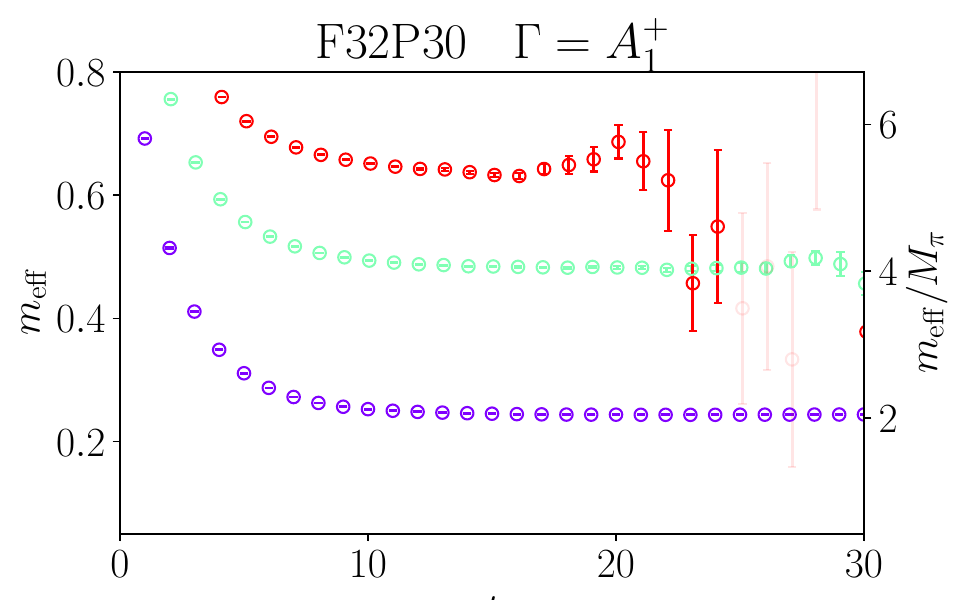}
\includegraphics[width=0.4\columnwidth]{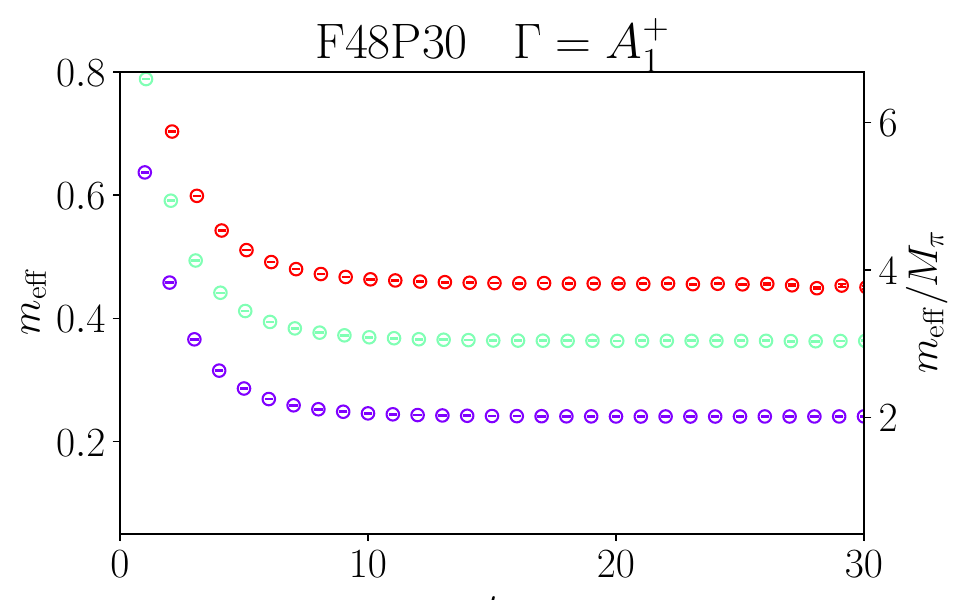}
\\
\includegraphics[width=0.4\columnwidth]{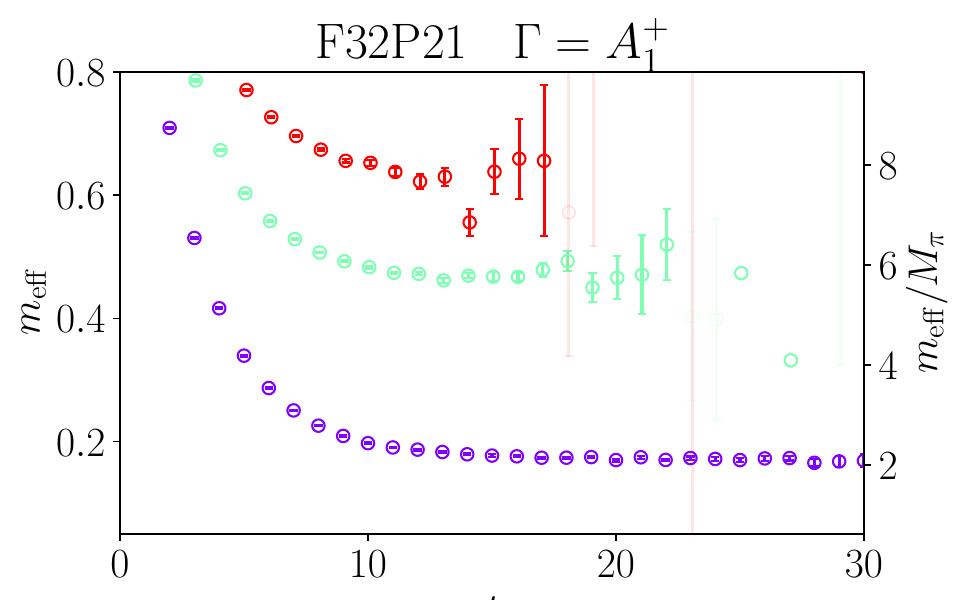}
\includegraphics[width=0.4\columnwidth]{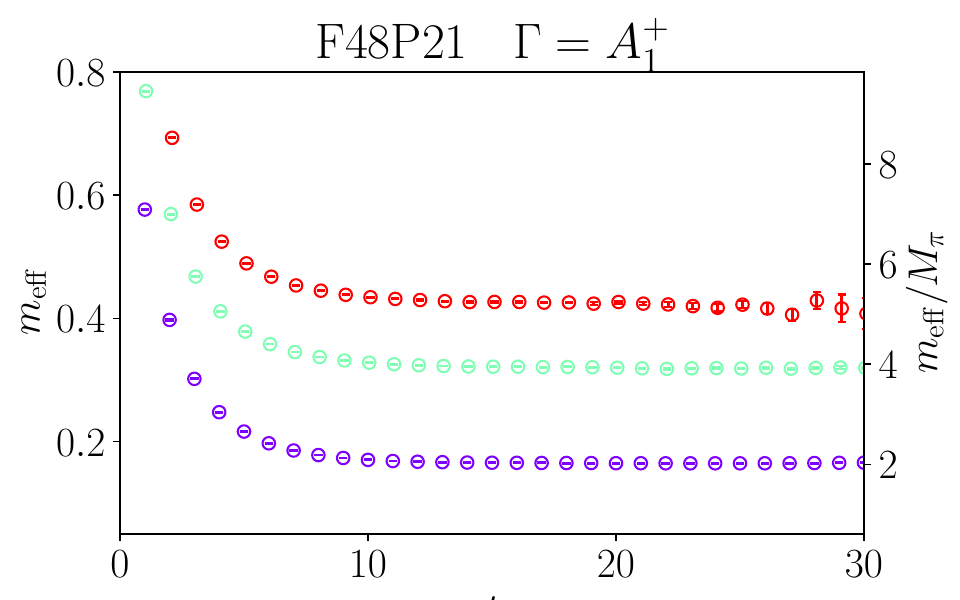}
\caption{Effective mass plots for the $I=2$ $\pi\pi$ channel.}
\label{fig:pipi-I=2-meff}
\end{figure}

\begin{figure}[htbp]
\centering
\includegraphics[width=0.4\columnwidth]{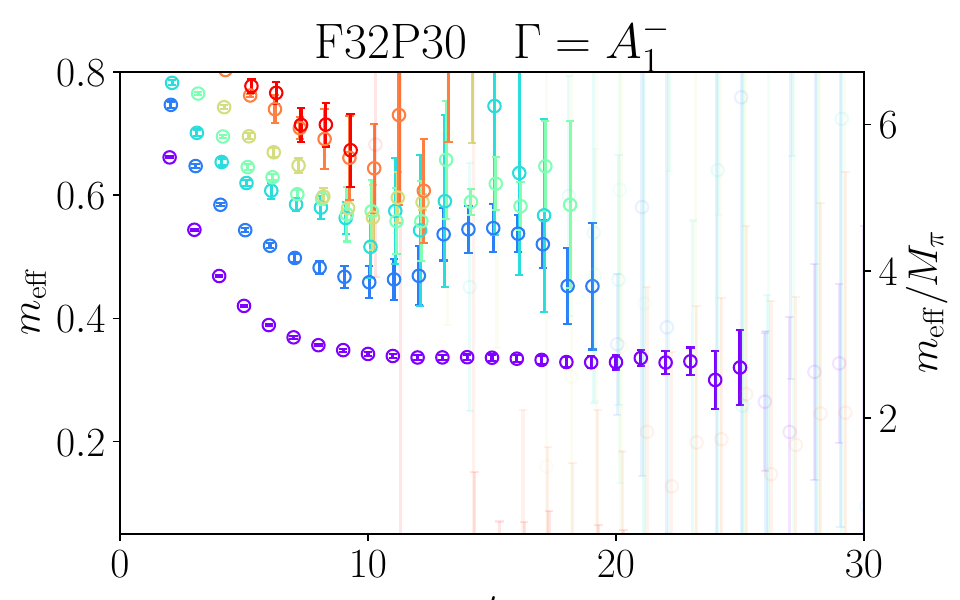}
\includegraphics[width=0.4\columnwidth]{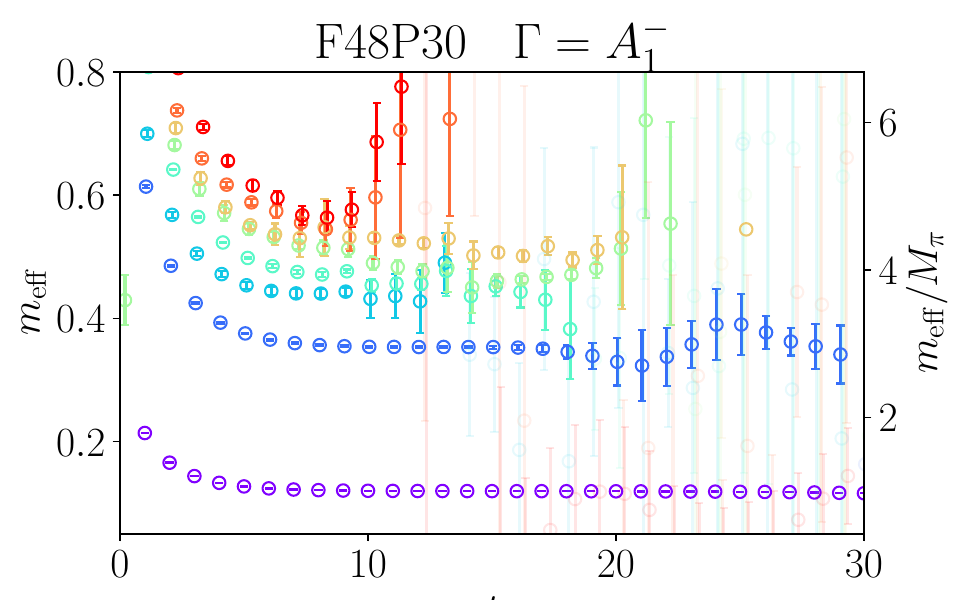}
\\
\includegraphics[width=0.4\columnwidth]{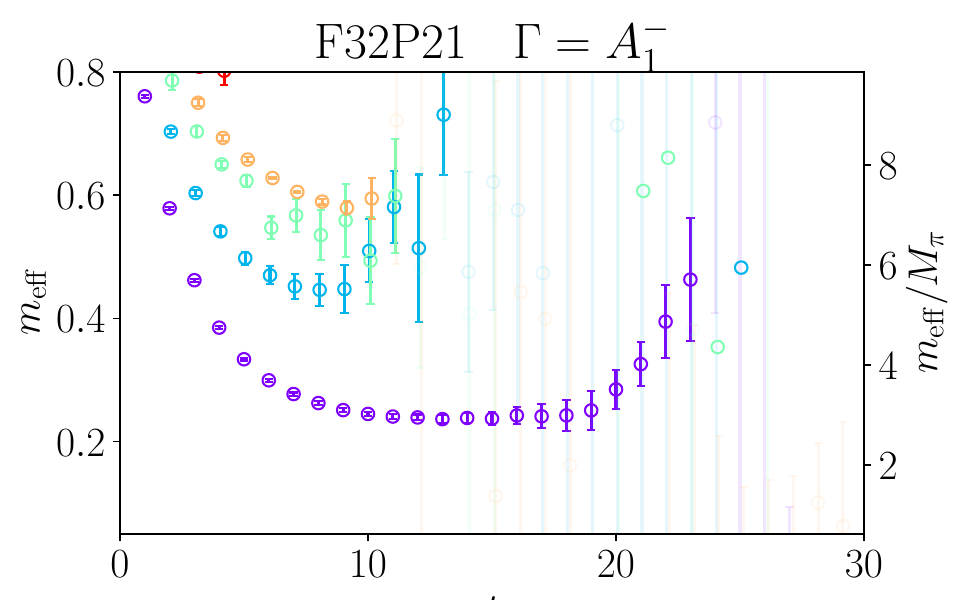}
\includegraphics[width=0.4\columnwidth]{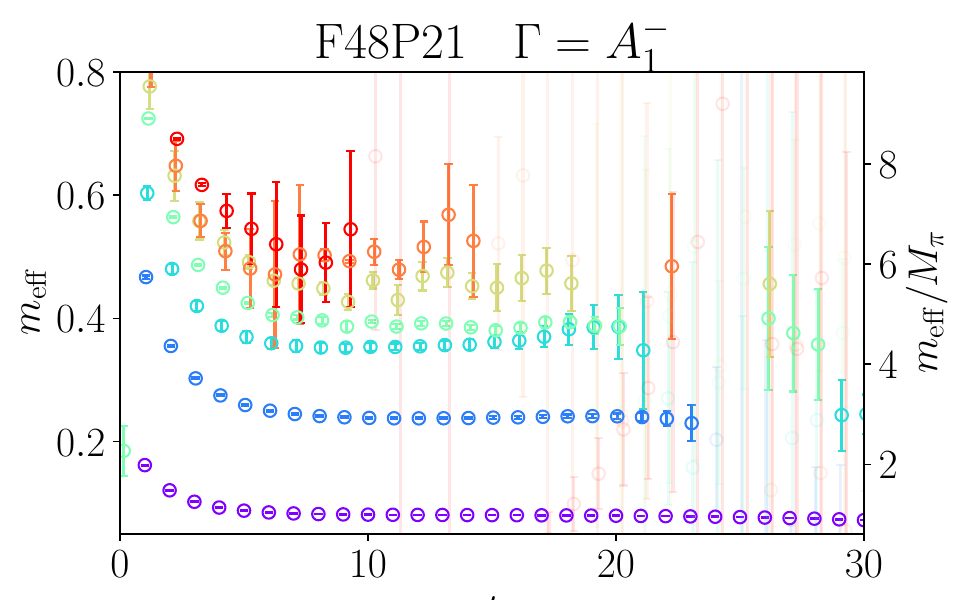}
\caption{Effective mass plots for the $I=1$ $\pi\pi\pi$ channel.}
\label{fig:pipipi-I=1-meff}
\end{figure}

Effective masses of the eigenvalues in the $\pi\pi$ and $\pi\pi\pi$ channels are shown in Figs.~\ref{fig:pipi-I=0-meff}, \ref{fig:pipi-I=1-meff}, \ref{fig:pipi-I=2-meff}, and \ref{fig:pipipi-I=1-meff}, respectively. To improve readability, data points with particularly large uncertainties are displayed with reduced opacity.

\begin{figure}[htbp]
\centering
\includegraphics[width=0.32\columnwidth]{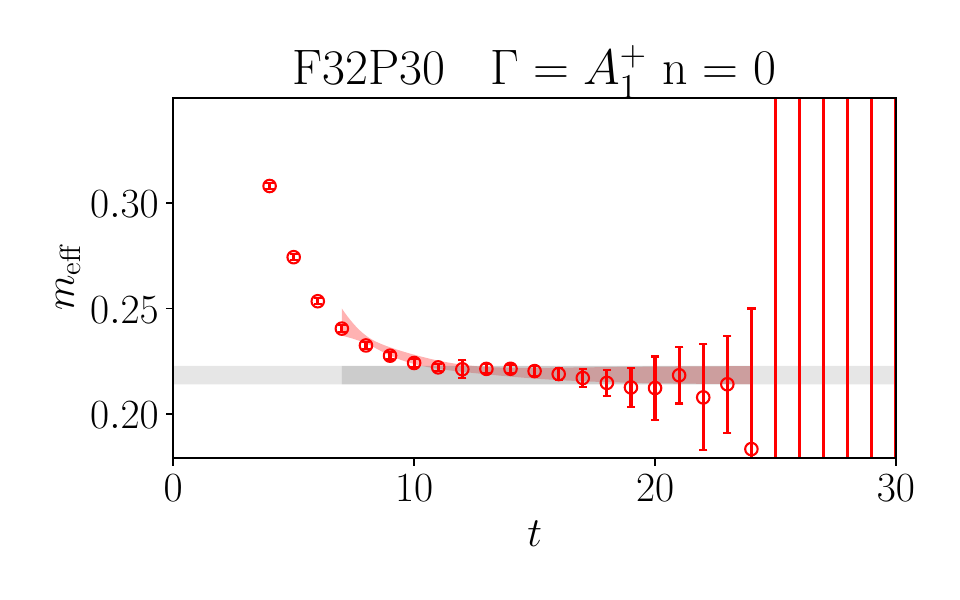}
\includegraphics[width=0.32\columnwidth]{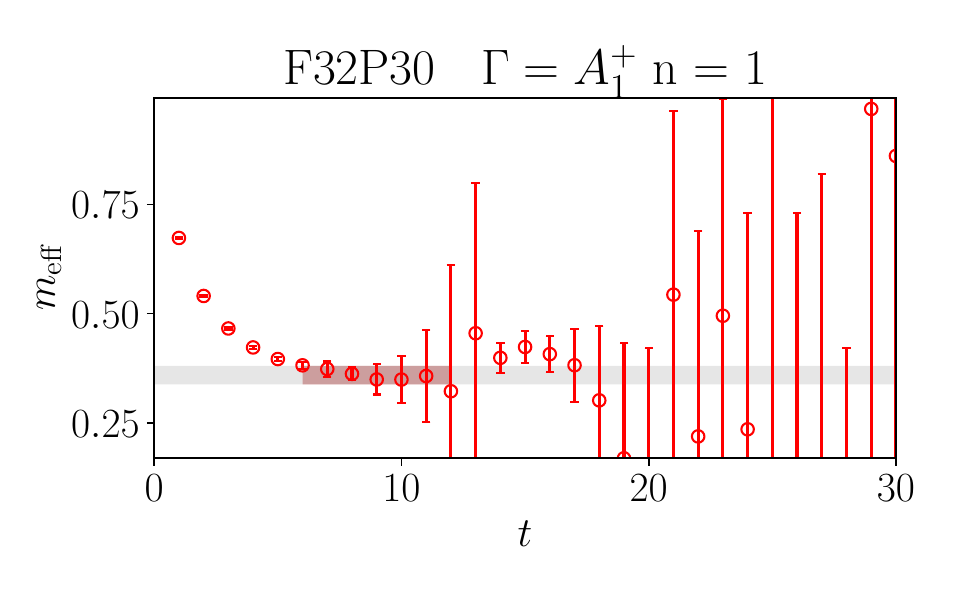}
\\
\includegraphics[width=0.32\columnwidth]{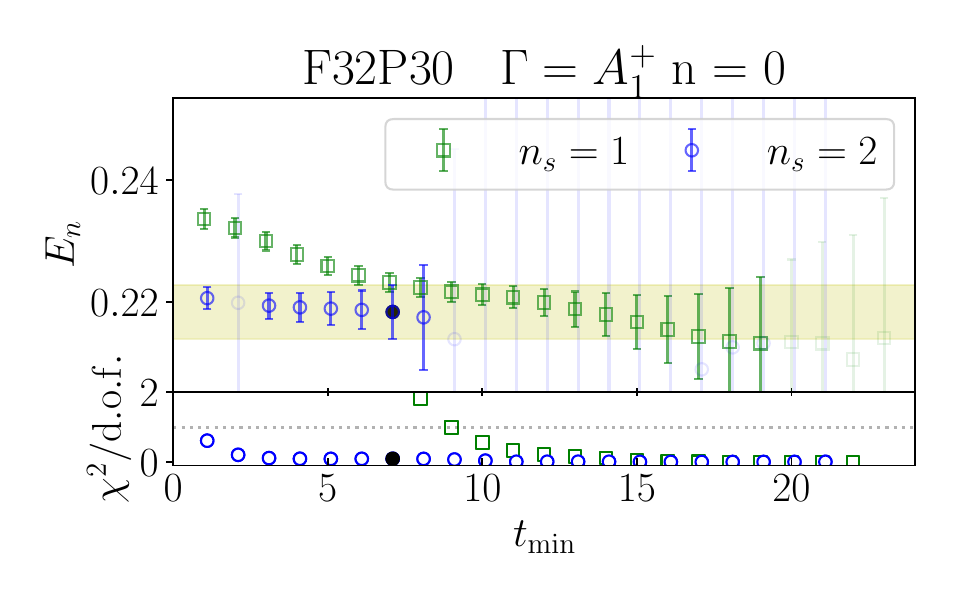}
\includegraphics[width=0.32\columnwidth]{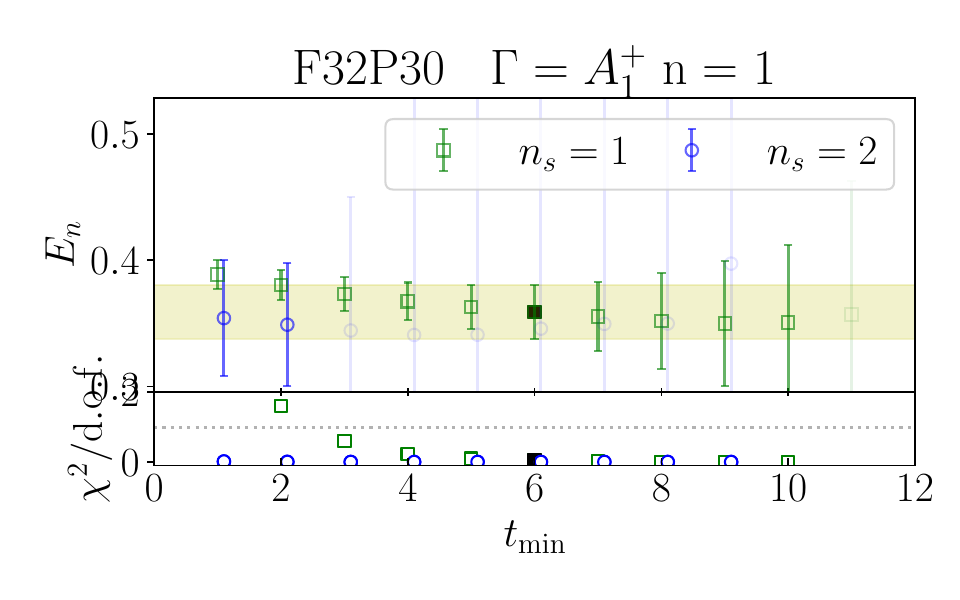}
\\
\includegraphics[width=0.32\columnwidth]{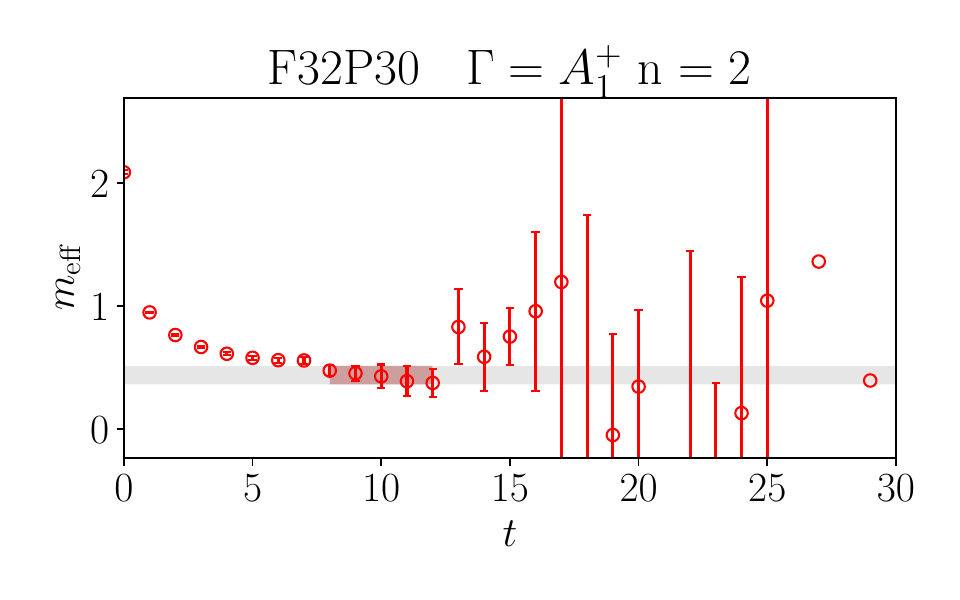}
\includegraphics[width=0.32\columnwidth]{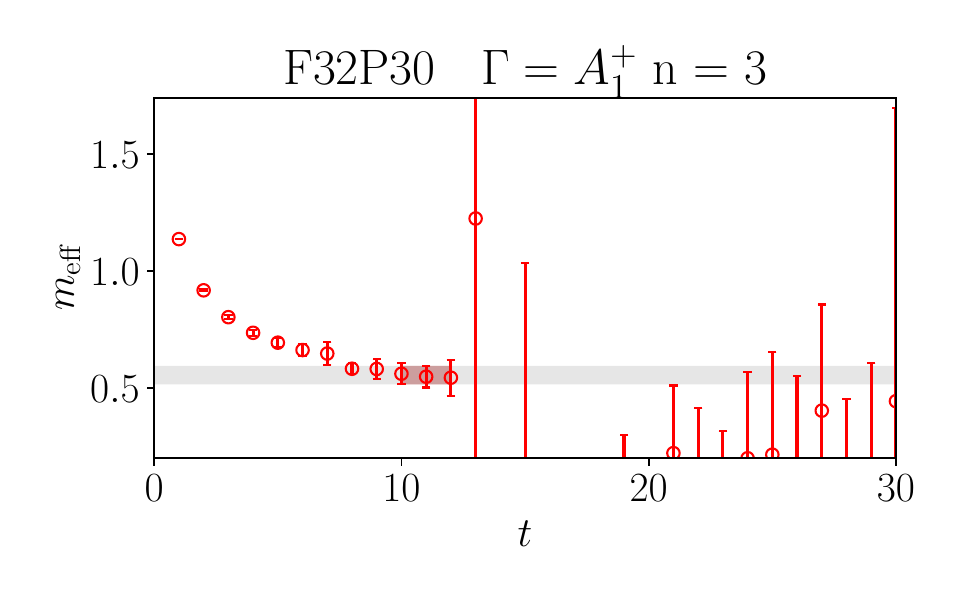}
\\
\includegraphics[width=0.32\columnwidth]{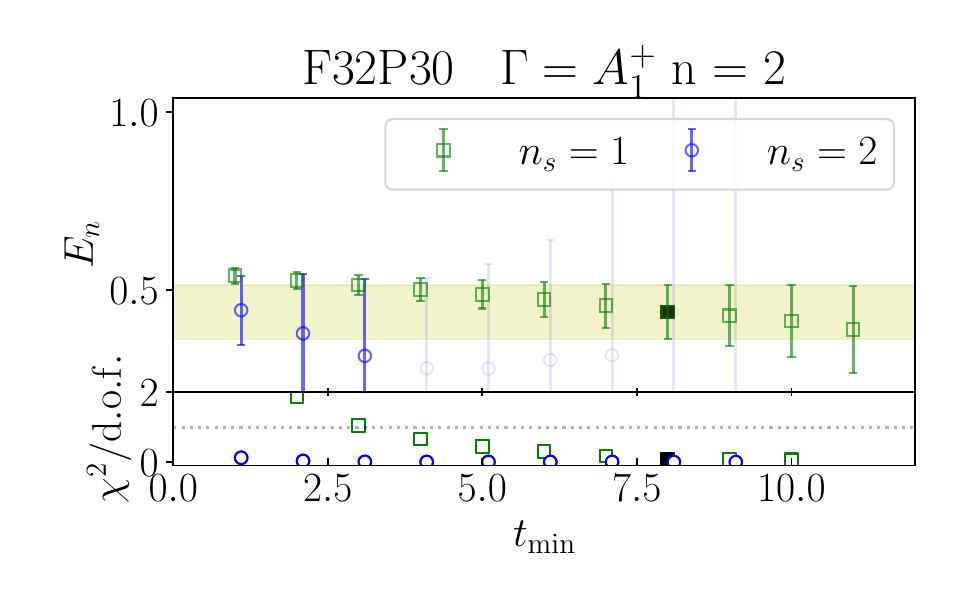}
\includegraphics[width=0.32\columnwidth]{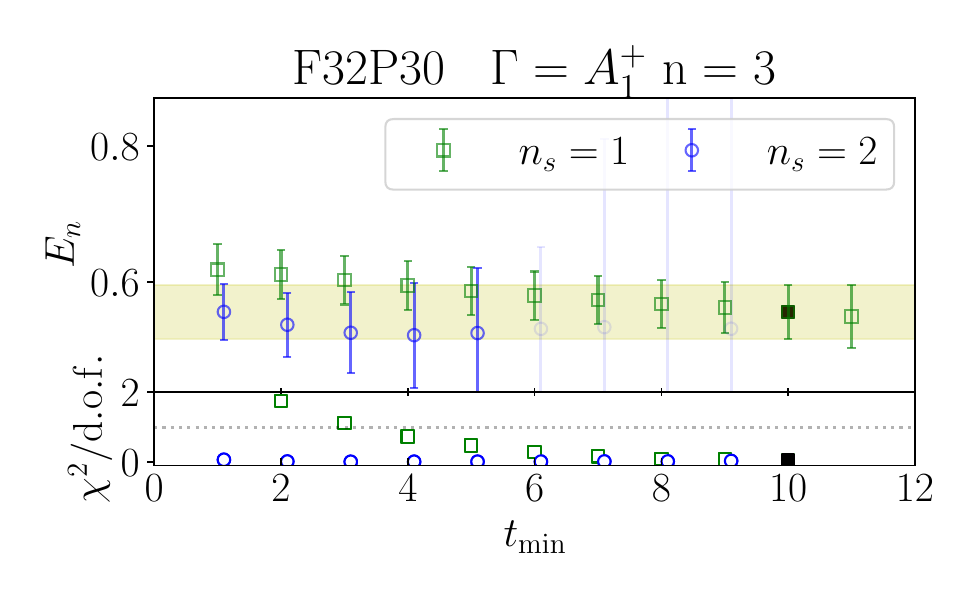}
\caption{Fitting for the $I=0$ $\pi\pi$ channel for ensemble F32P30. Each energy level fit consists of two plots: one above and one below. In the upper plot, the red band denotes the reconstructed effective mass, while the gray band represents the extracted energy. The lower plot shows the stability of the fit as a function of the fitting start time, with green and blue dots representing one- and two-state fits, respectively. The black error bar indicates the chosen start time. The lower plot also shows the $\chi^2/\mathrm{d.o.f.}$.}
\label{fig:pipi-I=0-fit-F32P30}
\end{figure}

\begin{figure}[htbp]
\centering
\includegraphics[width=0.32\columnwidth]{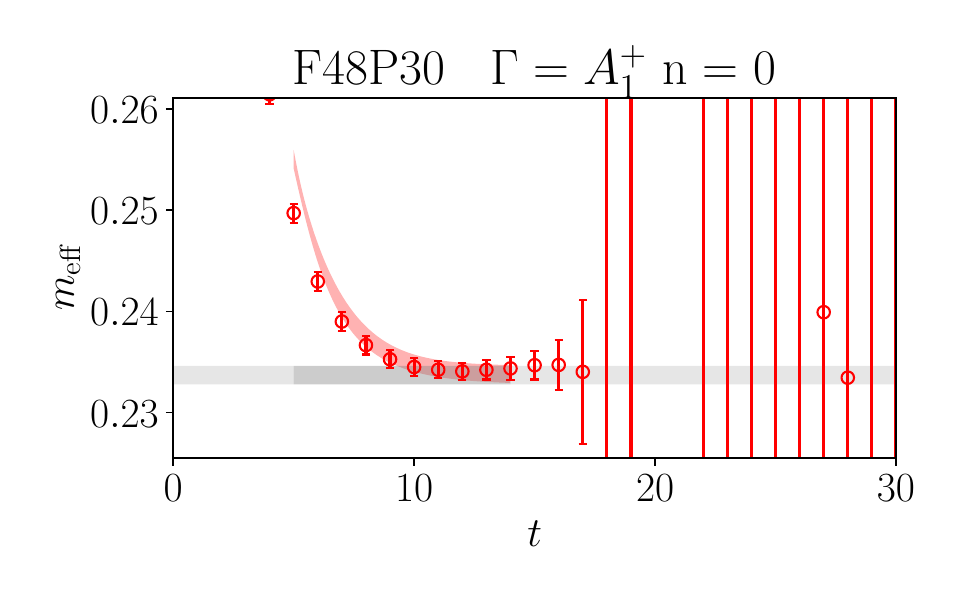}
\includegraphics[width=0.32\columnwidth]{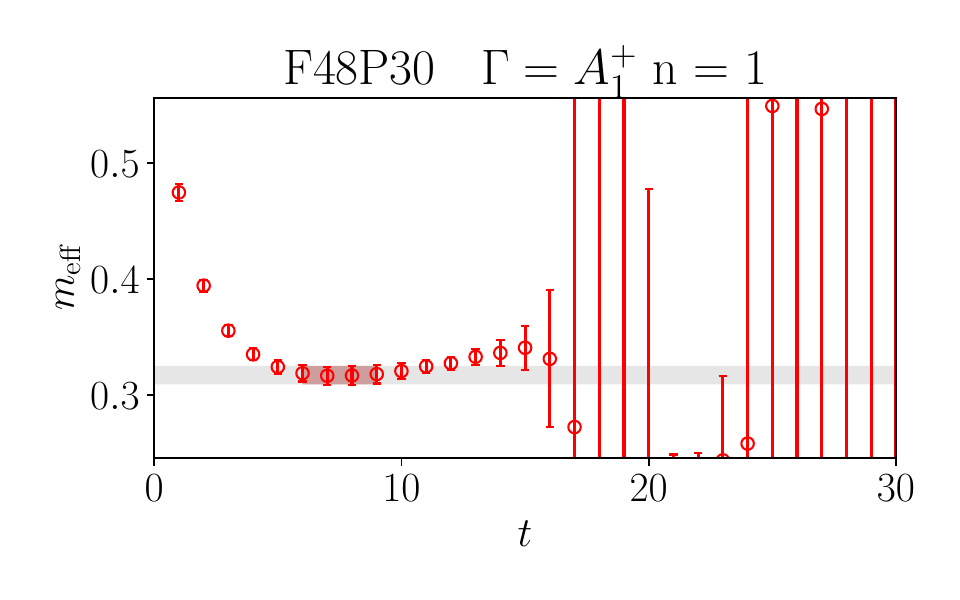}
\\
\includegraphics[width=0.32\columnwidth]{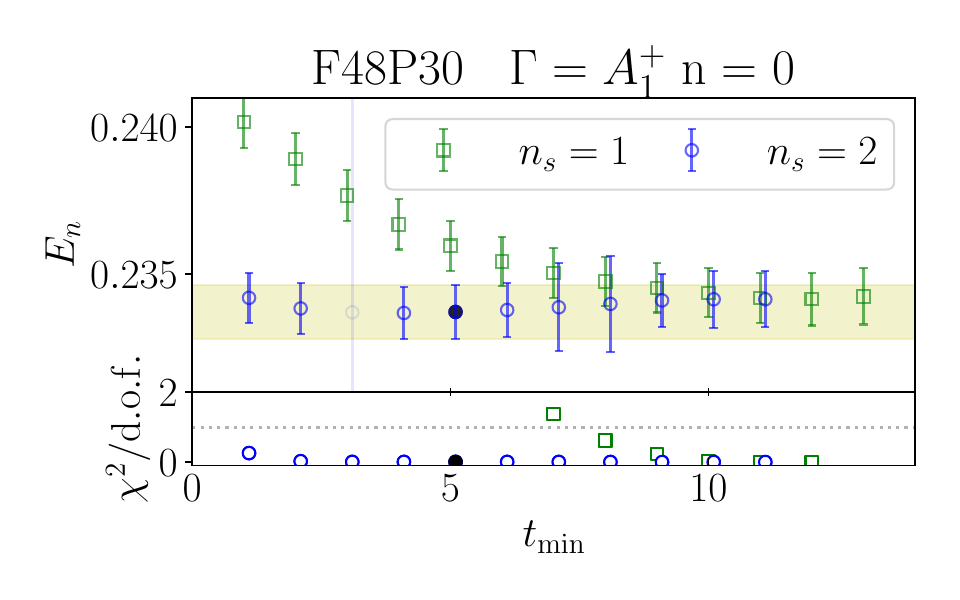}
\includegraphics[width=0.32\columnwidth]{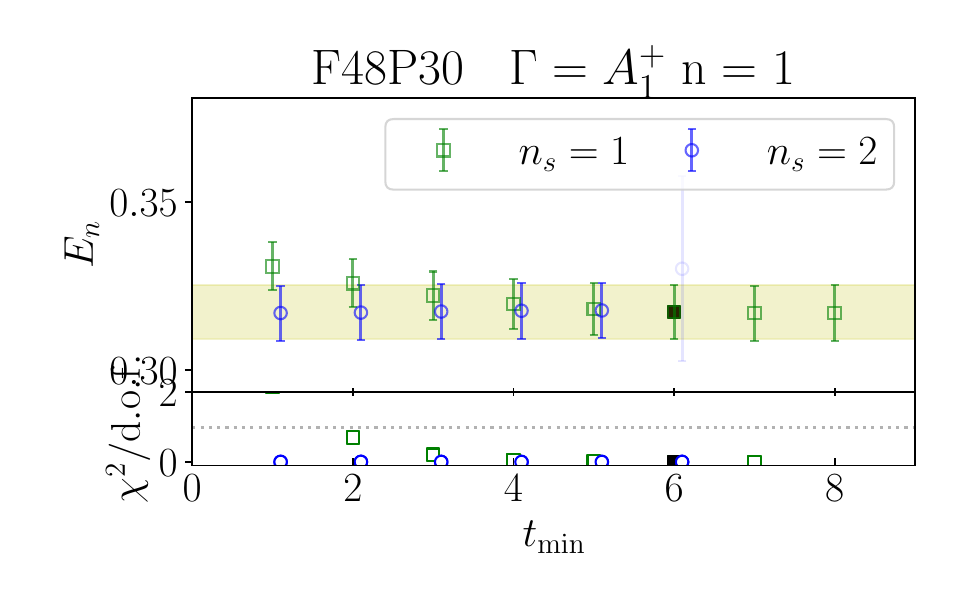}
\\
\includegraphics[width=0.32\columnwidth]{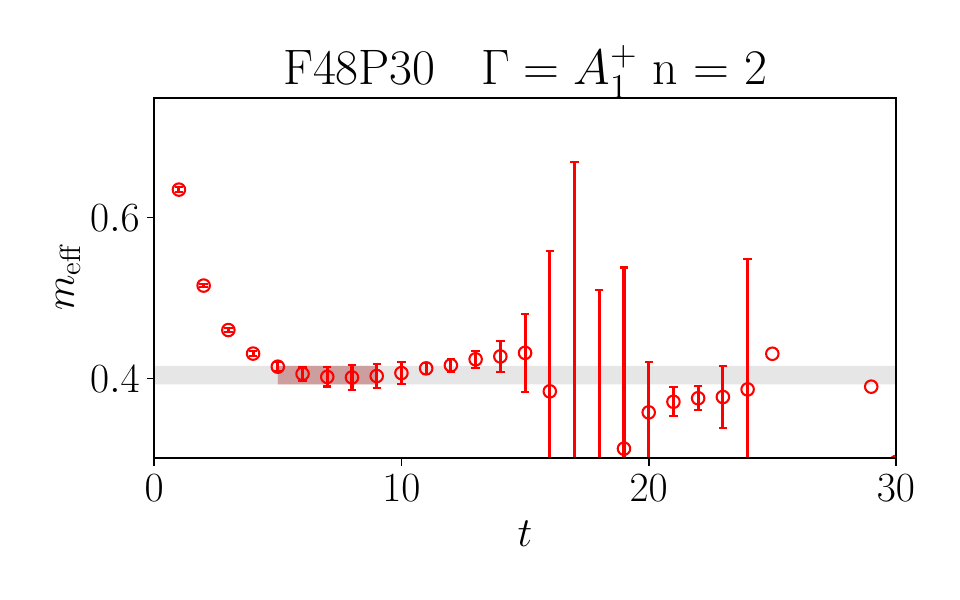}
\includegraphics[width=0.32\columnwidth]{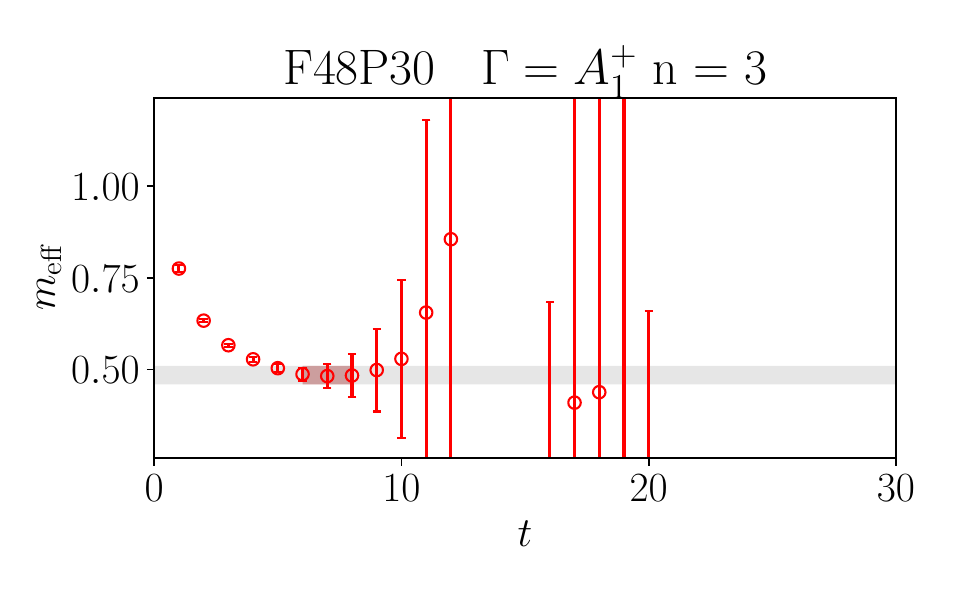}
\\
\includegraphics[width=0.32\columnwidth]{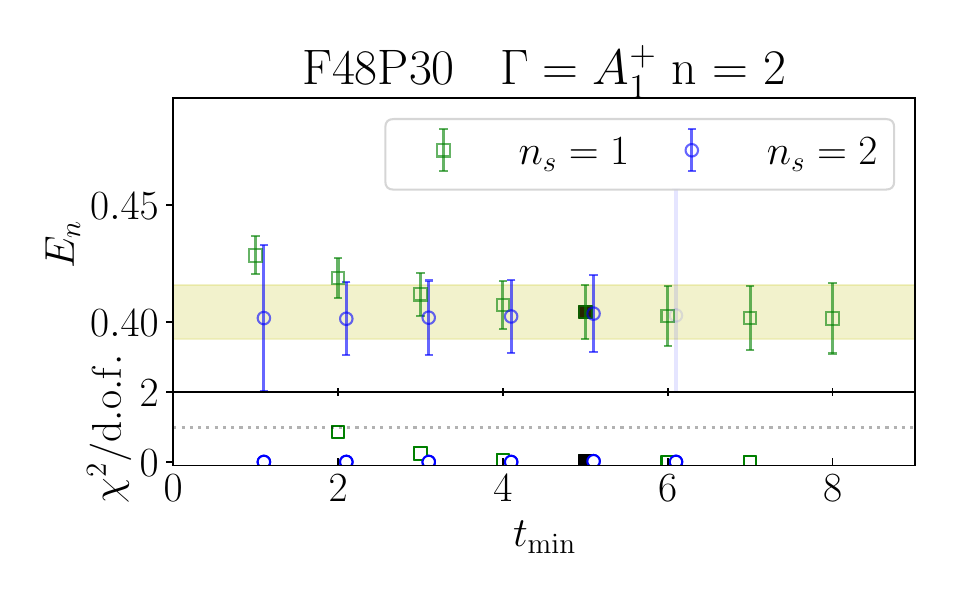}
\includegraphics[width=0.32\columnwidth]{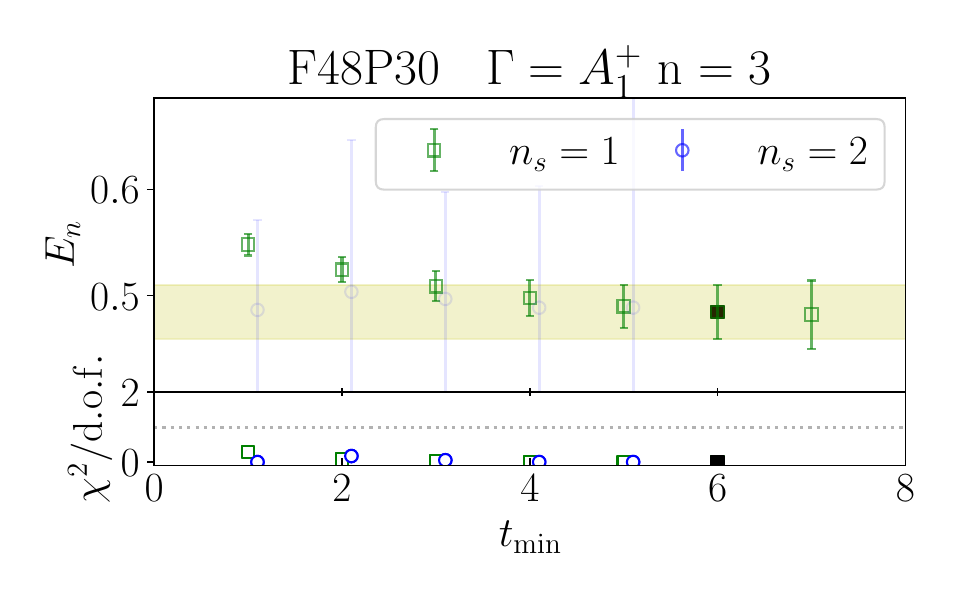}
\caption{Fitting for the $I=0$ $\pi\pi$ channel for ensemble F48P30.}
\label{fig:pipi-I=0-fit-F48P30}
\end{figure}

\begin{figure}[htbp]
\centering
\includegraphics[width=0.32\columnwidth]{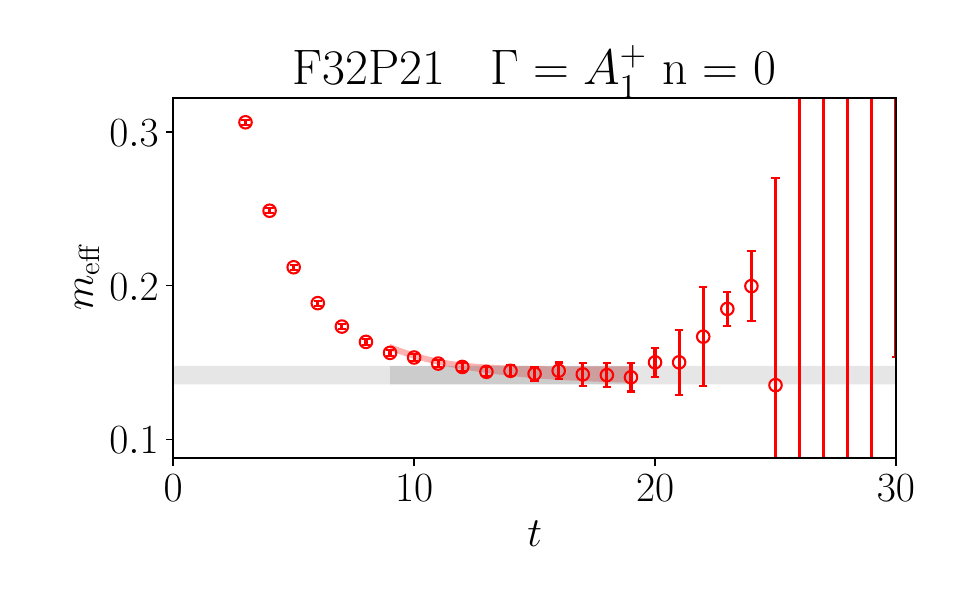}
\includegraphics[width=0.32\columnwidth]{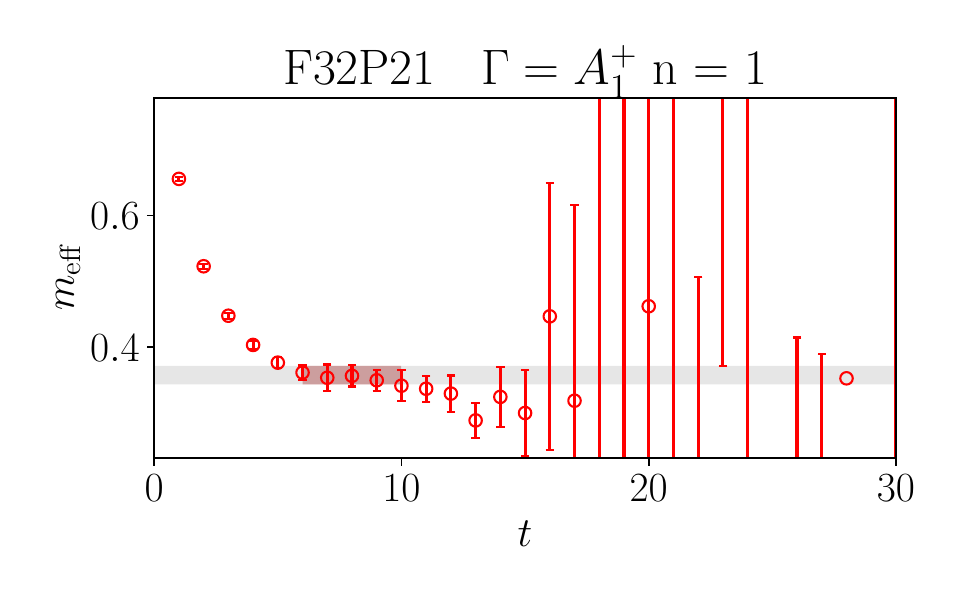}
\\
\includegraphics[width=0.32\columnwidth]{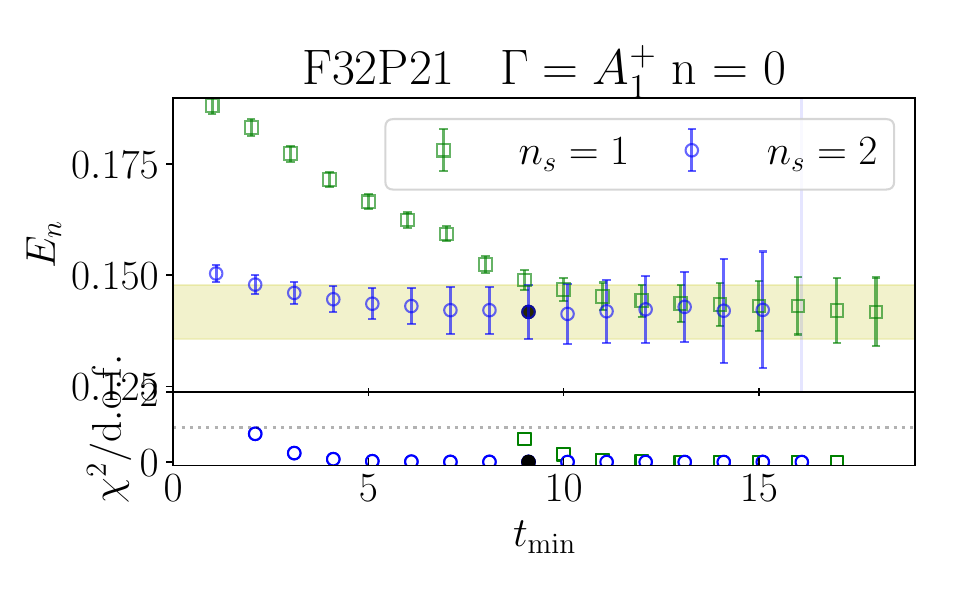}
\includegraphics[width=0.32\columnwidth]{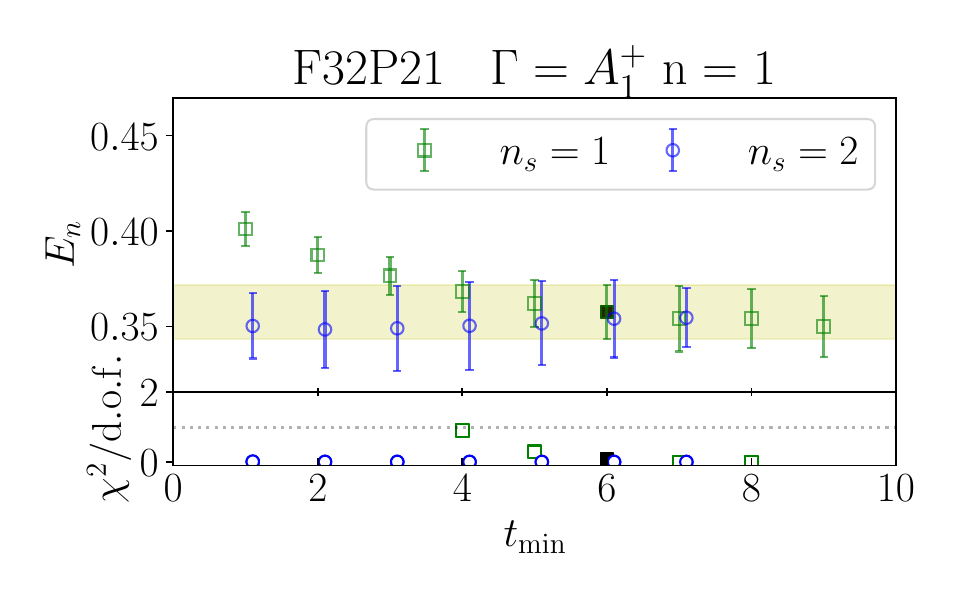}
\\
\includegraphics[width=0.32\columnwidth]{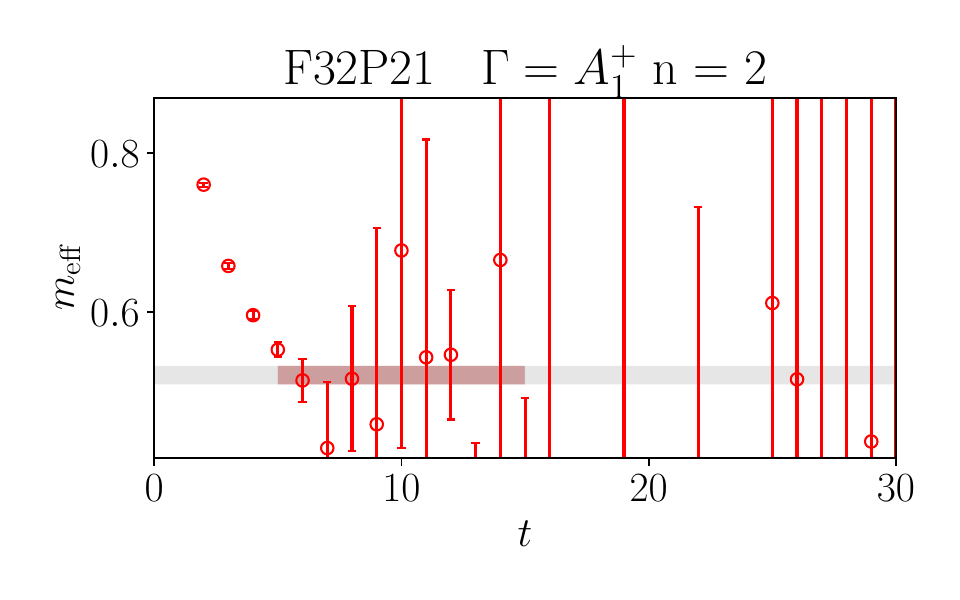}
\includegraphics[width=0.32\columnwidth]{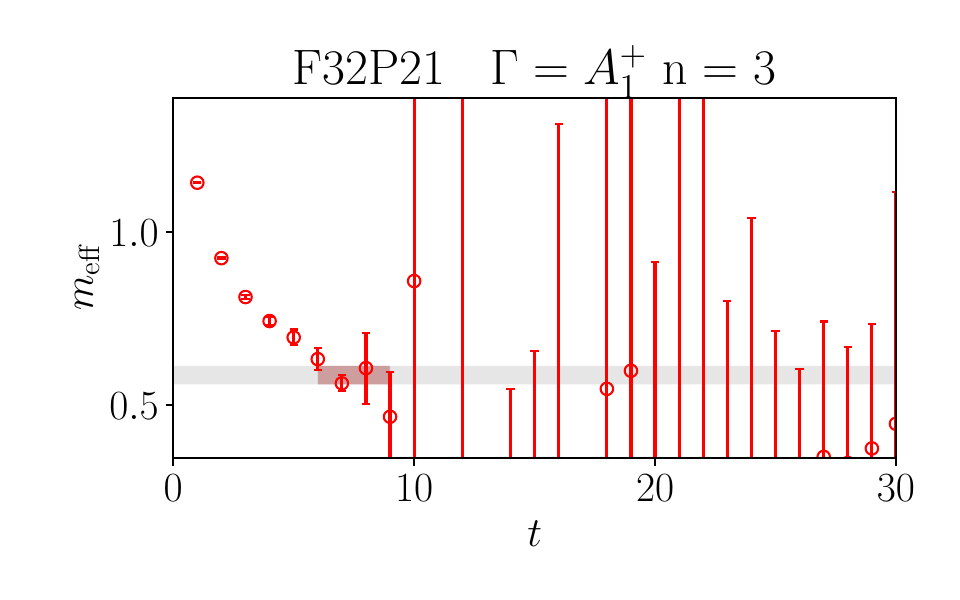}
\\
\includegraphics[width=0.32\columnwidth]{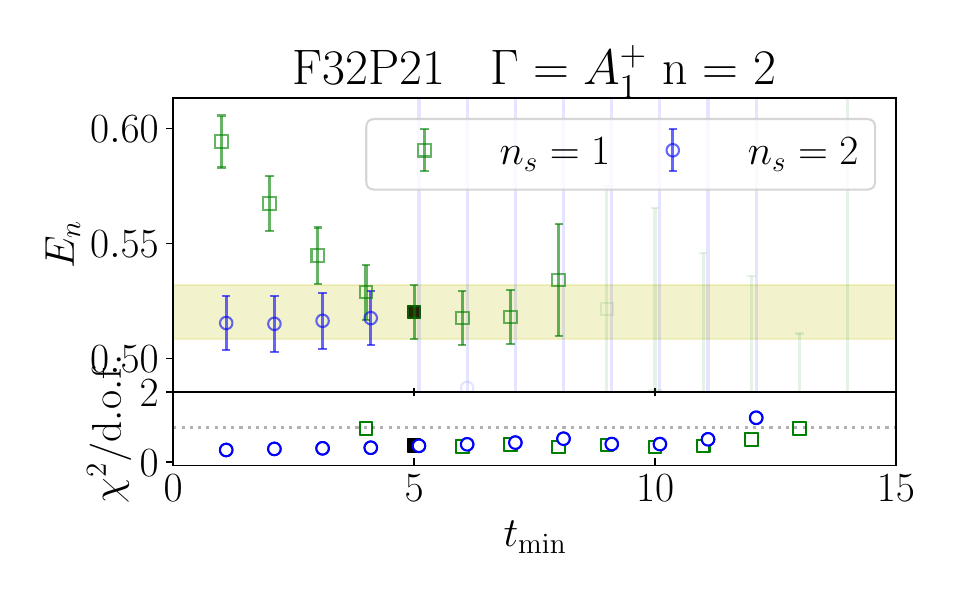}
\includegraphics[width=0.32\columnwidth]{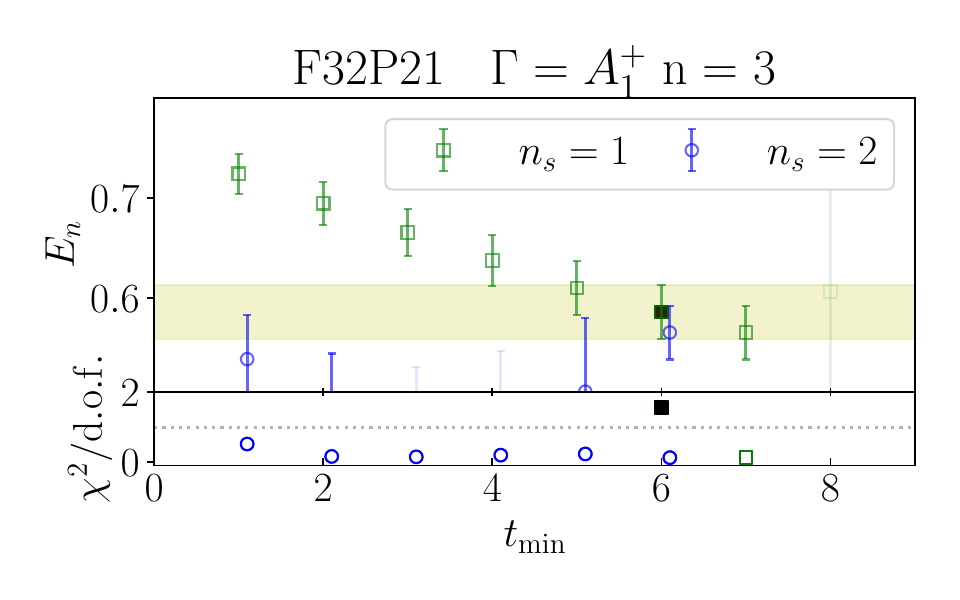}
\caption{Fitting for the $I=0$ $\pi\pi$ channel for ensemble F32P21.}
\label{fig:pipi-I=0-fit-F32P21}
\end{figure}

\begin{figure}[htbp]
\centering
\includegraphics[width=0.32\columnwidth]{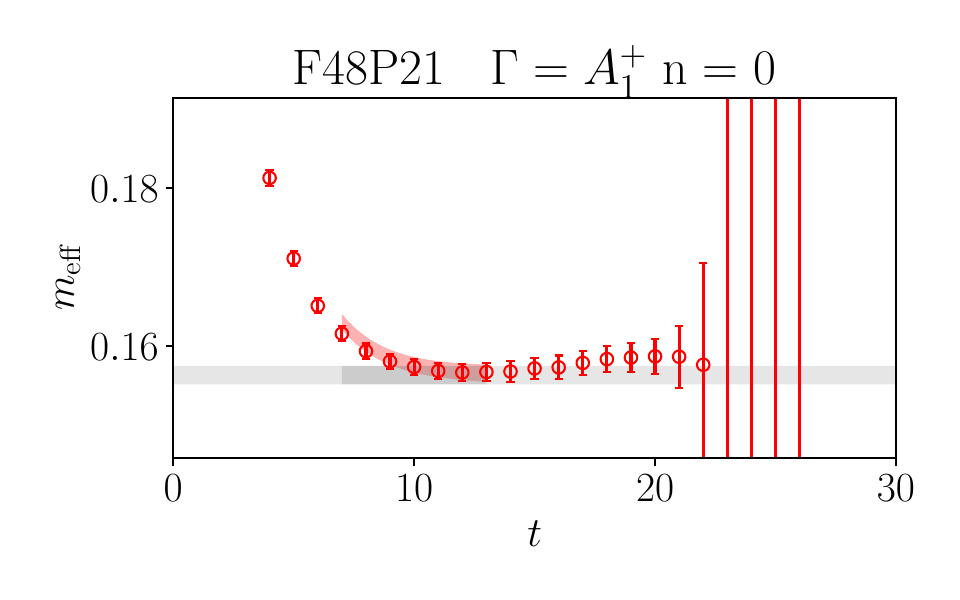}
\includegraphics[width=0.32\columnwidth]{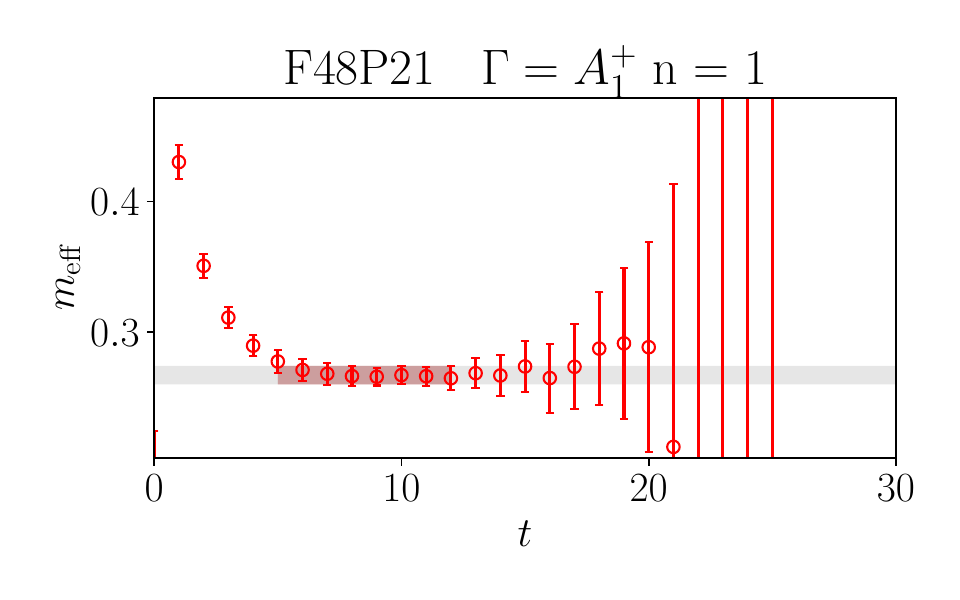}
\\
\includegraphics[width=0.32\columnwidth]{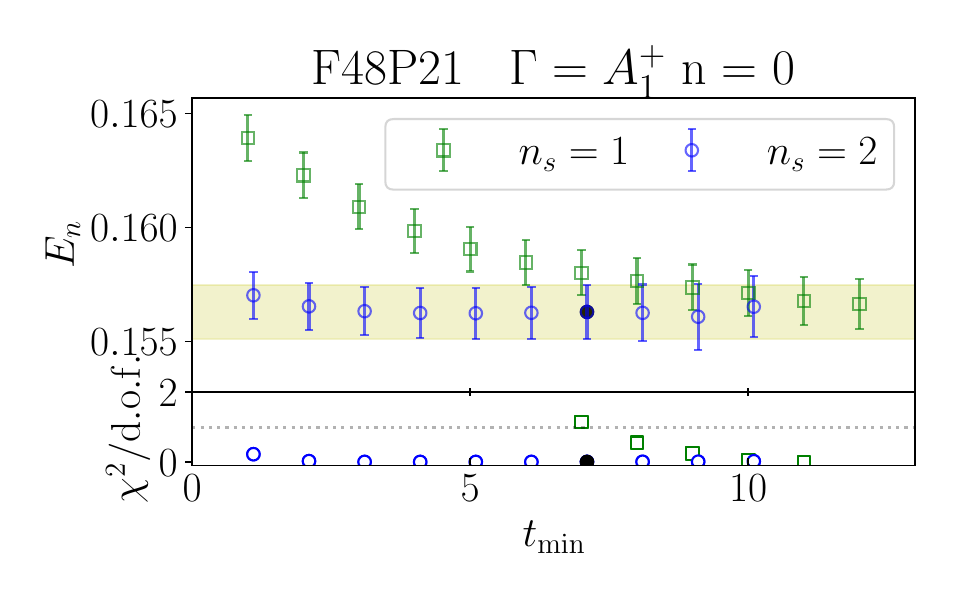}
\includegraphics[width=0.32\columnwidth]{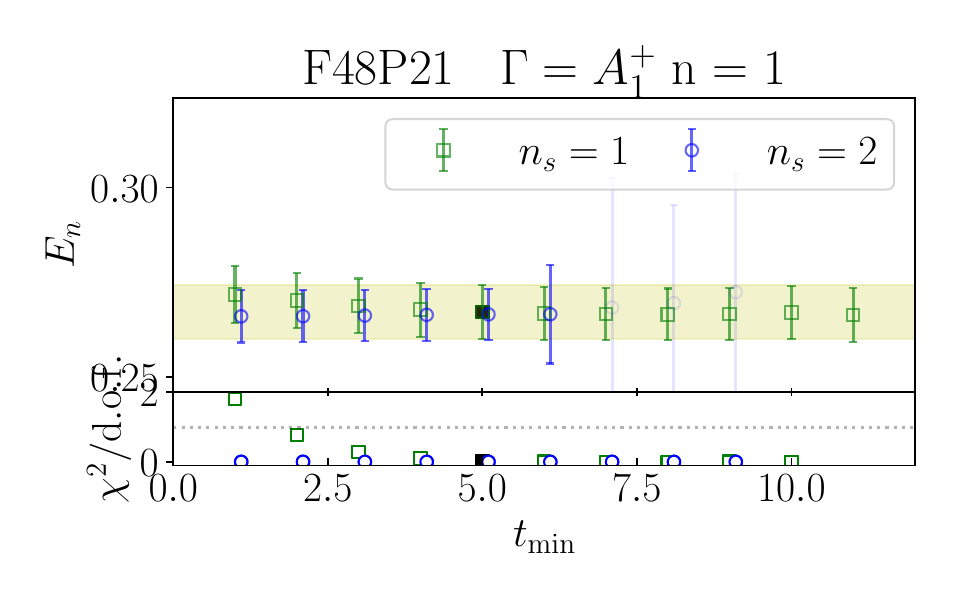}
\\
\includegraphics[width=0.32\columnwidth]{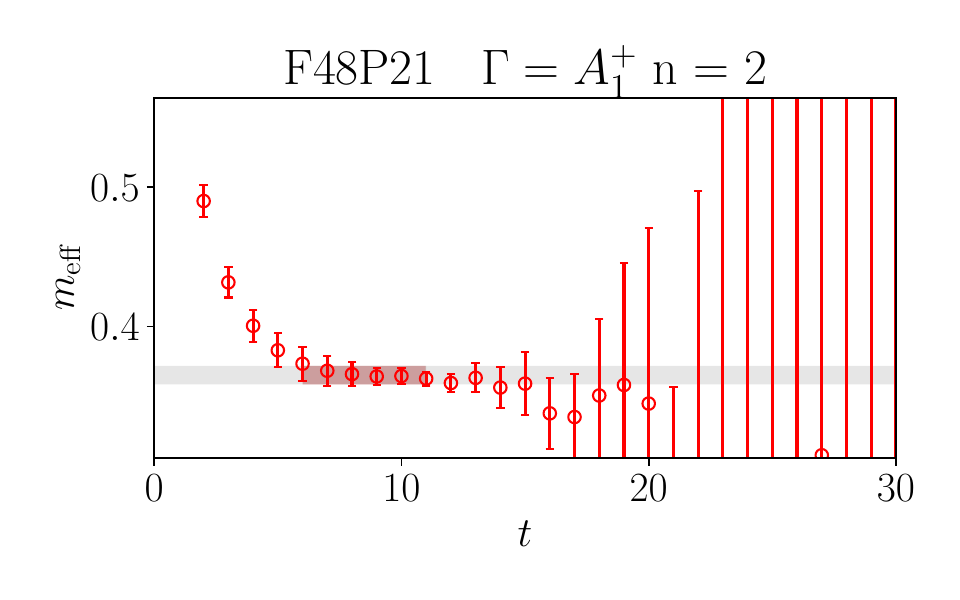}
\includegraphics[width=0.32\columnwidth]{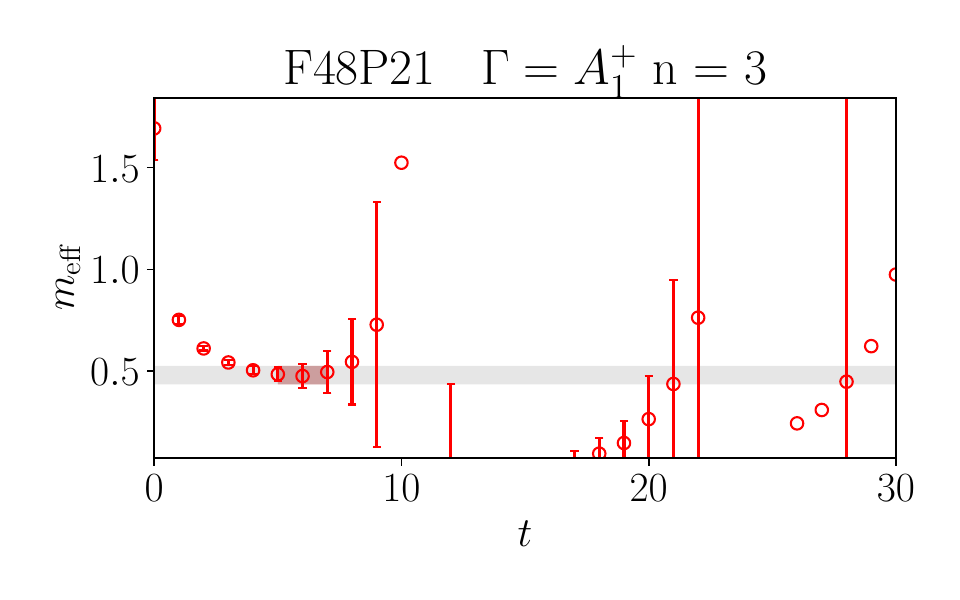}
\\
\includegraphics[width=0.32\columnwidth]{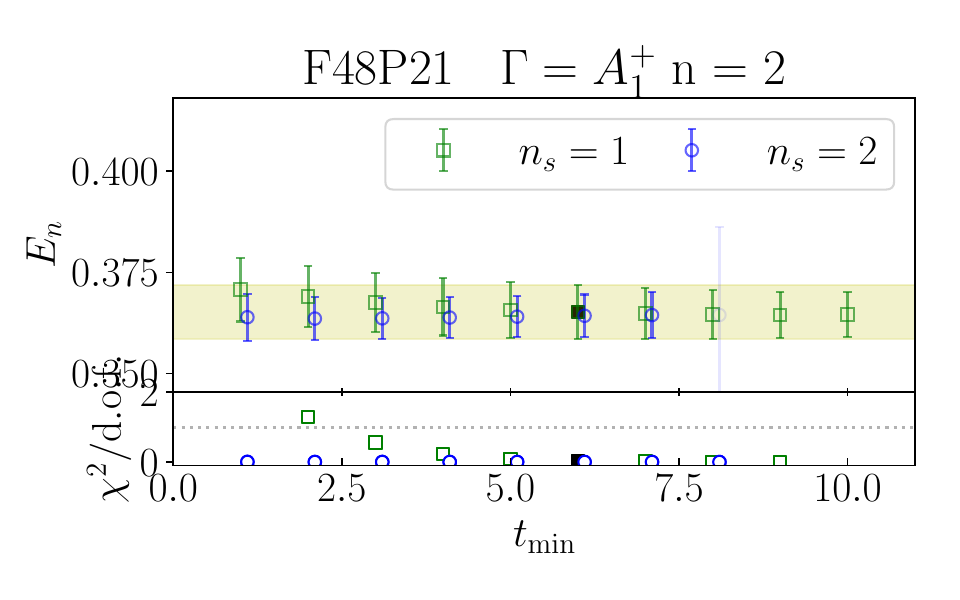}
\includegraphics[width=0.32\columnwidth]{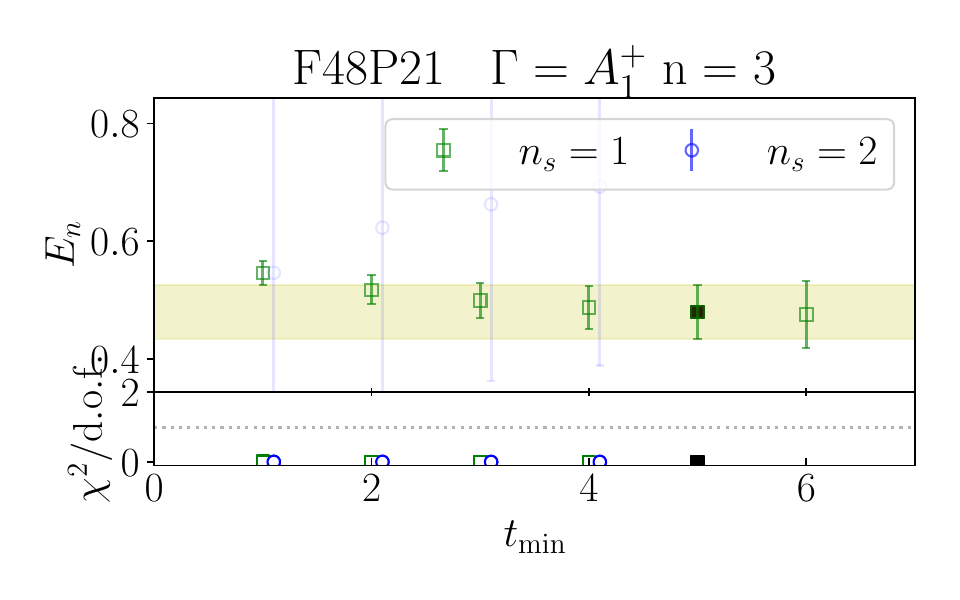}
\caption{Fitting for the $I=0$ $\pi\pi$ channel for ensemble F48P21.}
\label{fig:pipi-I=0-fit-F48P21}
\end{figure}

\begin{figure}[htbp]
\centering
\includegraphics[width=0.32\columnwidth]{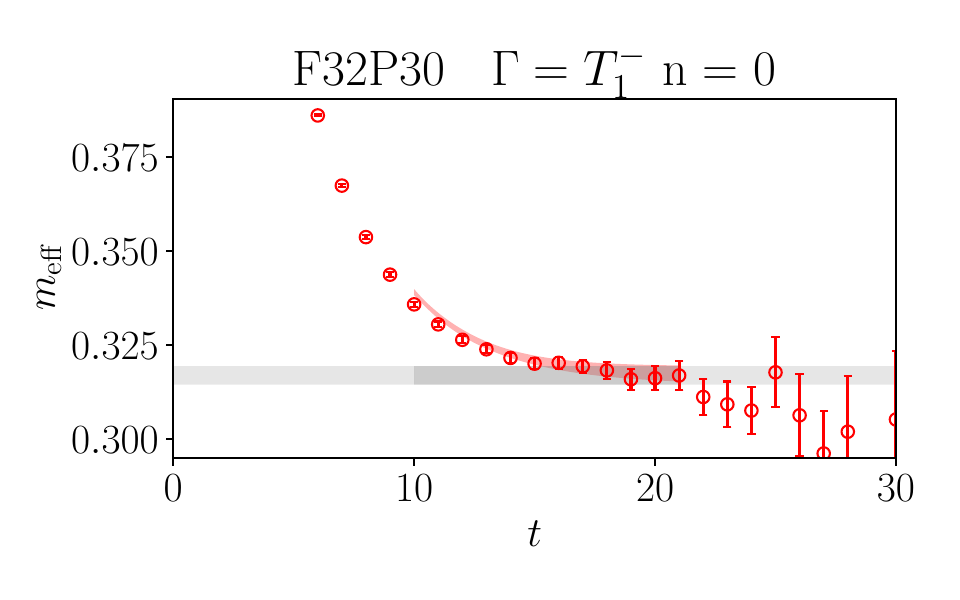}
\includegraphics[width=0.32\columnwidth]{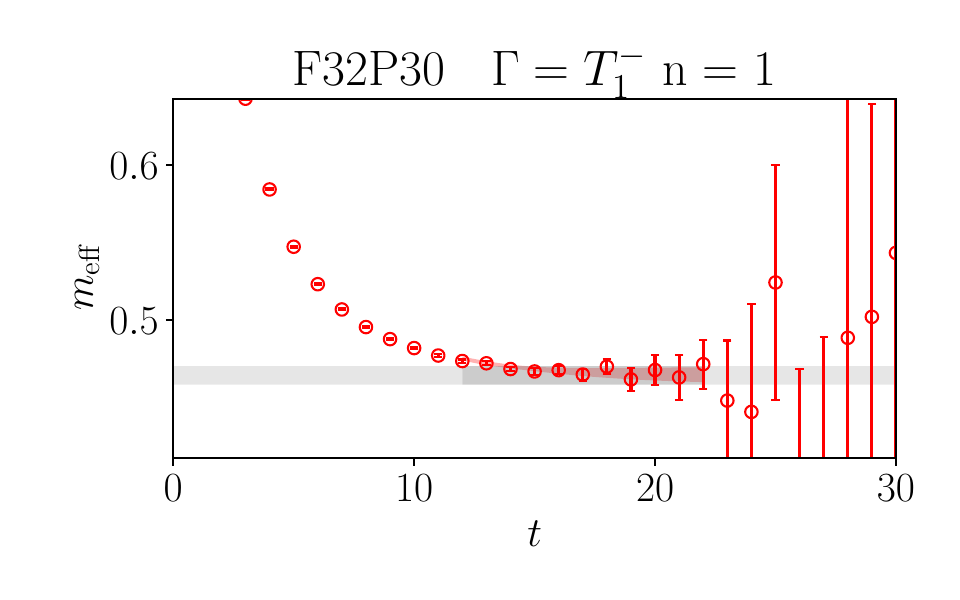}
\includegraphics[width=0.32\columnwidth]{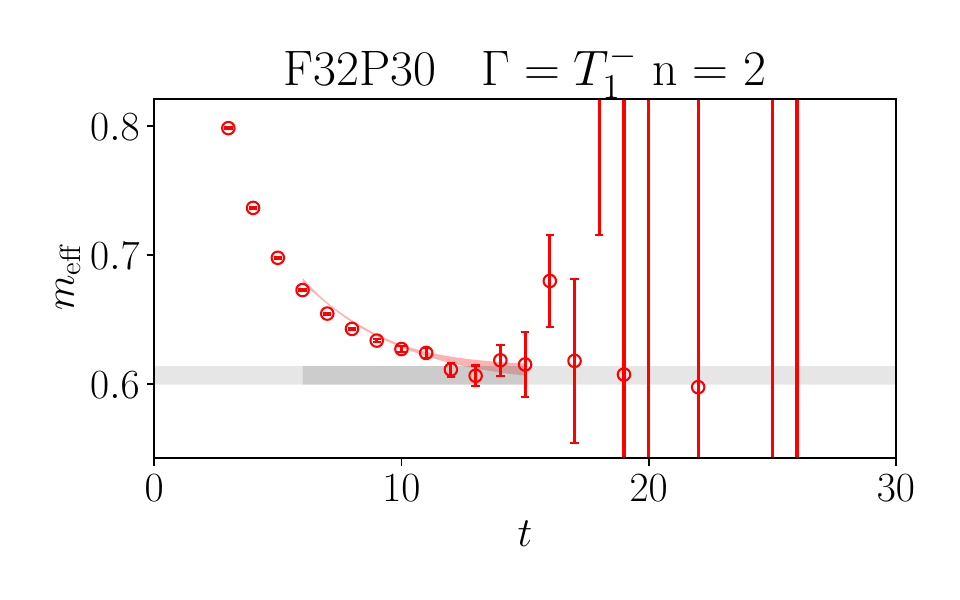}
\\
\includegraphics[width=0.32\columnwidth]{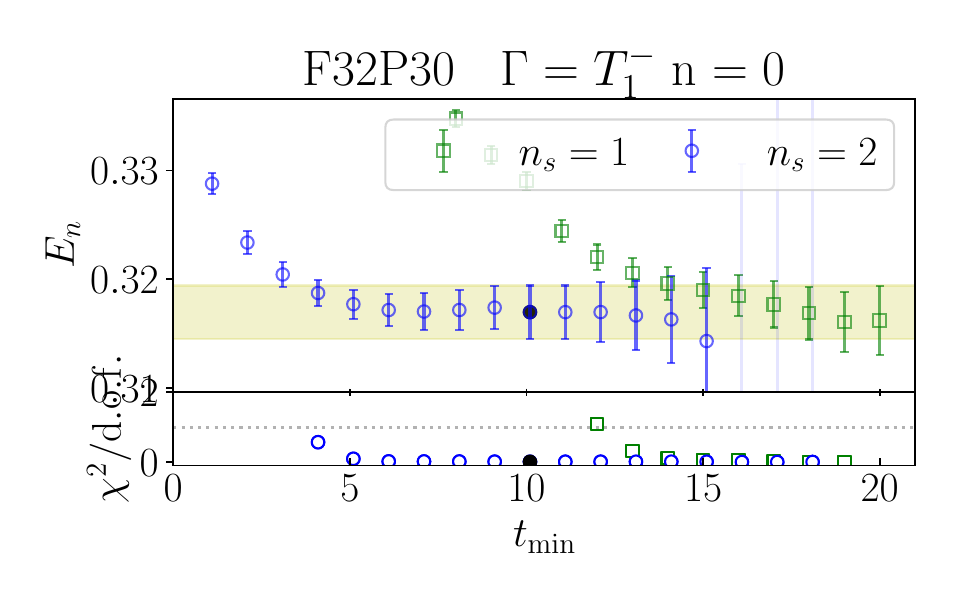}
\includegraphics[width=0.32\columnwidth]{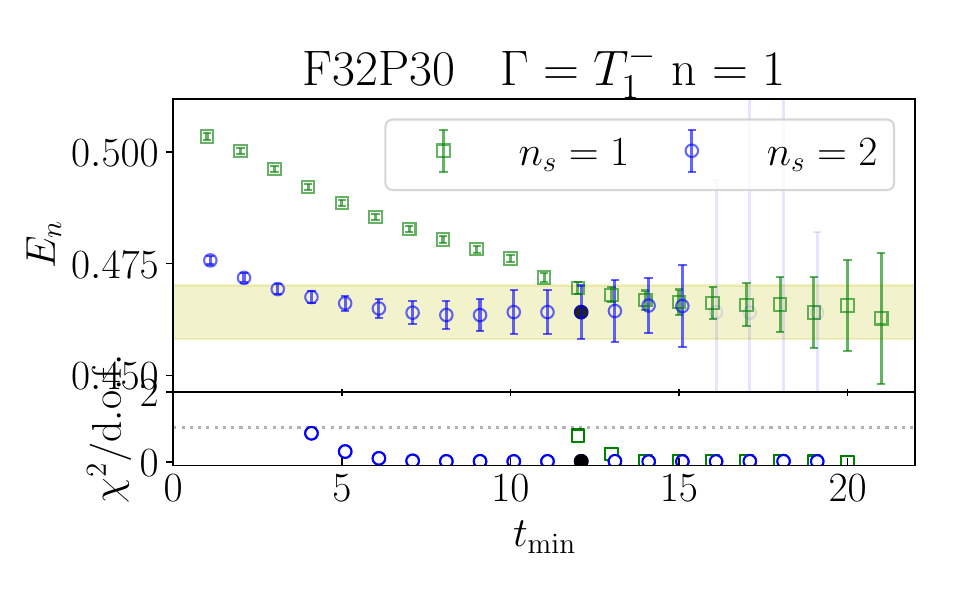}
\includegraphics[width=0.32\columnwidth]{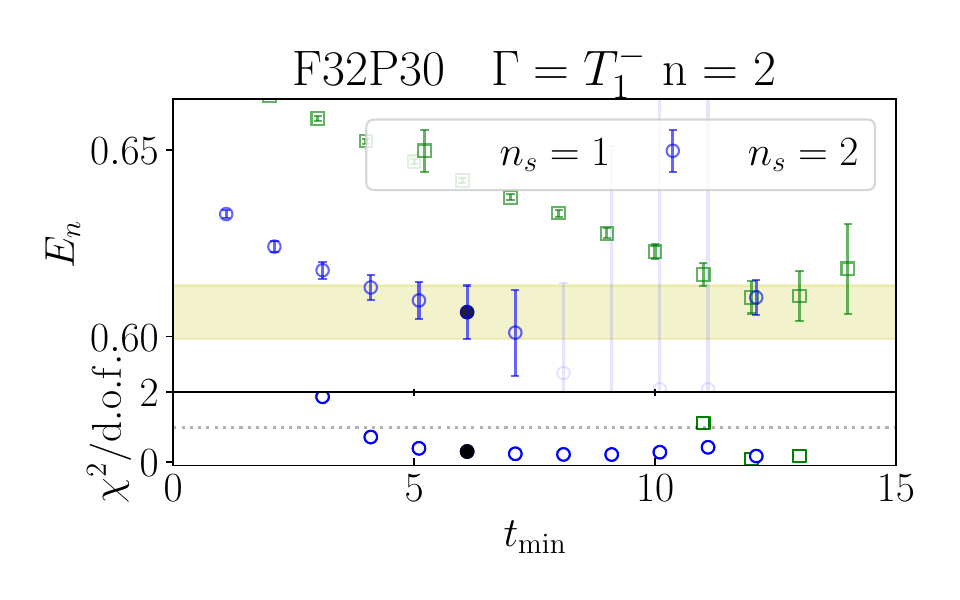}
\caption{Fitting for the $I=1$ $\pi\pi$ channel for ensemble F32P30.}
\label{fig:pipi-I=1-fit-F32P30}
\end{figure}

\begin{figure}[htbp]
\centering
\includegraphics[width=0.32\columnwidth]{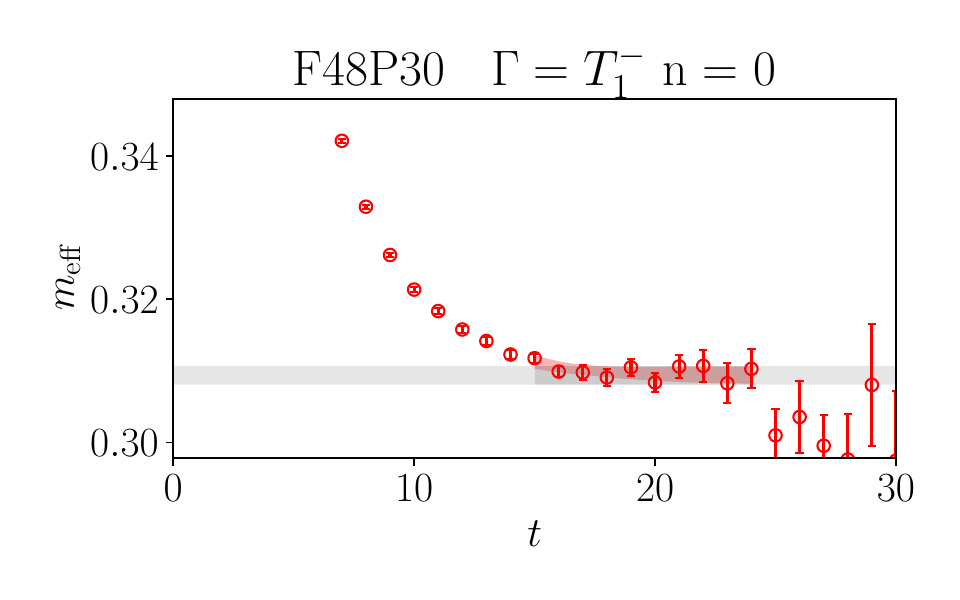}
\includegraphics[width=0.32\columnwidth]{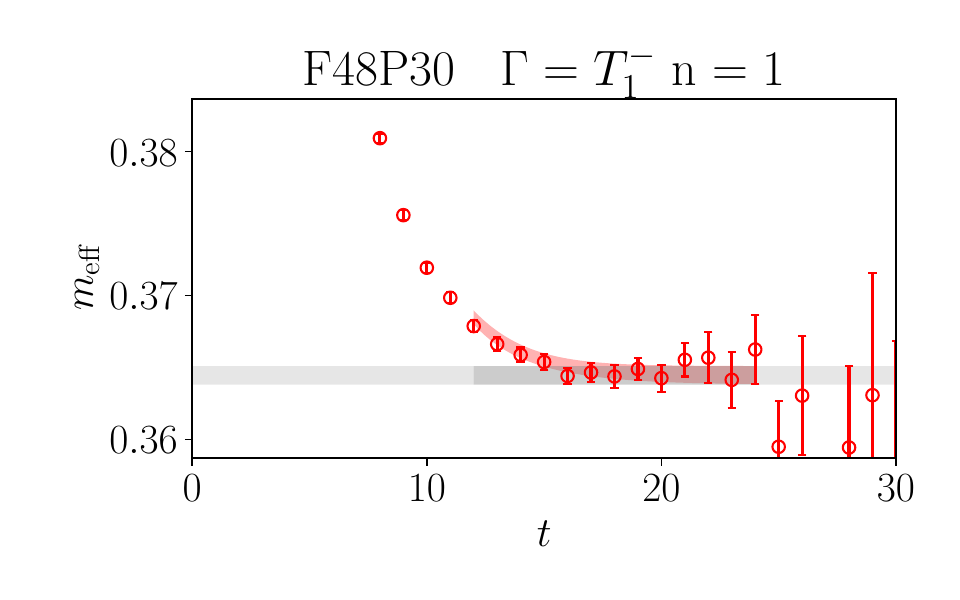}
\includegraphics[width=0.32\columnwidth]{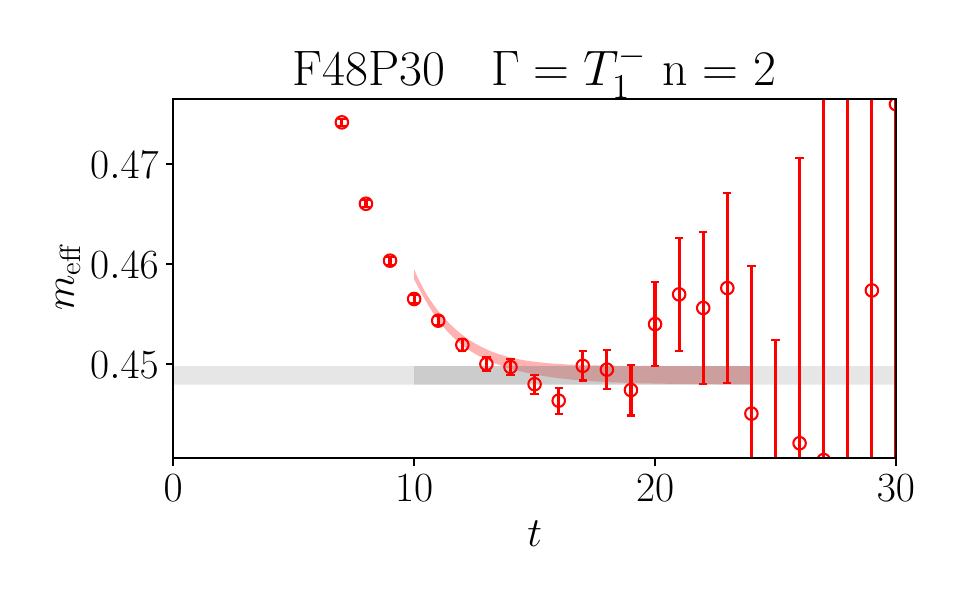}
\\
\includegraphics[width=0.32\columnwidth]{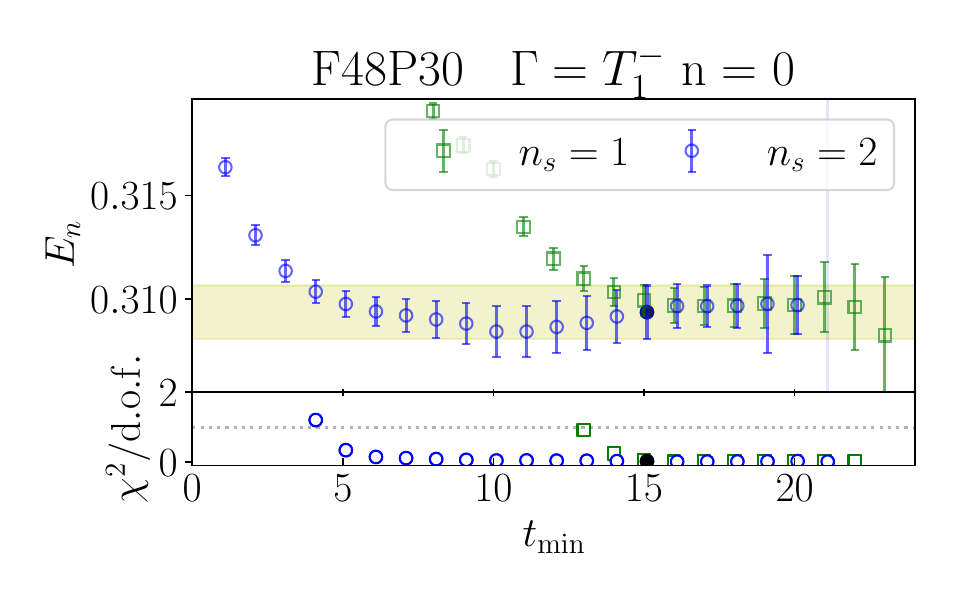}
\includegraphics[width=0.32\columnwidth]{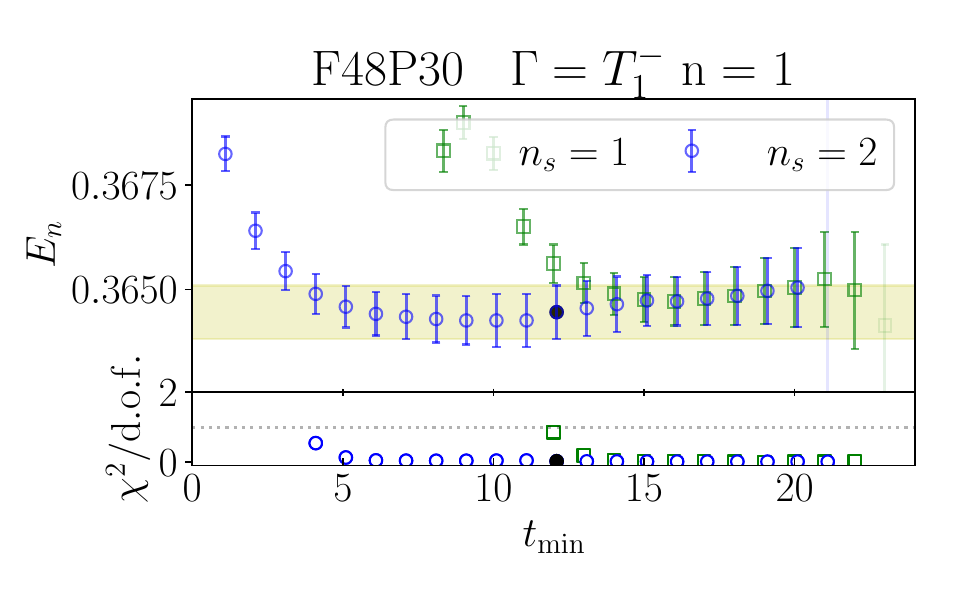}
\includegraphics[width=0.32\columnwidth]{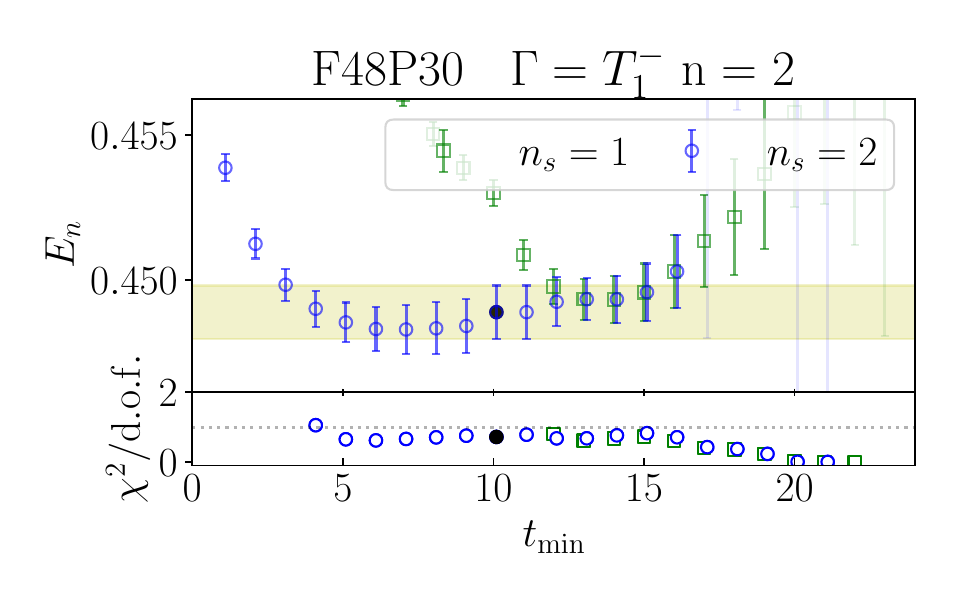}
\caption{Fitting for the $I=1$ $\pi\pi$ channel for ensemble F48P30.}
\label{fig:pipi-I=1-fit-F48P30}
\end{figure}

\begin{figure}[htbp]
\centering
\includegraphics[width=0.32\columnwidth]{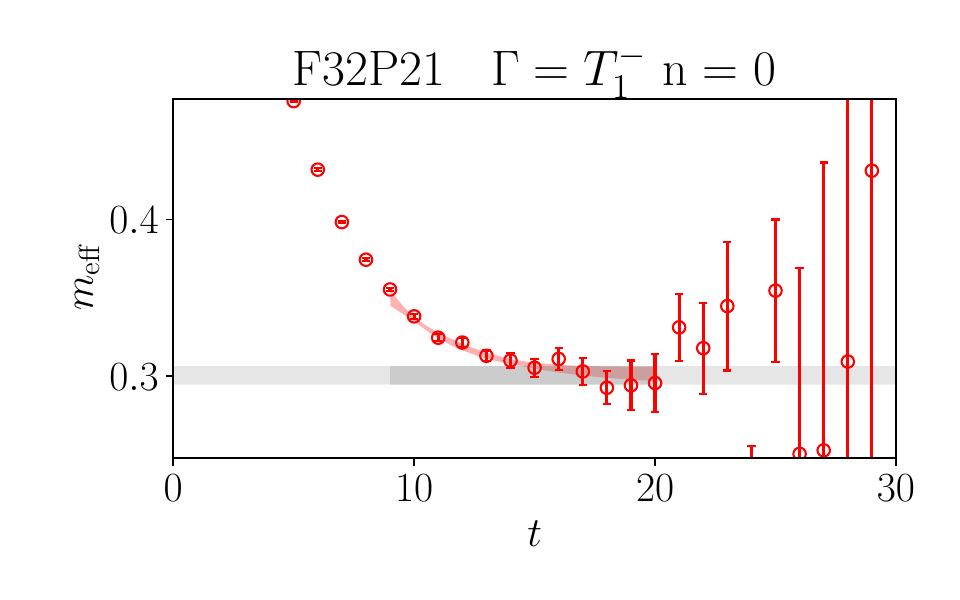}
\includegraphics[width=0.32\columnwidth]{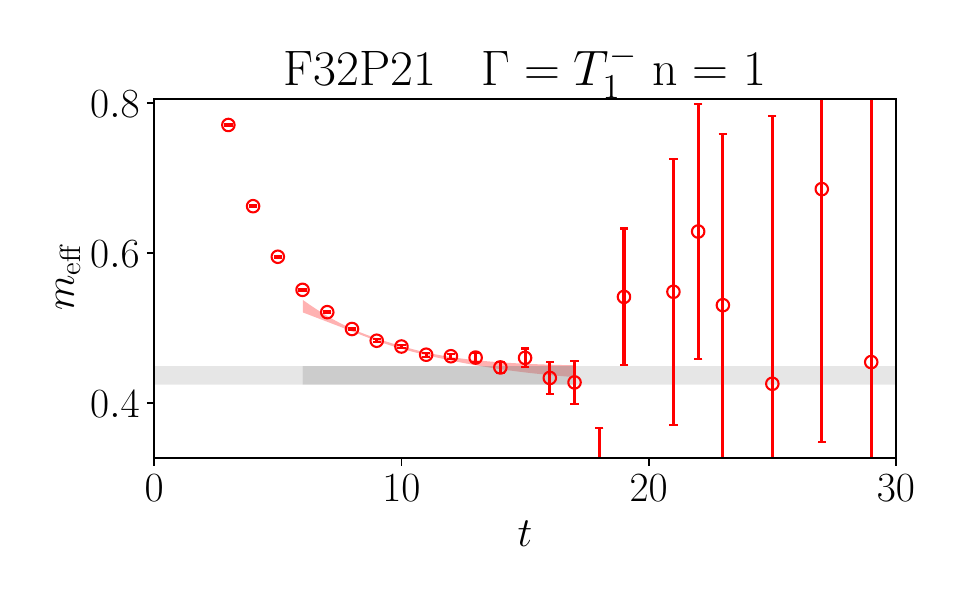}
\includegraphics[width=0.32\columnwidth]{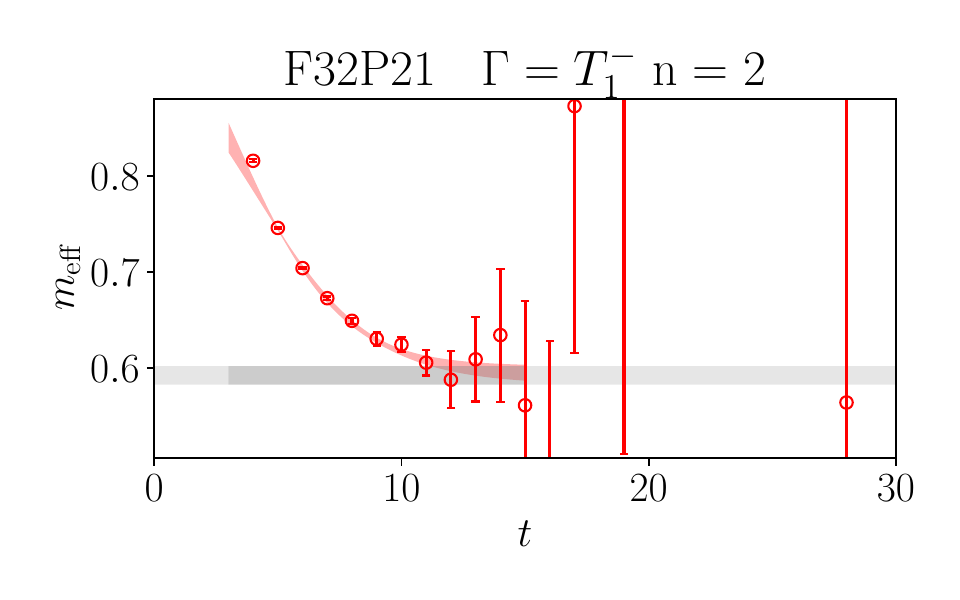}
\\
\includegraphics[width=0.32\columnwidth]{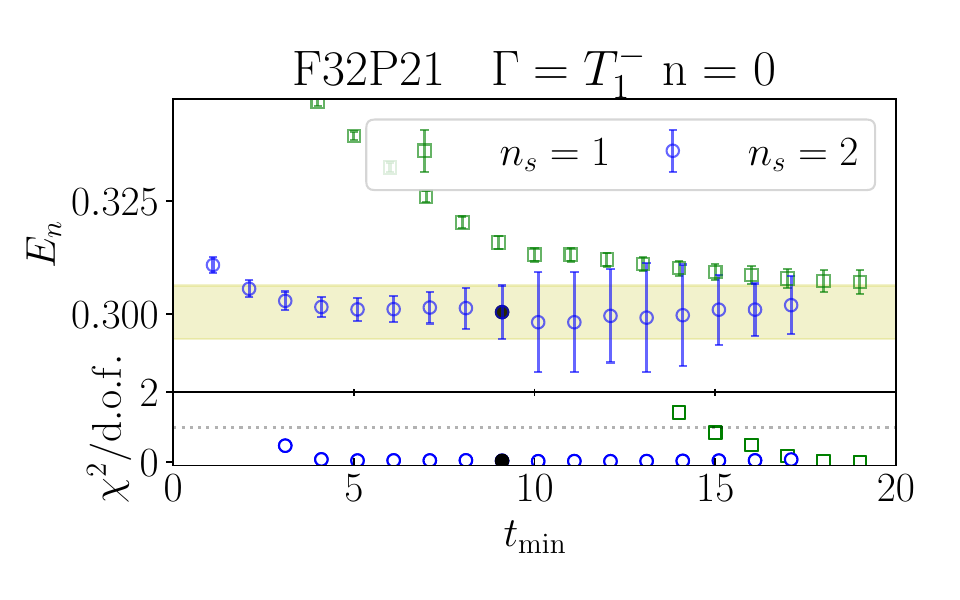}
\includegraphics[width=0.32\columnwidth]{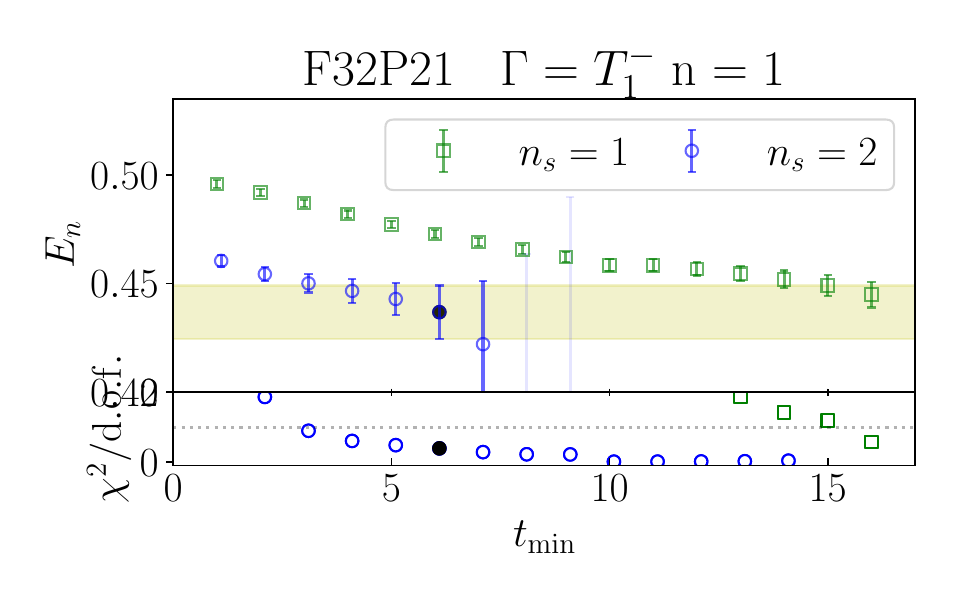}
\includegraphics[width=0.32\columnwidth]{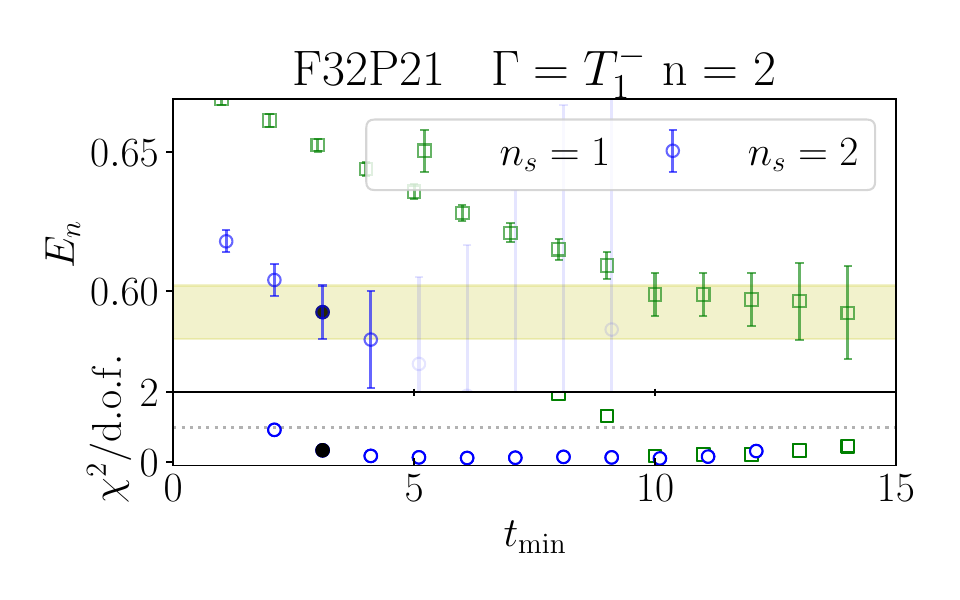}
\caption{Fitting for the $I=1$ $\pi\pi$ channel for ensemble F32P21.}
\label{fig:pipi-I=1-fit-F32P21}
\end{figure}

\begin{figure}[htbp]
\centering
\includegraphics[width=0.32\columnwidth]{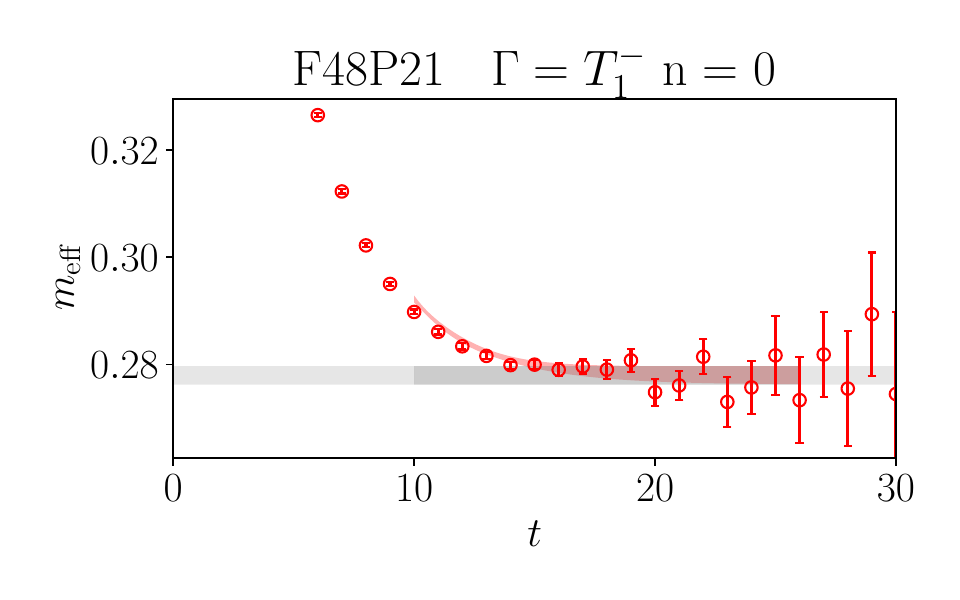}
\includegraphics[width=0.32\columnwidth]{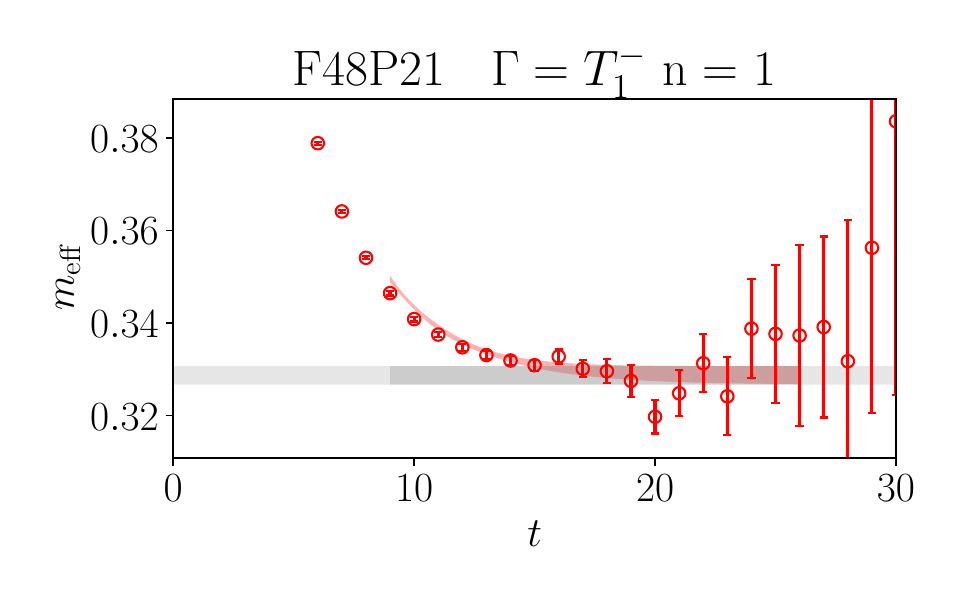}
\includegraphics[width=0.32\columnwidth]{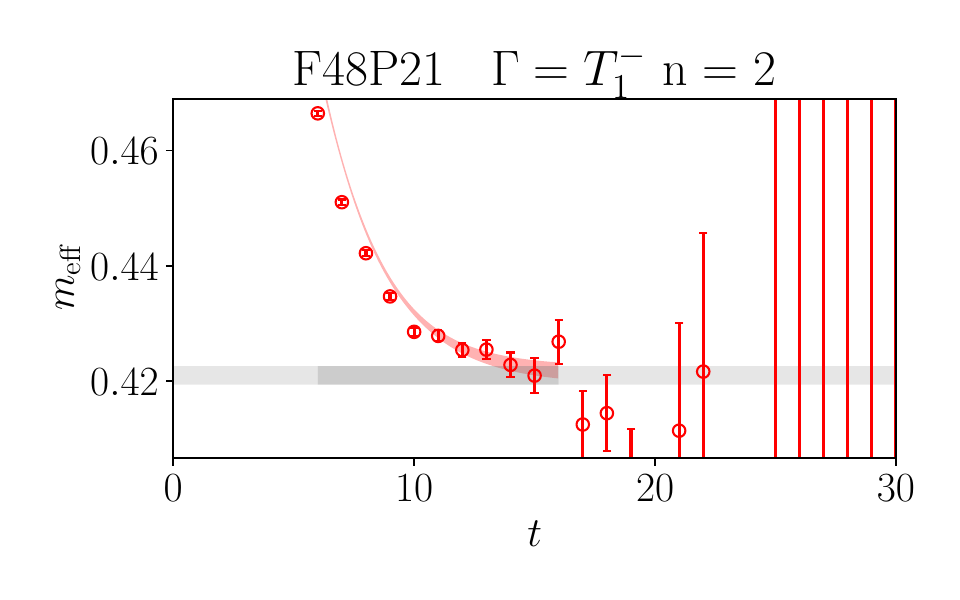}
\\
\includegraphics[width=0.32\columnwidth]{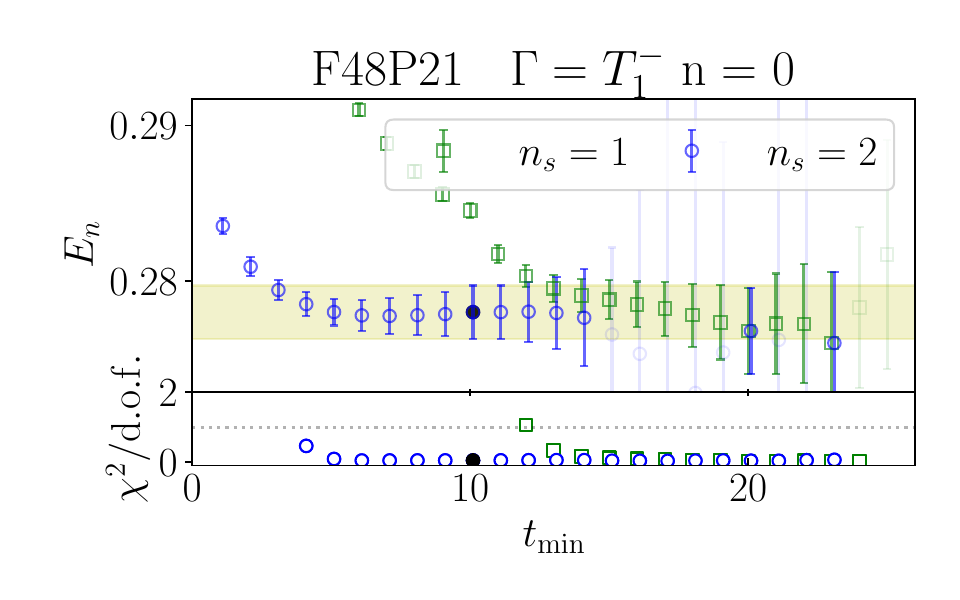}
\includegraphics[width=0.32\columnwidth]{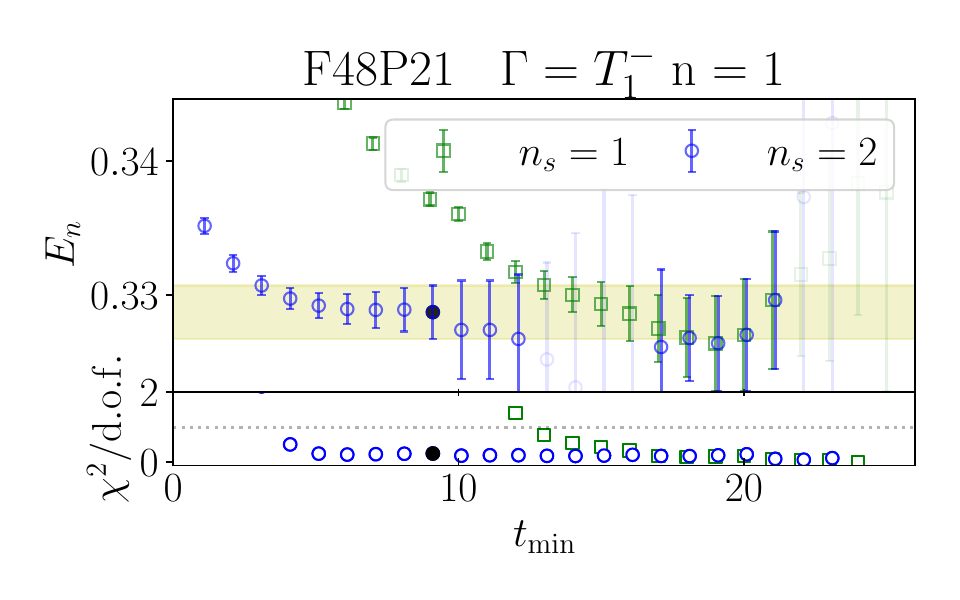}
\includegraphics[width=0.32\columnwidth]{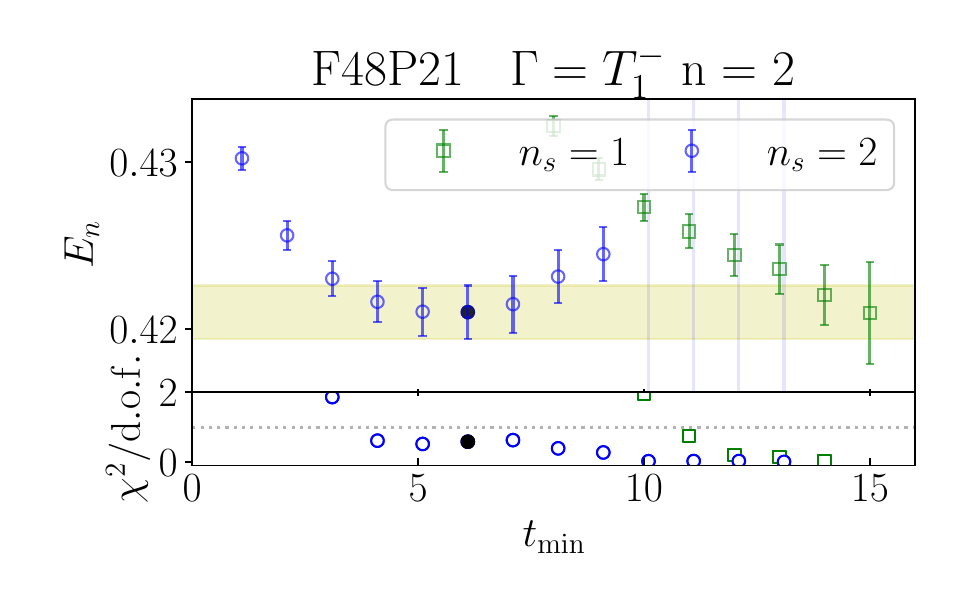}
\caption{Fitting for the $I=1$ $\pi\pi$ channel for ensemble F48P21.}
\label{fig:pipi-I=1-fit-F48P21}
\end{figure}

\begin{figure}[htbp]
\centering
\includegraphics[width=0.32\columnwidth]{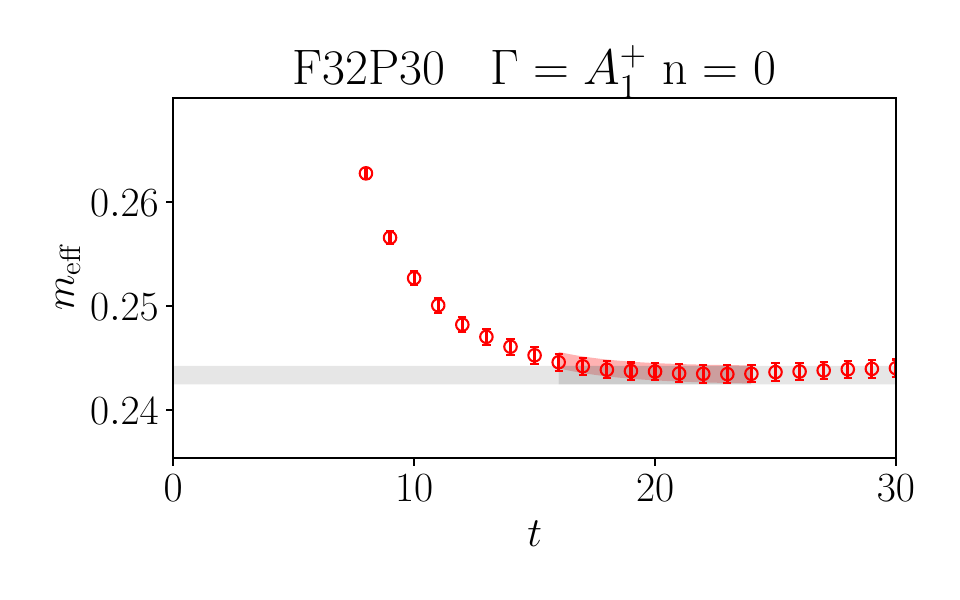}
\includegraphics[width=0.32\columnwidth]{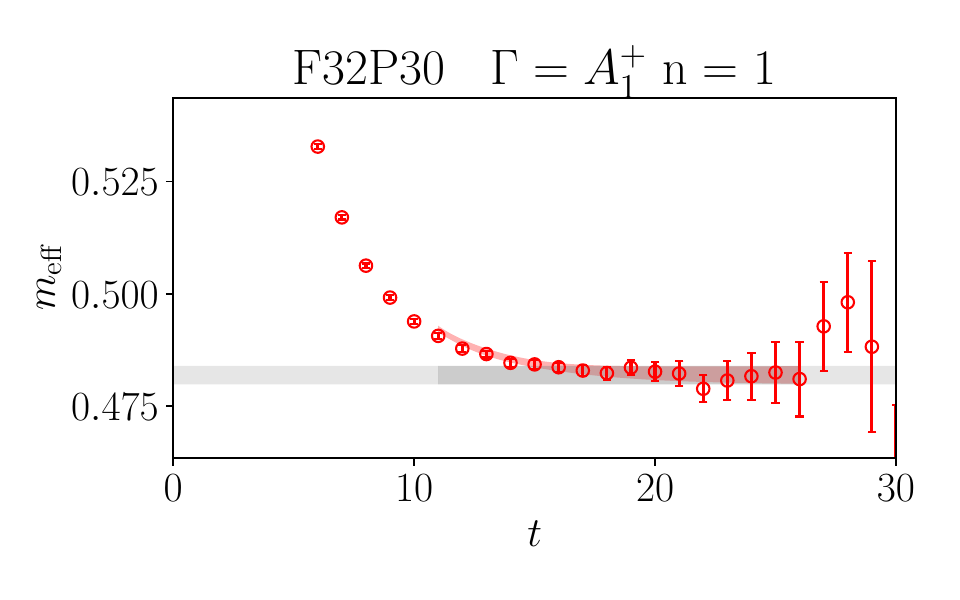}
\includegraphics[width=0.32\columnwidth]{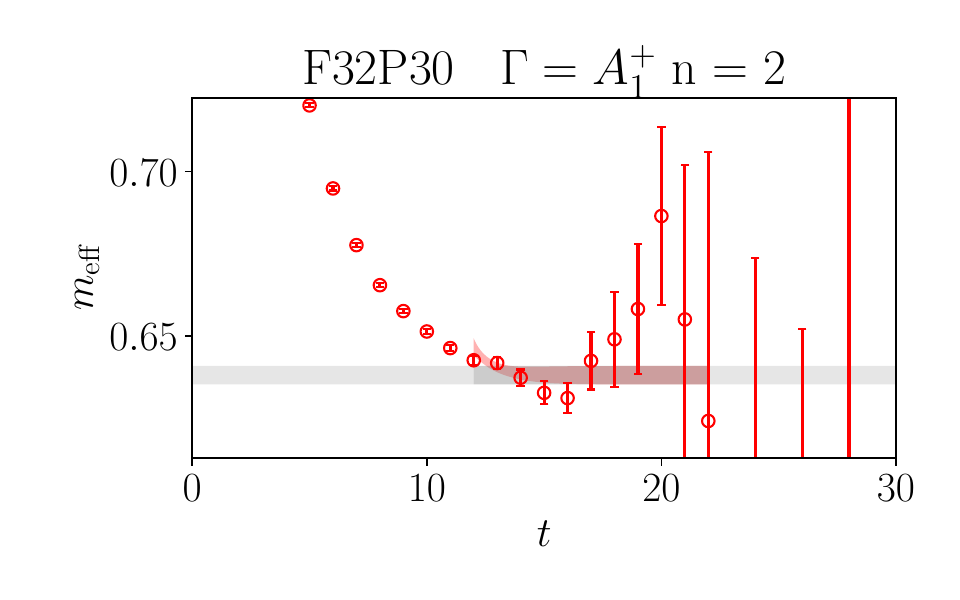}
\\
\includegraphics[width=0.32\columnwidth]{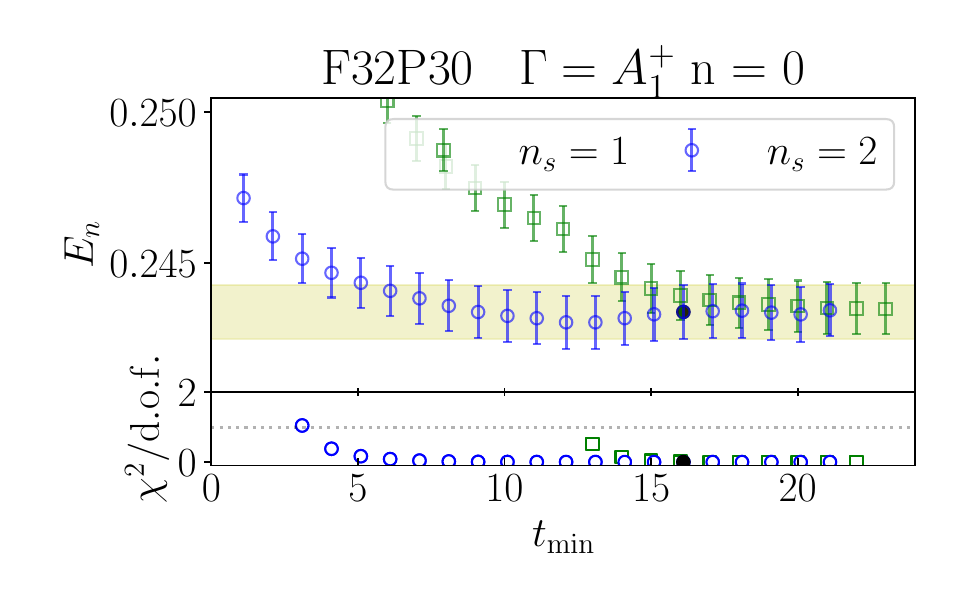}
\includegraphics[width=0.32\columnwidth]{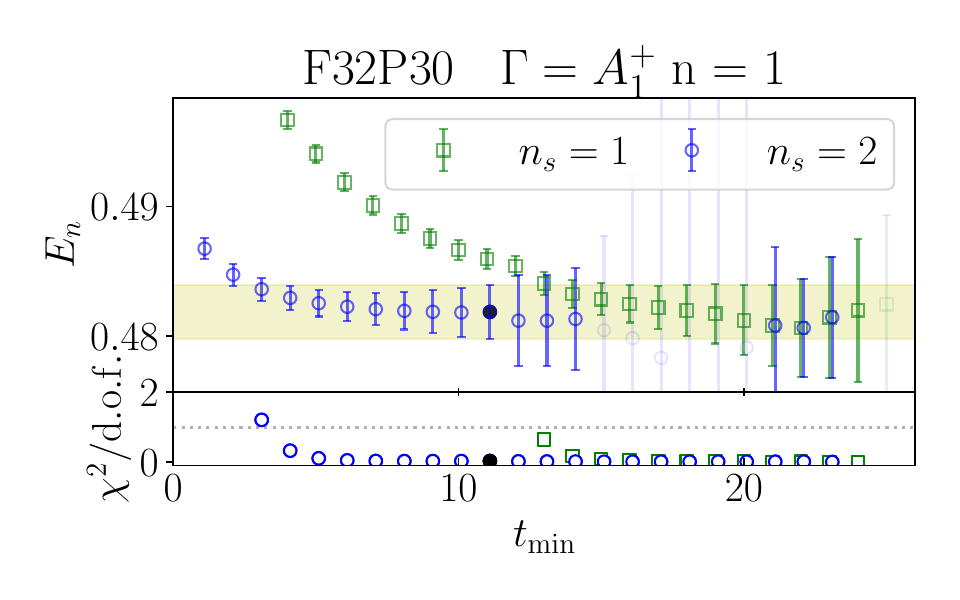}
\includegraphics[width=0.32\columnwidth]{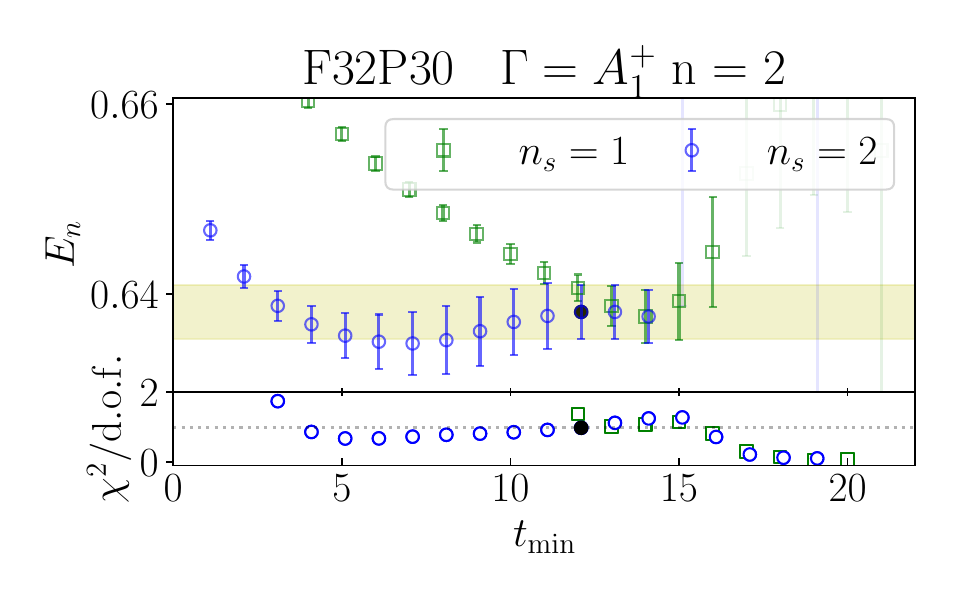}
\caption{Fitting for the $I=2$ $\pi\pi$ channel for ensemble F32P30.}
\label{fig:pipi-I=2-fit-F32P30}
\end{figure}

\begin{figure}[htbp]
\centering
\includegraphics[width=0.32\columnwidth]{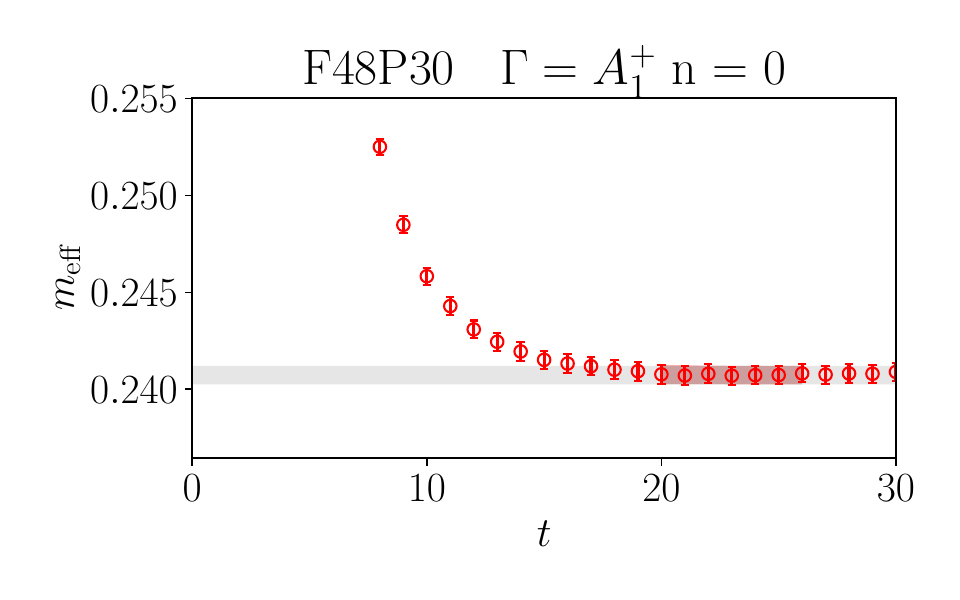}
\includegraphics[width=0.32\columnwidth]{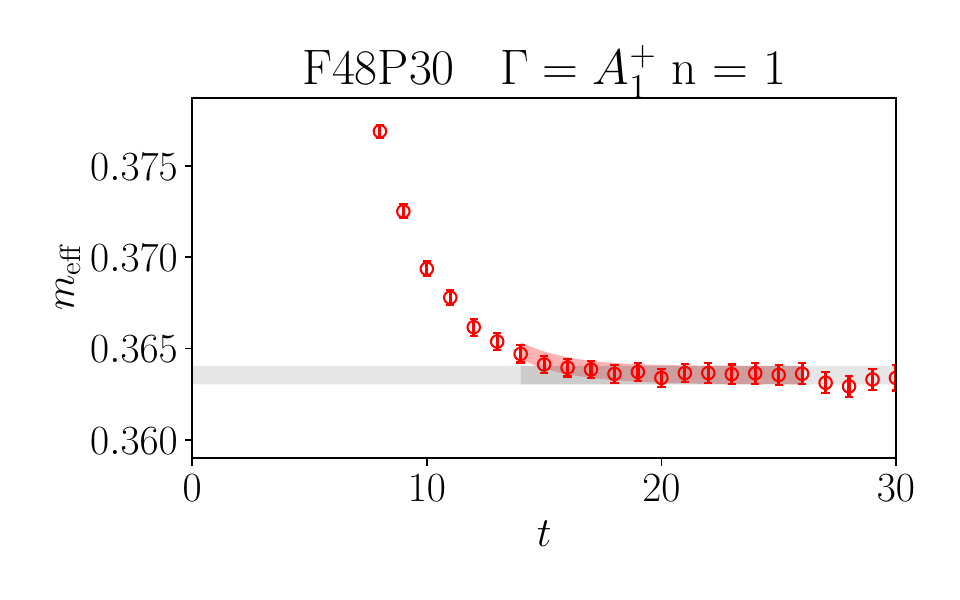}
\includegraphics[width=0.32\columnwidth]{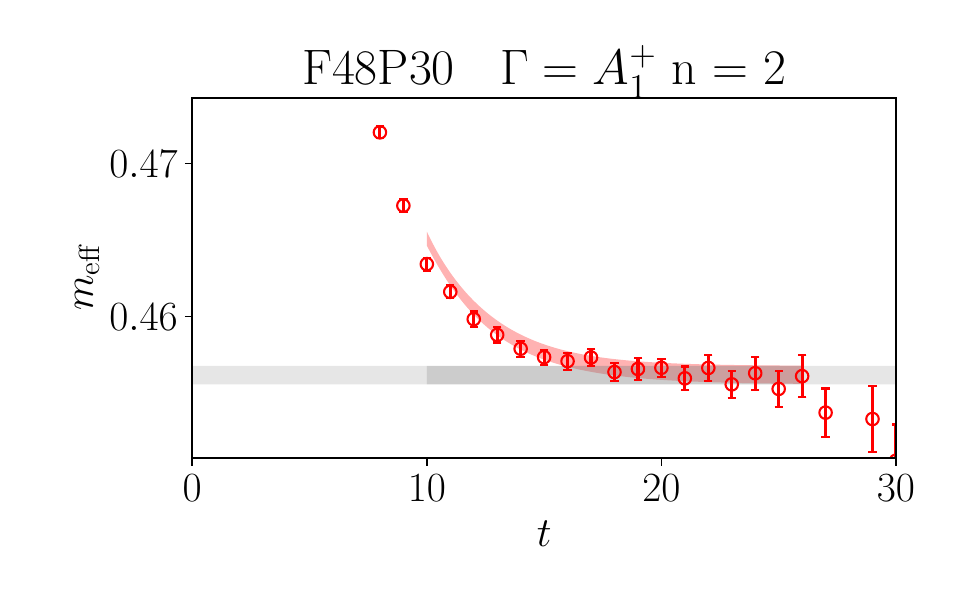}
\\
\includegraphics[width=0.32\columnwidth]{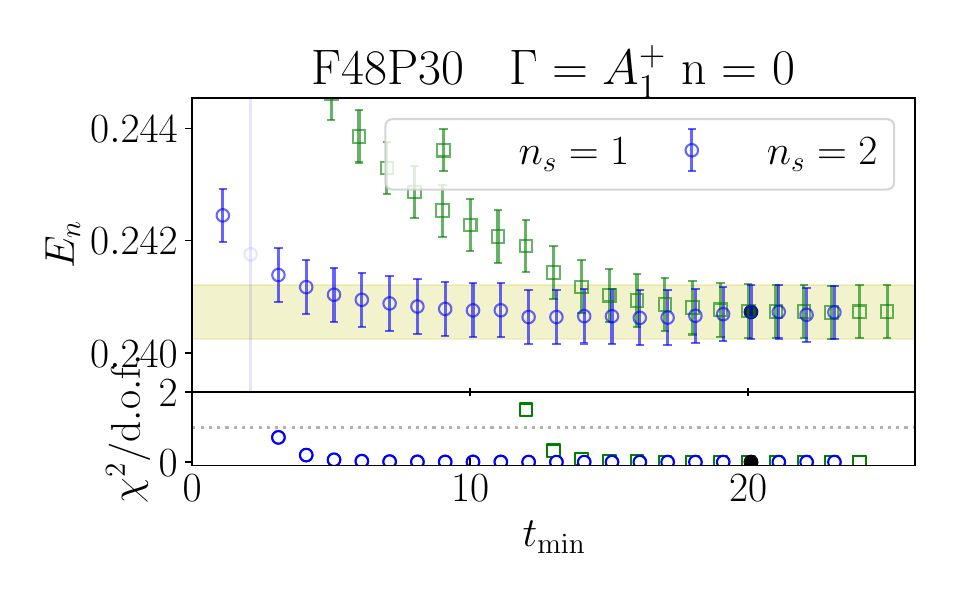}
\includegraphics[width=0.32\columnwidth]{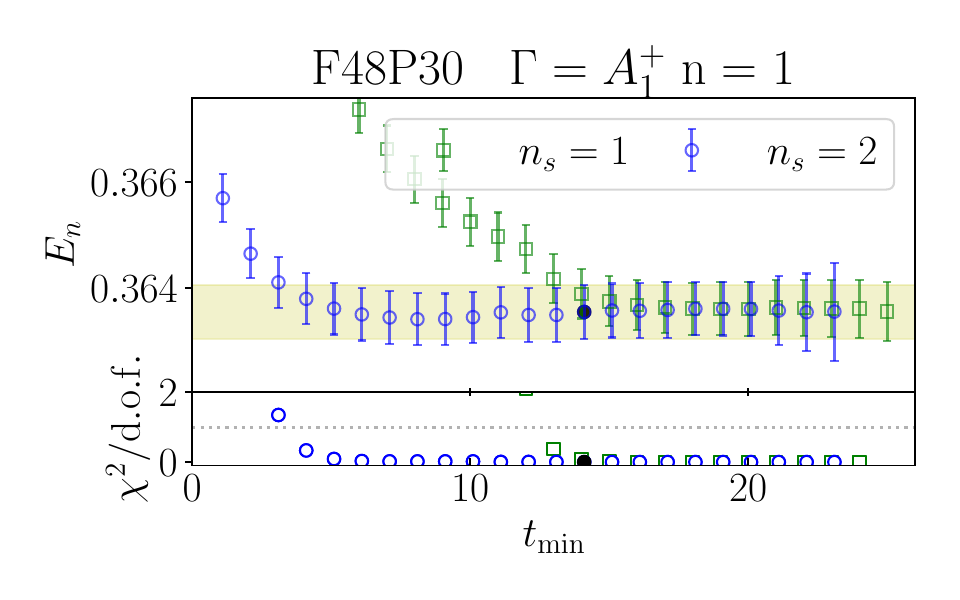}
\includegraphics[width=0.32\columnwidth]{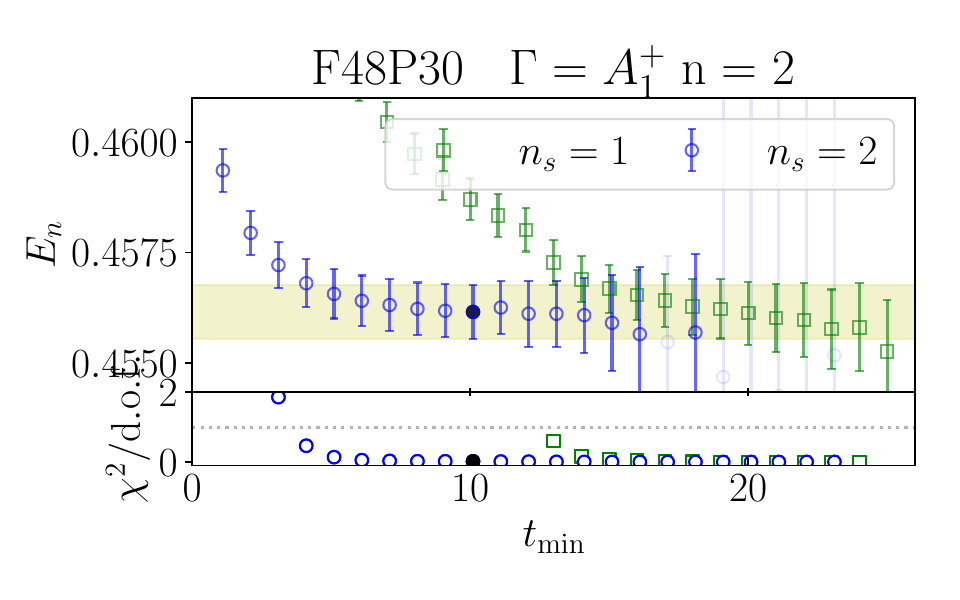}
\caption{Fitting for the $I=2$ $\pi\pi$ channel for ensemble F48P30.}
\label{fig:pipi-I=2-fit-F48P30}
\end{figure}

\begin{figure}[htbp]
\centering
\includegraphics[width=0.32\columnwidth]{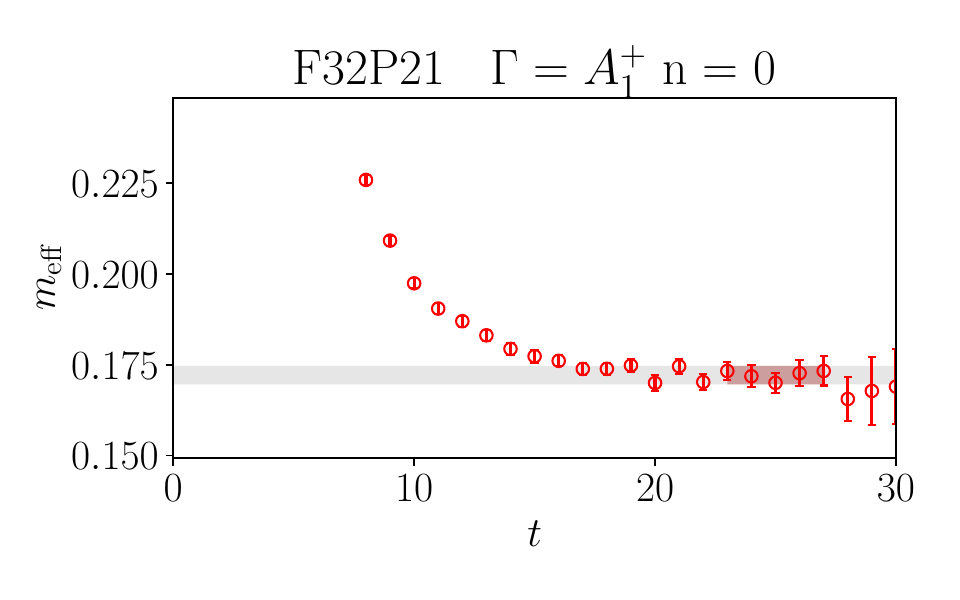}
\includegraphics[width=0.32\columnwidth]{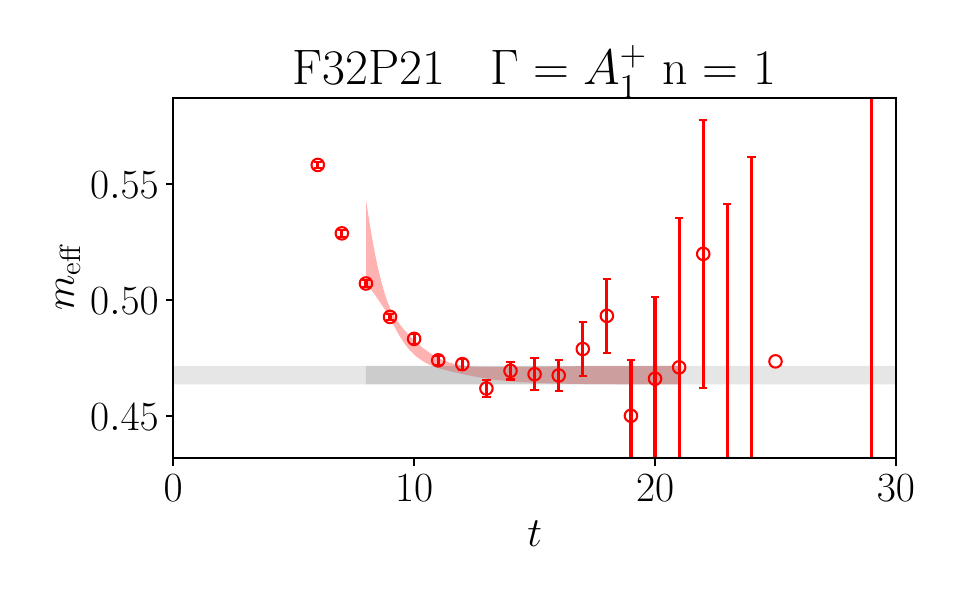}
\includegraphics[width=0.32\columnwidth]{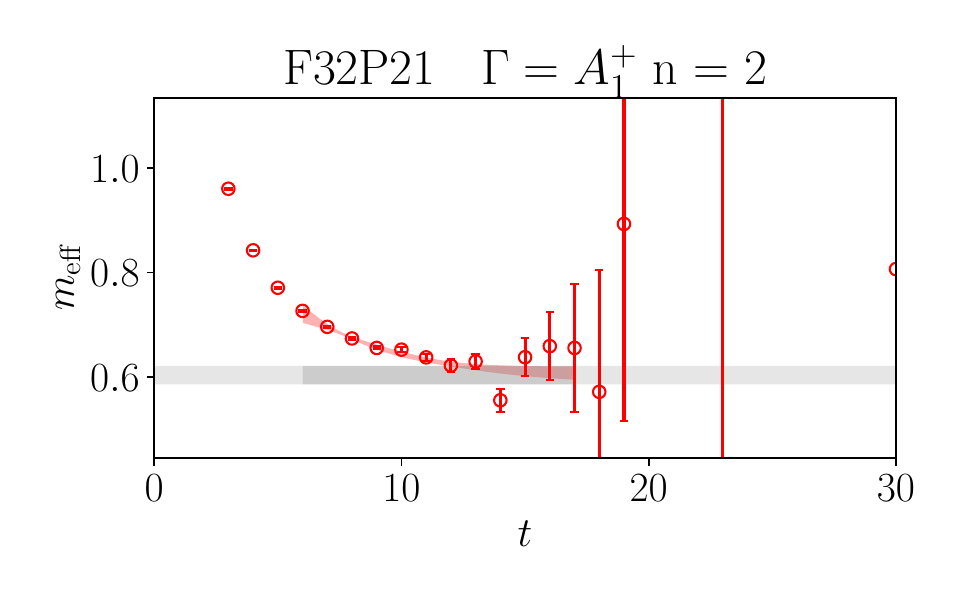}
\\
\includegraphics[width=0.32\columnwidth]{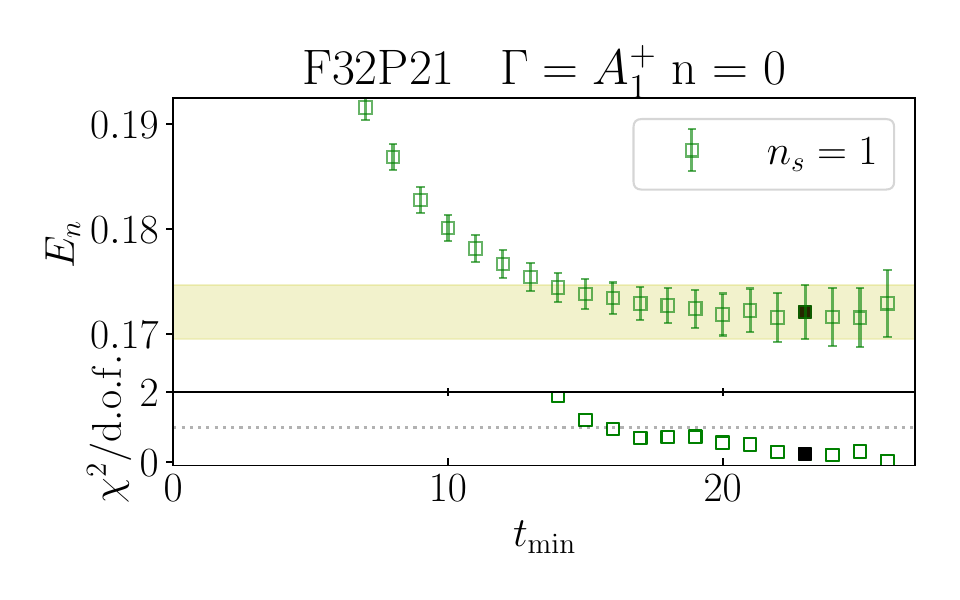}
\includegraphics[width=0.32\columnwidth]{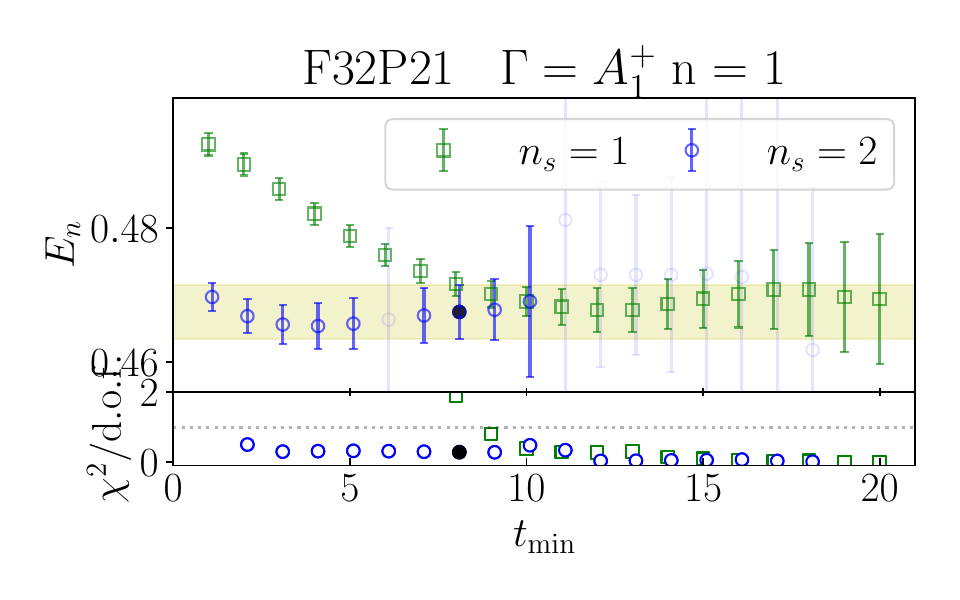}
\includegraphics[width=0.32\columnwidth]{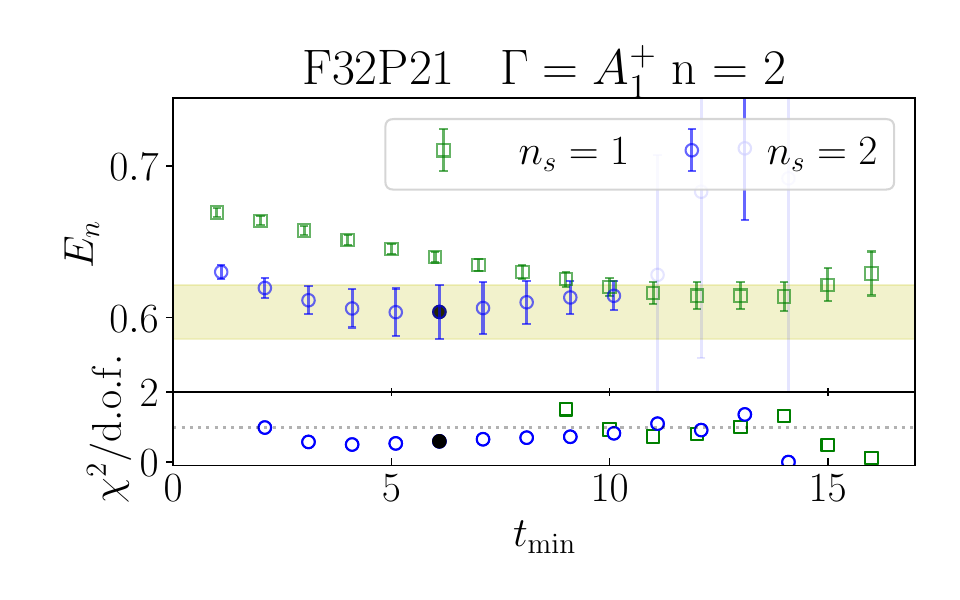}
\caption{Fitting for the $I=2$ $\pi\pi$ channel for ensemble F32P21.}
\label{fig:pipi-I=2-fit-F32P21}
\end{figure}

\begin{figure}[htbp]
\centering
\includegraphics[width=0.32\columnwidth]{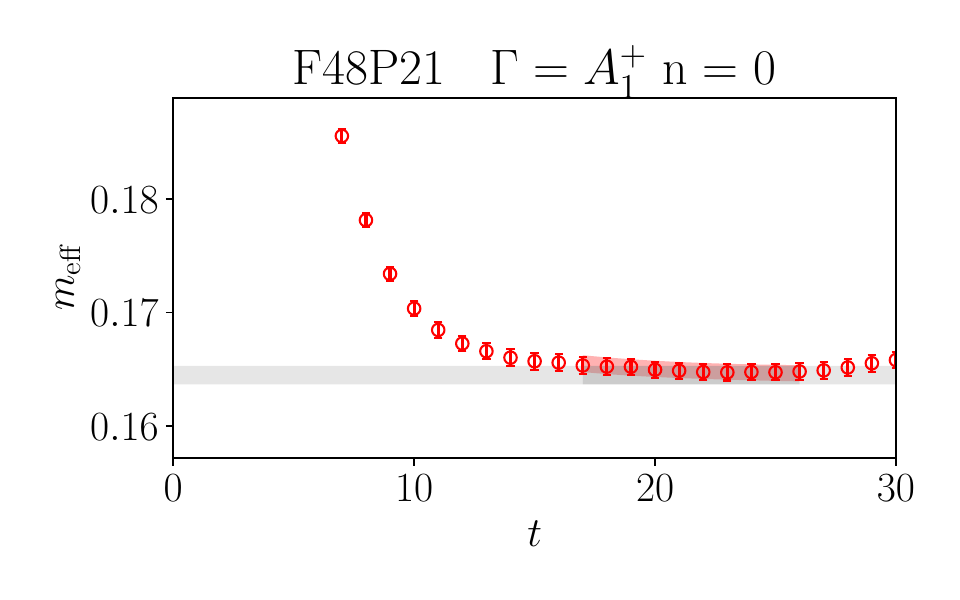}
\includegraphics[width=0.32\columnwidth]{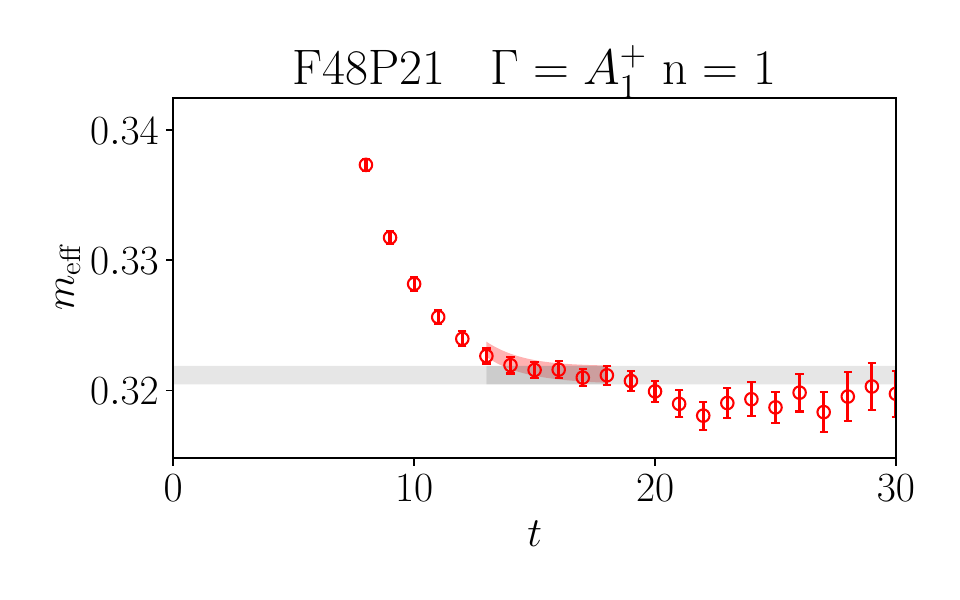}
\includegraphics[width=0.32\columnwidth]{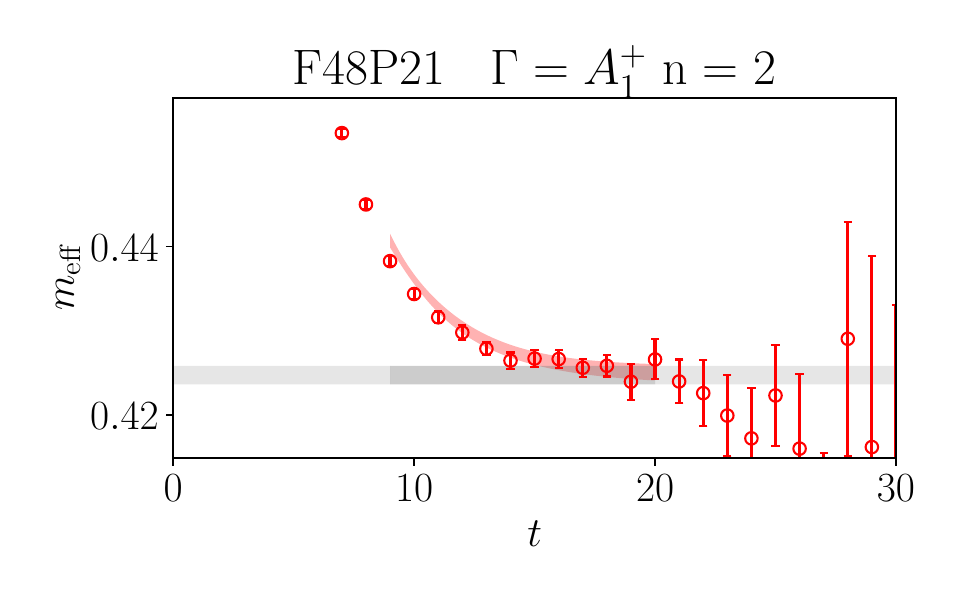}
\\
\includegraphics[width=0.32\columnwidth]{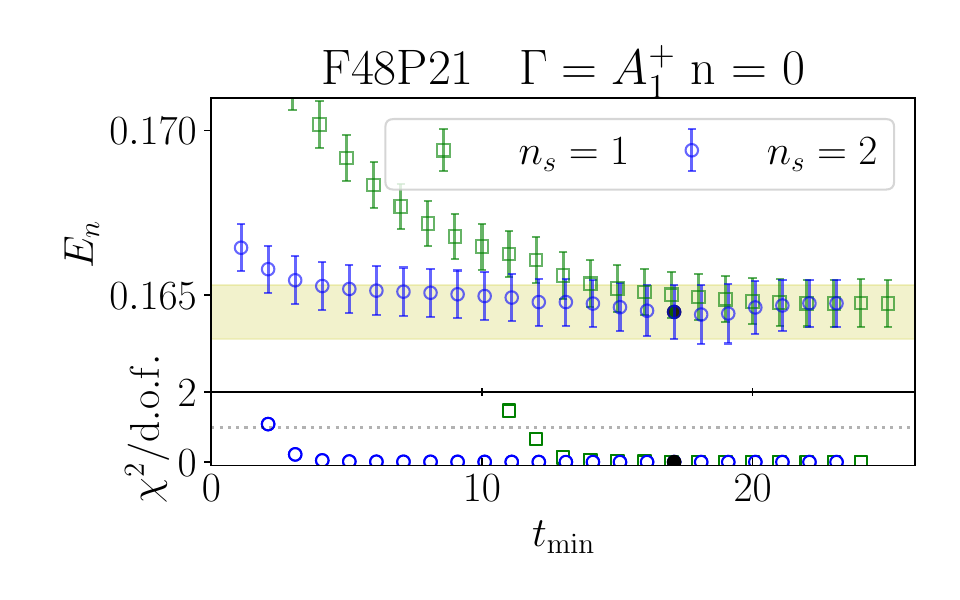}
\includegraphics[width=0.32\columnwidth]{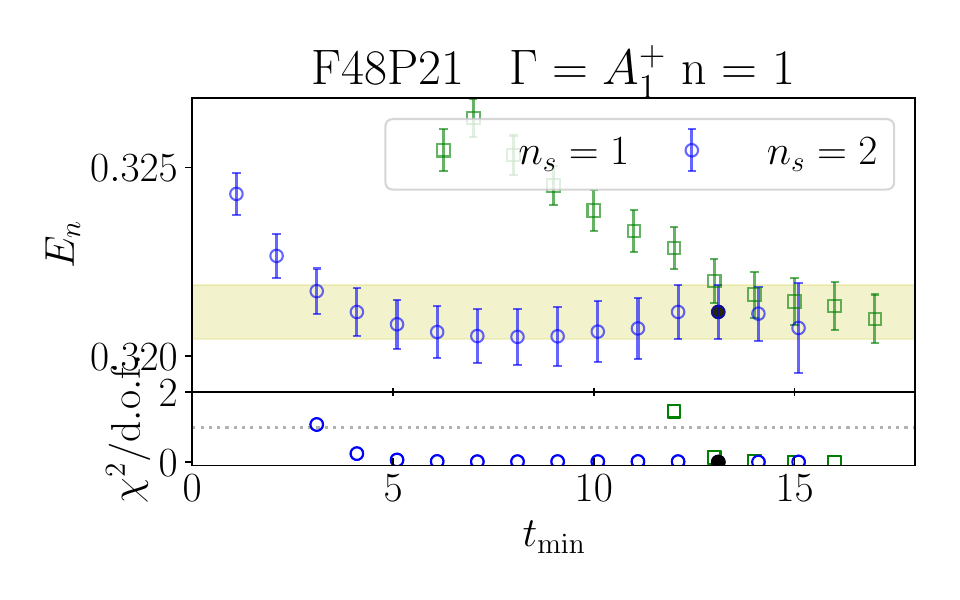}
\includegraphics[width=0.32\columnwidth]{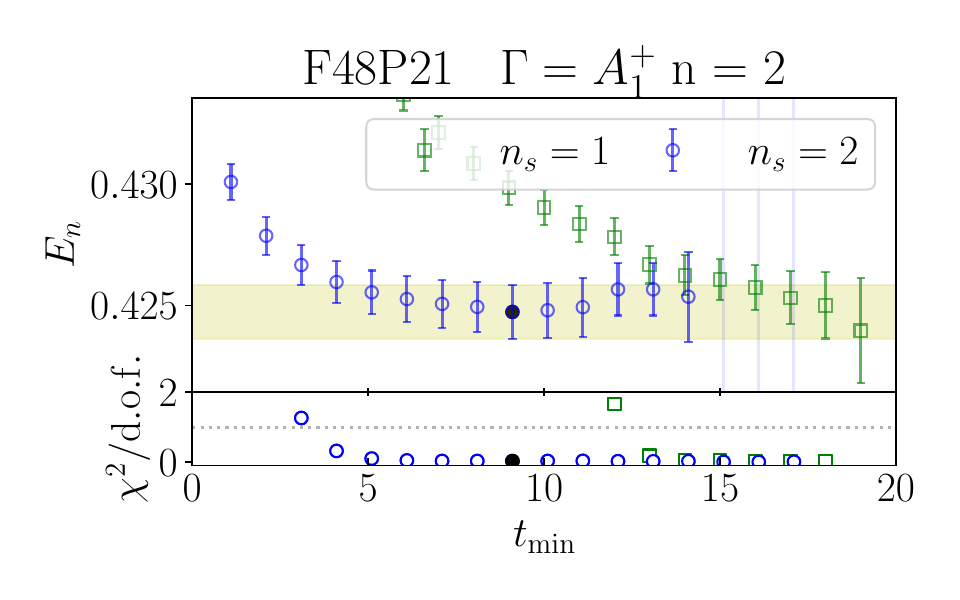}
\caption{Fitting for the $I=2$ $\pi\pi$ channel for ensemble F48P21.}
\label{fig:pipi-I=2-fit-F48P21}
\end{figure}

\begin{figure}[htbp]
\centering
\includegraphics[width=0.32\columnwidth]{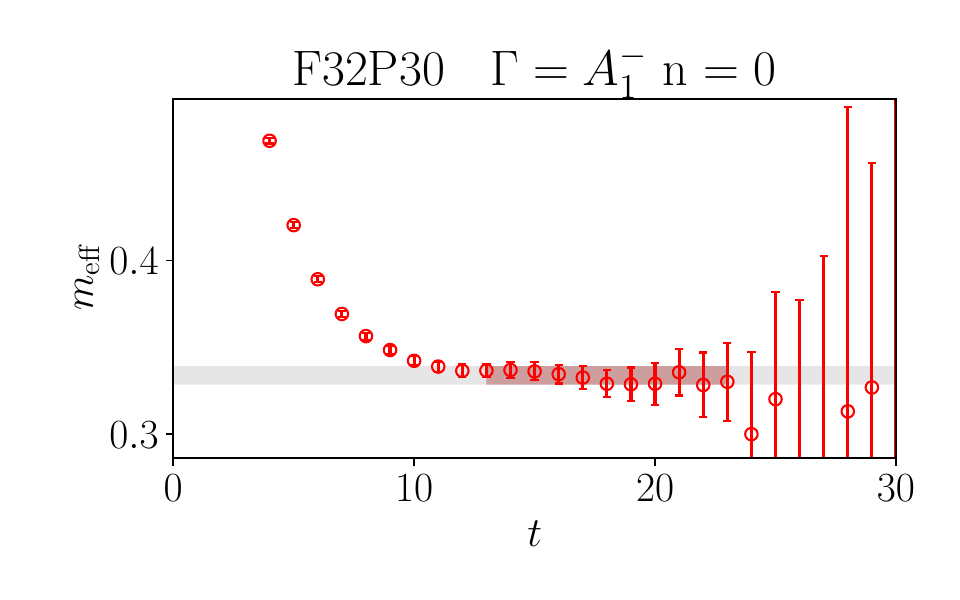}
\includegraphics[width=0.32\columnwidth]{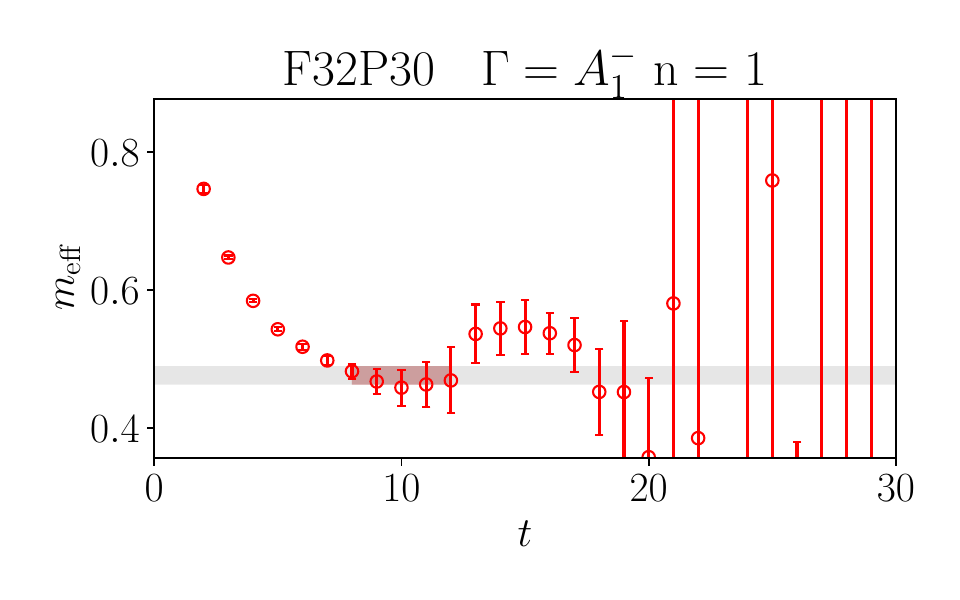}
\includegraphics[width=0.32\columnwidth]{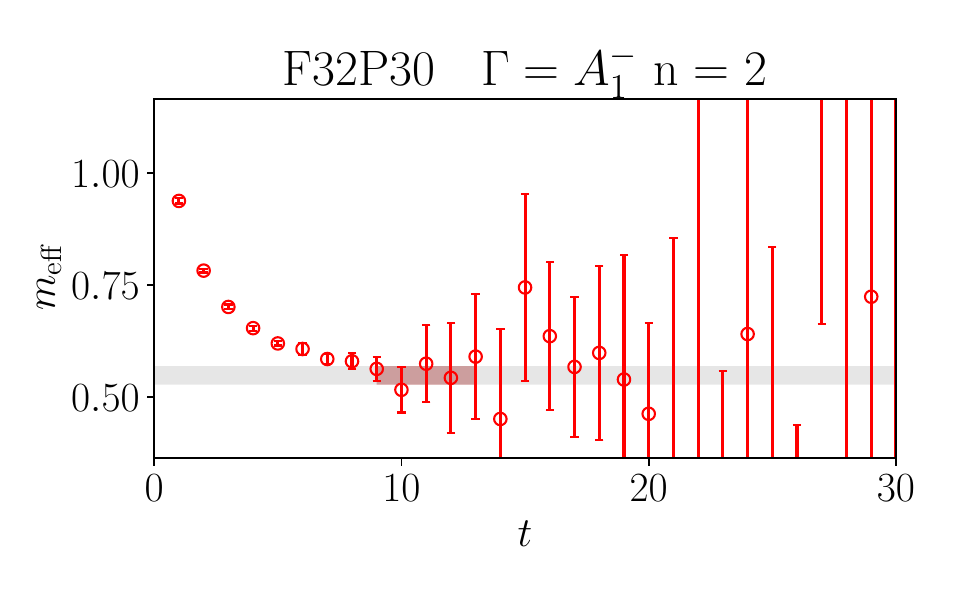}
\\
\includegraphics[width=0.32\columnwidth]{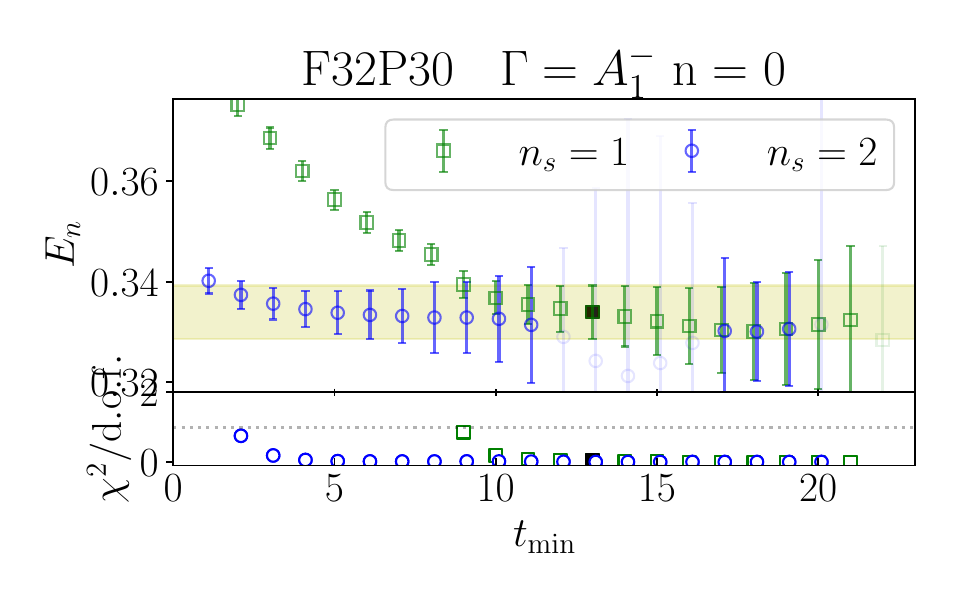}
\includegraphics[width=0.32\columnwidth]{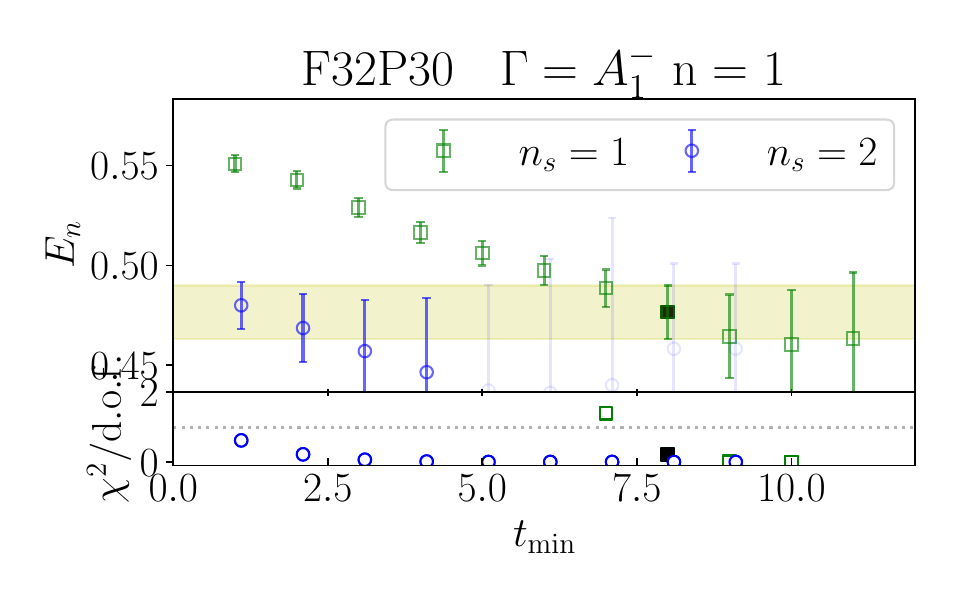}
\includegraphics[width=0.32\columnwidth]{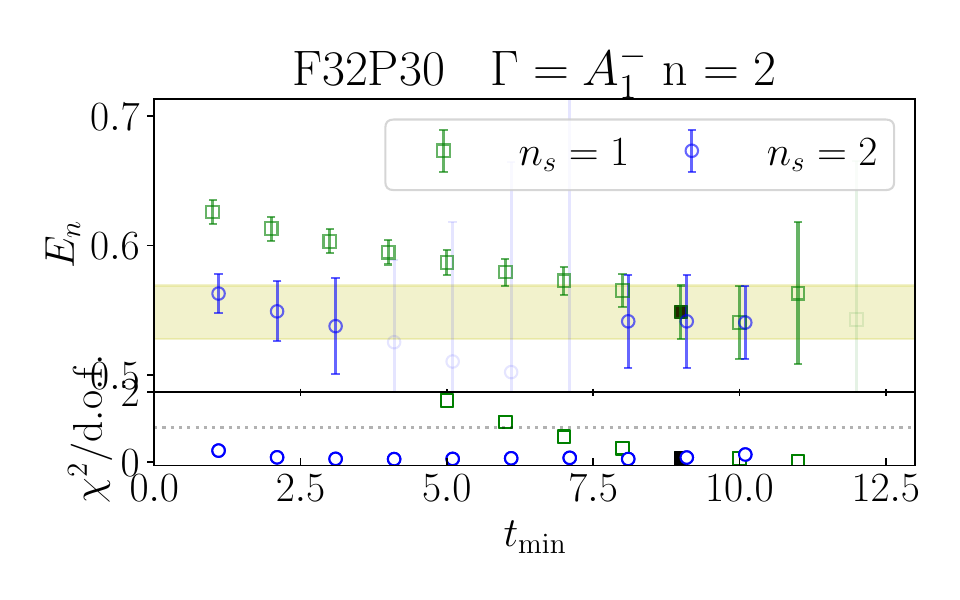}
\\
\includegraphics[width=0.32\columnwidth]{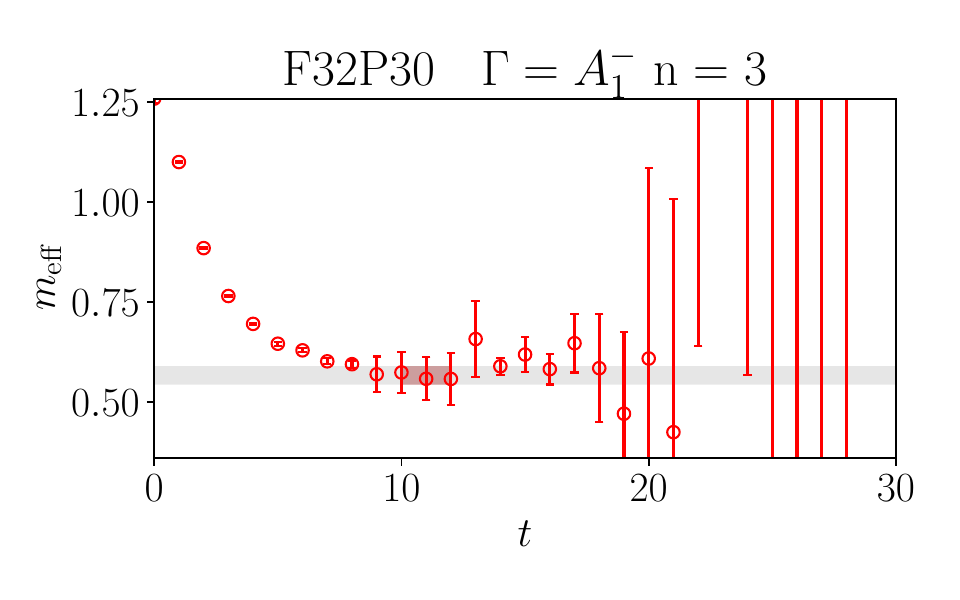}
\includegraphics[width=0.32\columnwidth]{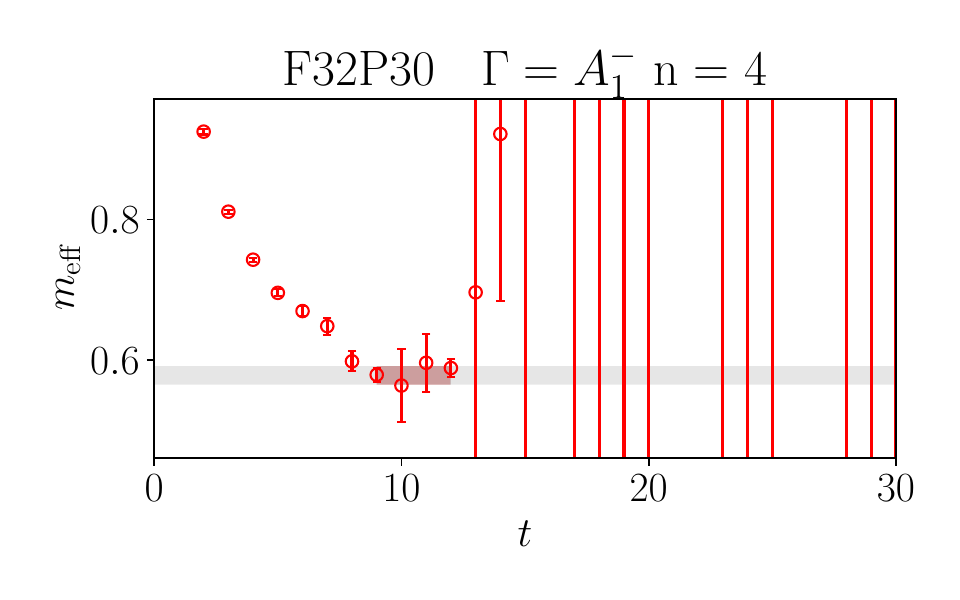}
\includegraphics[width=0.32\columnwidth]{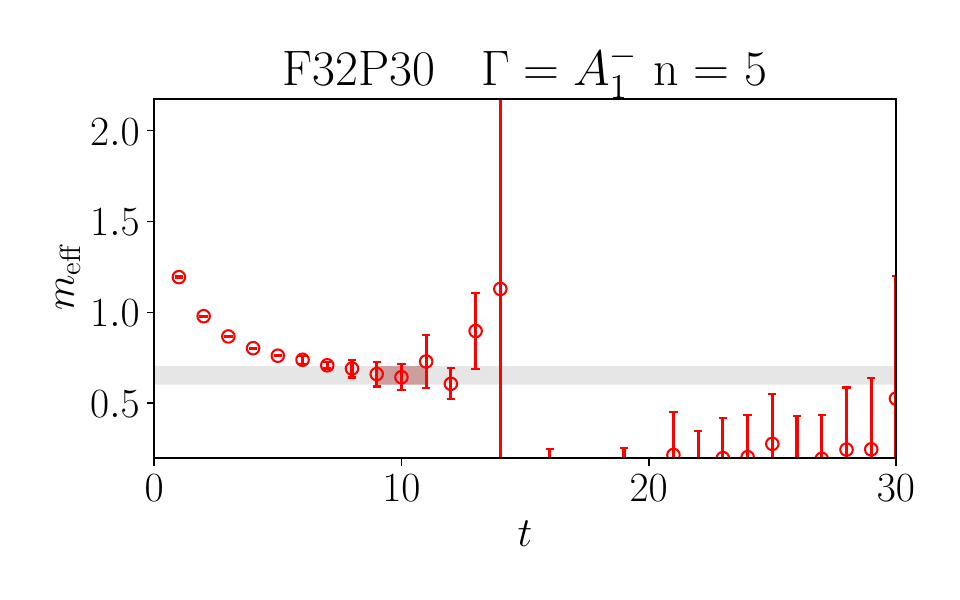}
\\
\includegraphics[width=0.32\columnwidth]{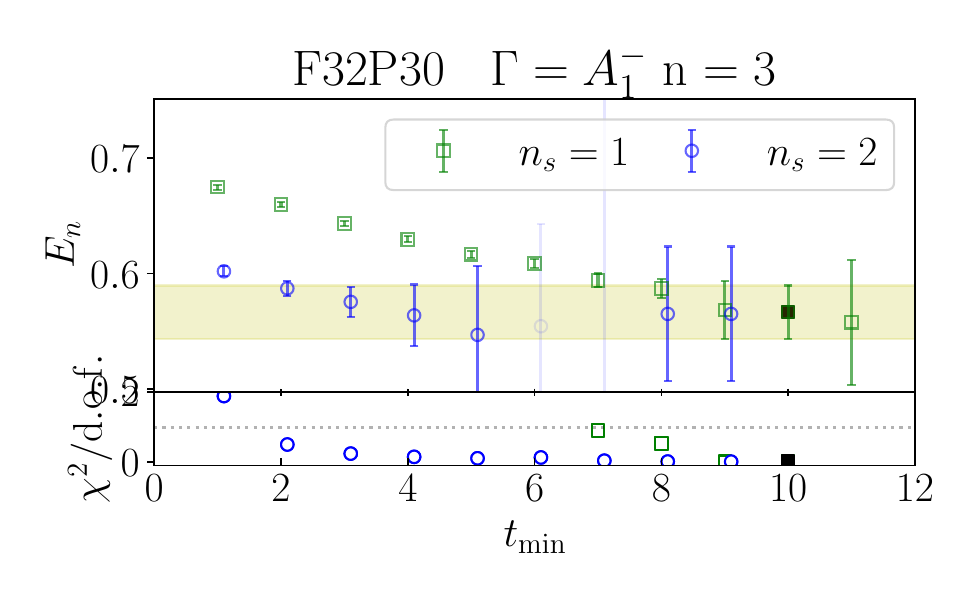}
\includegraphics[width=0.32\columnwidth]{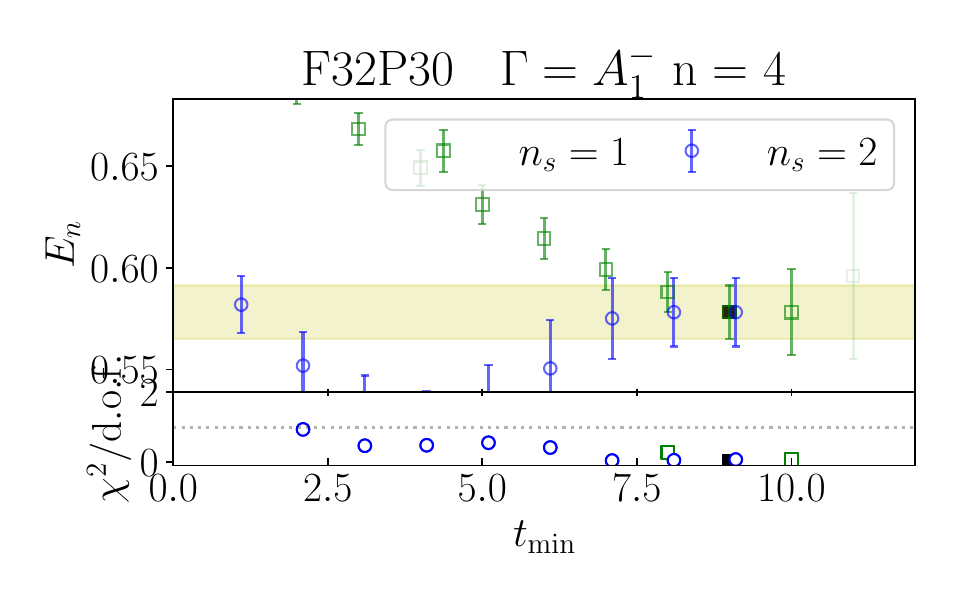}
\includegraphics[width=0.32\columnwidth]{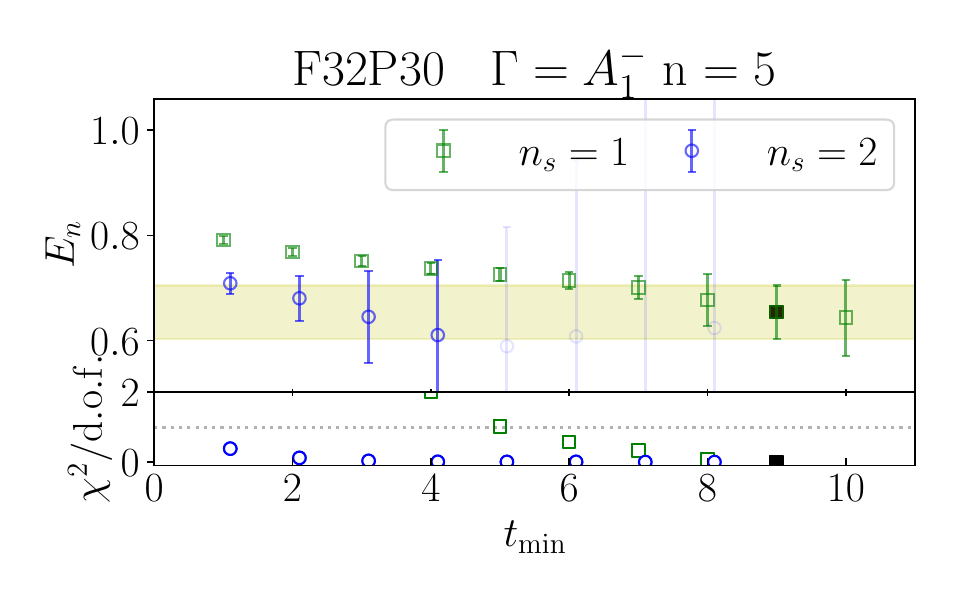}
\\
\includegraphics[width=0.32\columnwidth]{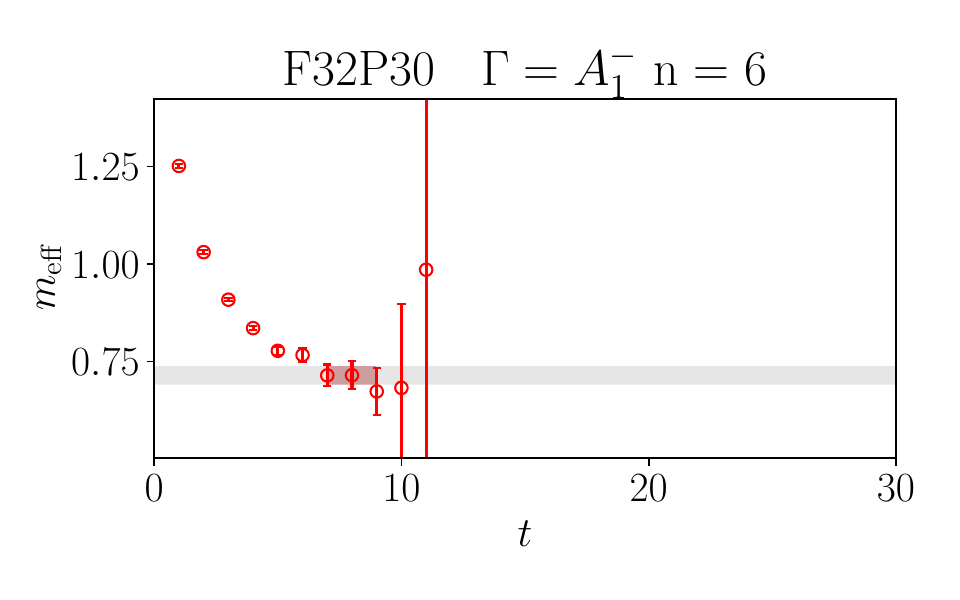}
\\
\includegraphics[width=0.32\columnwidth]{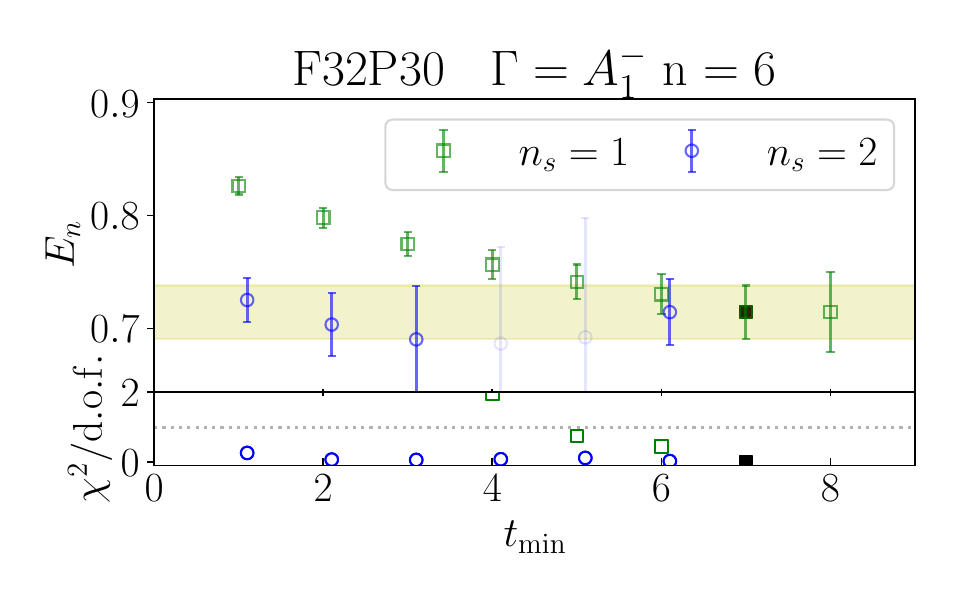}
\caption{Fitting for the $I=1$ $\pi\pi\pi$ channel for ensemble F32P30.}
\label{fig:pipipi-I=1-fit-F32P30}
\end{figure}

\begin{figure}[htbp]
\centering
\includegraphics[width=0.32\columnwidth]{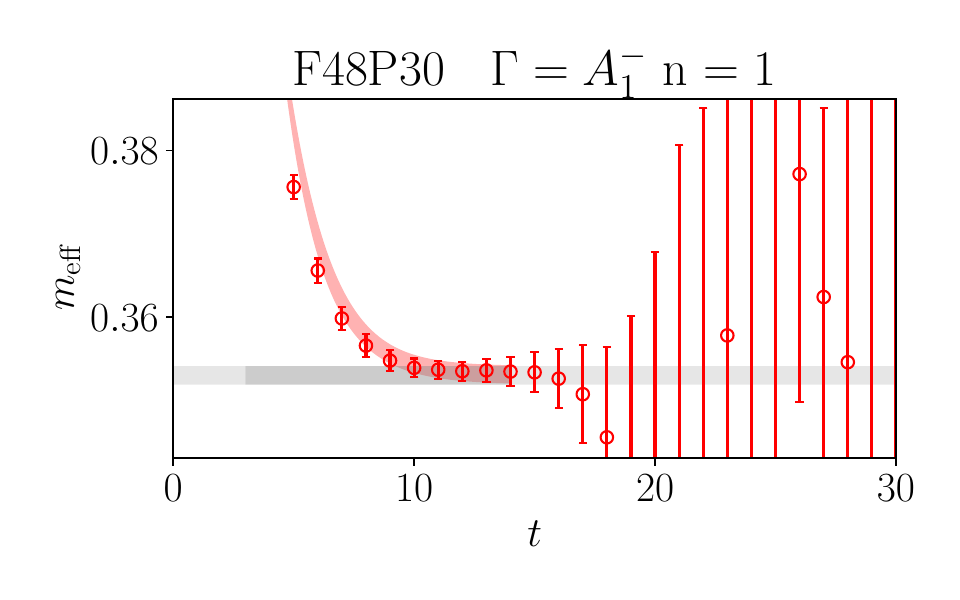}
\includegraphics[width=0.32\columnwidth]{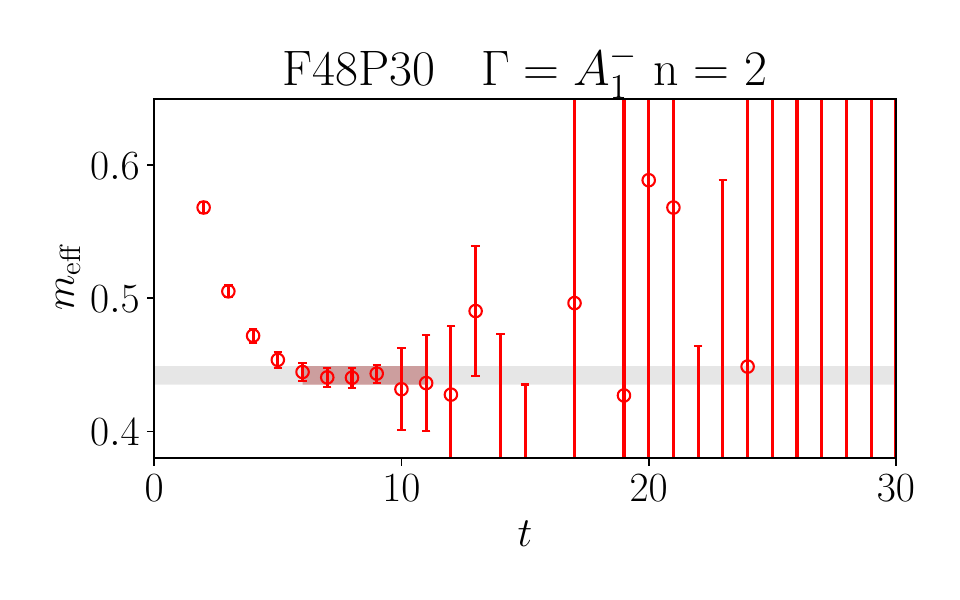}
\includegraphics[width=0.32\columnwidth]{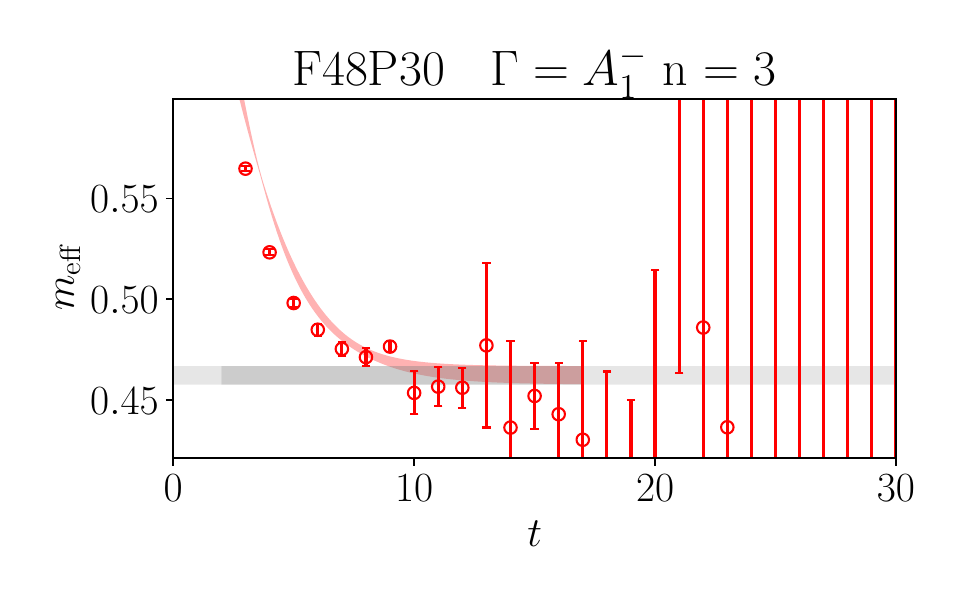}
\\
\includegraphics[width=0.32\columnwidth]{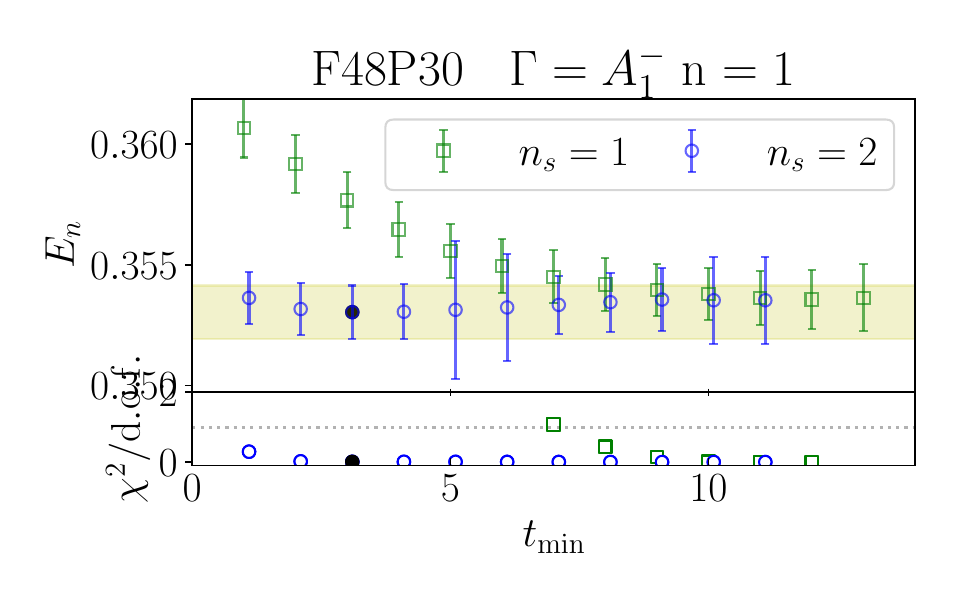}
\includegraphics[width=0.32\columnwidth]{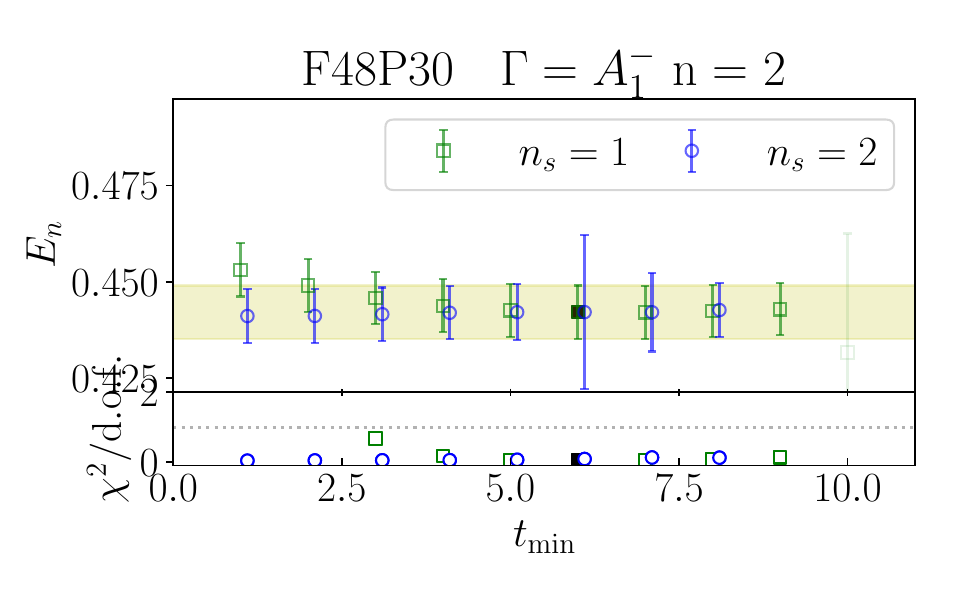}
\includegraphics[width=0.32\columnwidth]{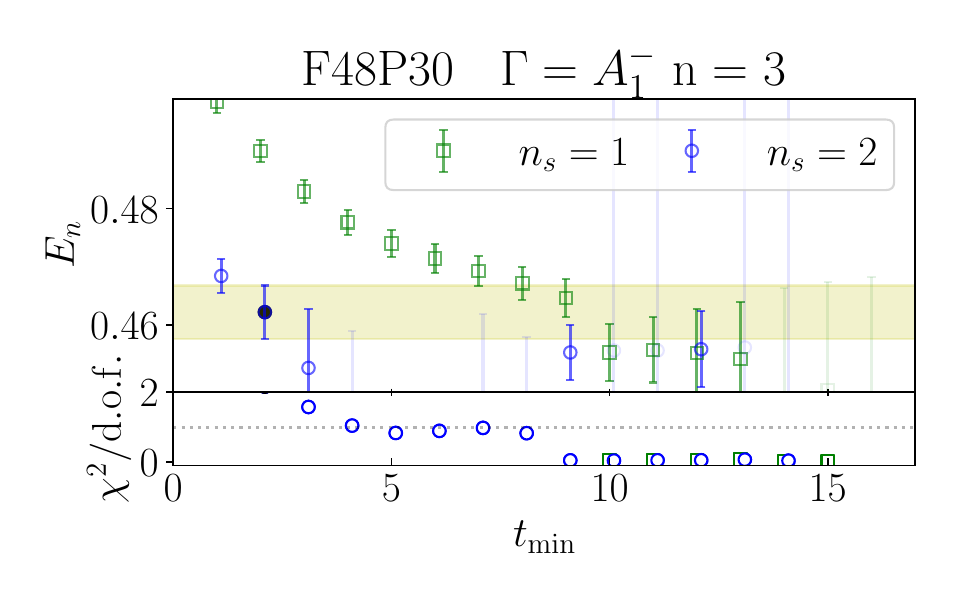}
\\
\includegraphics[width=0.32\columnwidth]{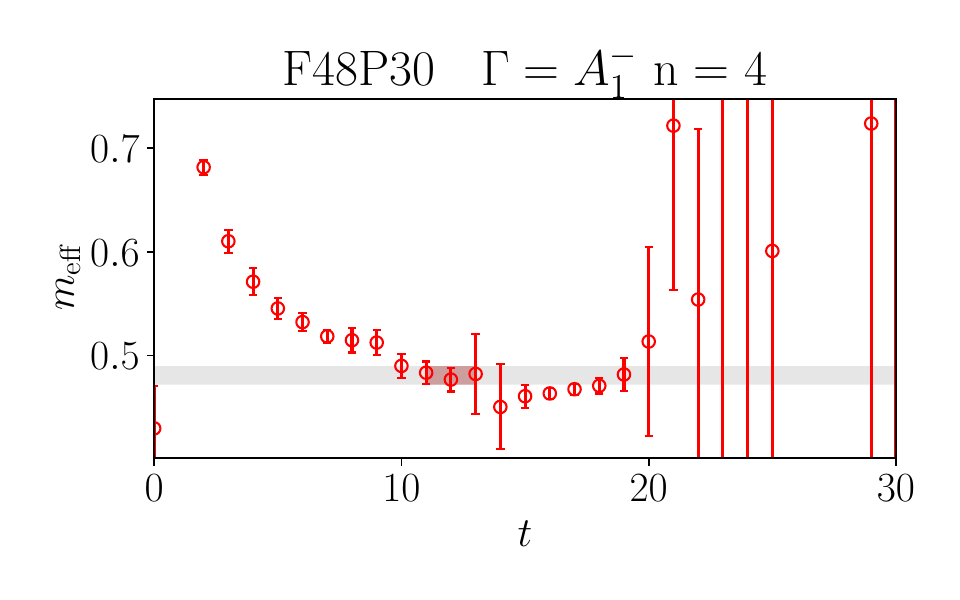}
\includegraphics[width=0.32\columnwidth]{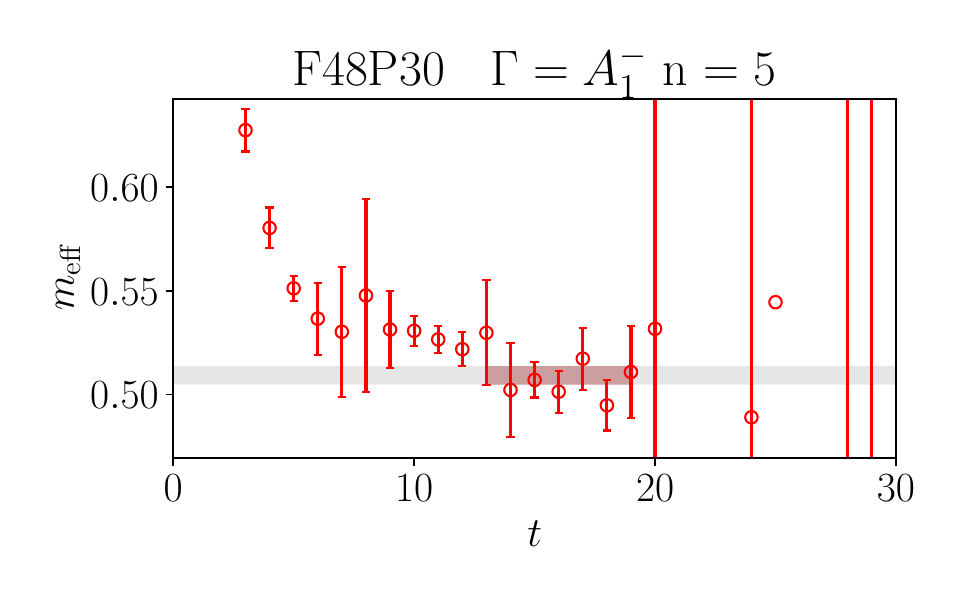}
\includegraphics[width=0.32\columnwidth]{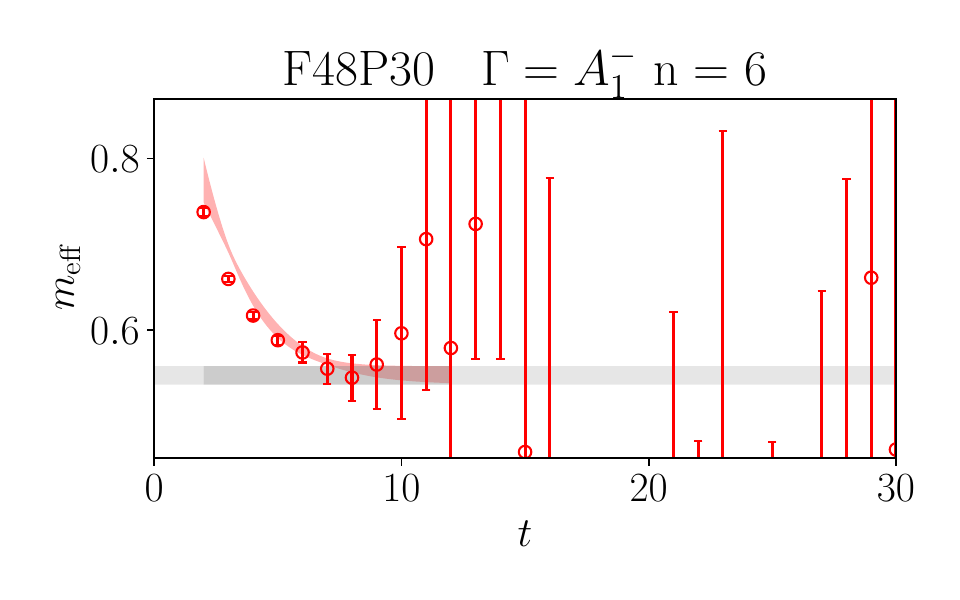}
\\
\includegraphics[width=0.32\columnwidth]{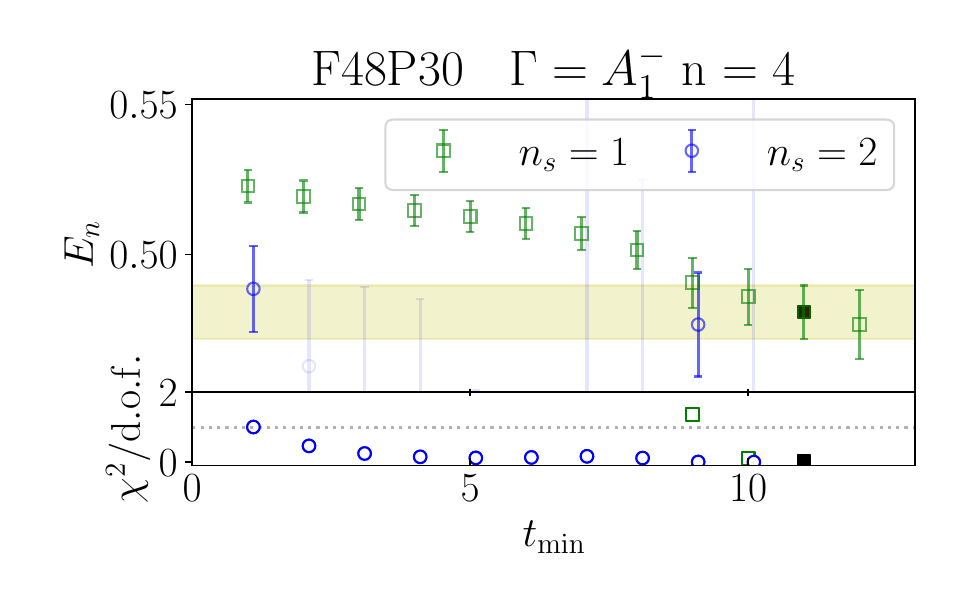}
\includegraphics[width=0.32\columnwidth]{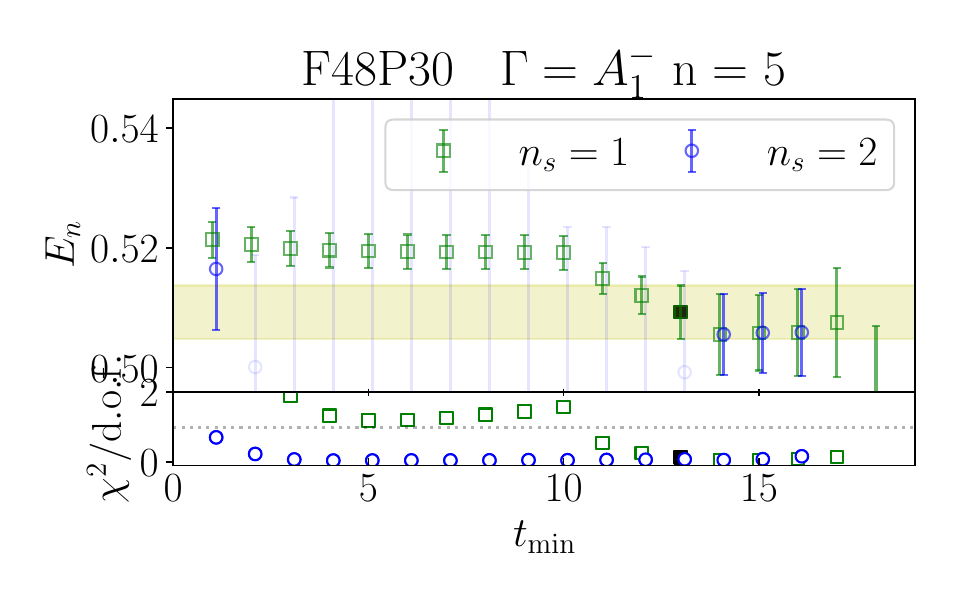}
\includegraphics[width=0.32\columnwidth]{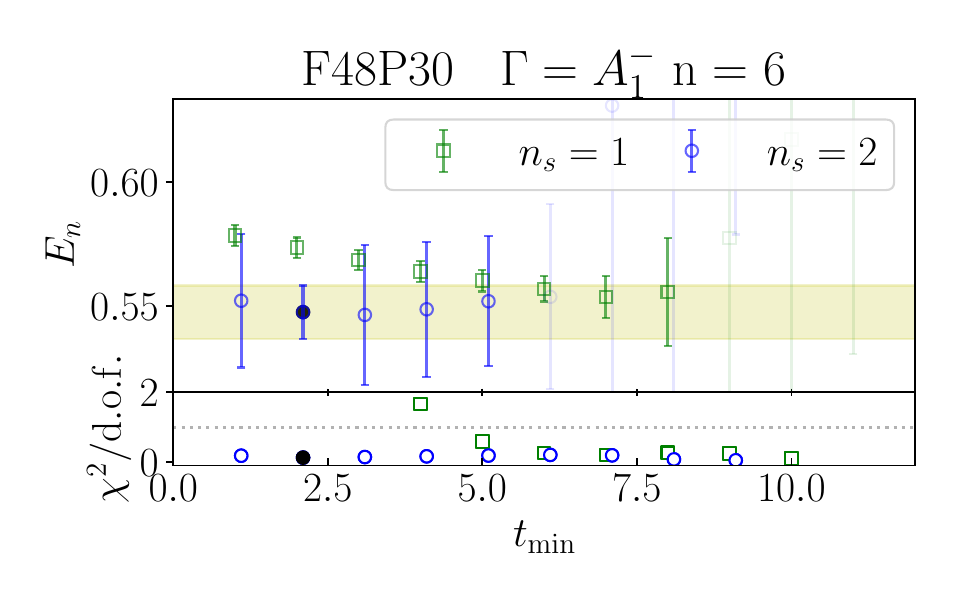}
\\
\includegraphics[width=0.32\columnwidth]{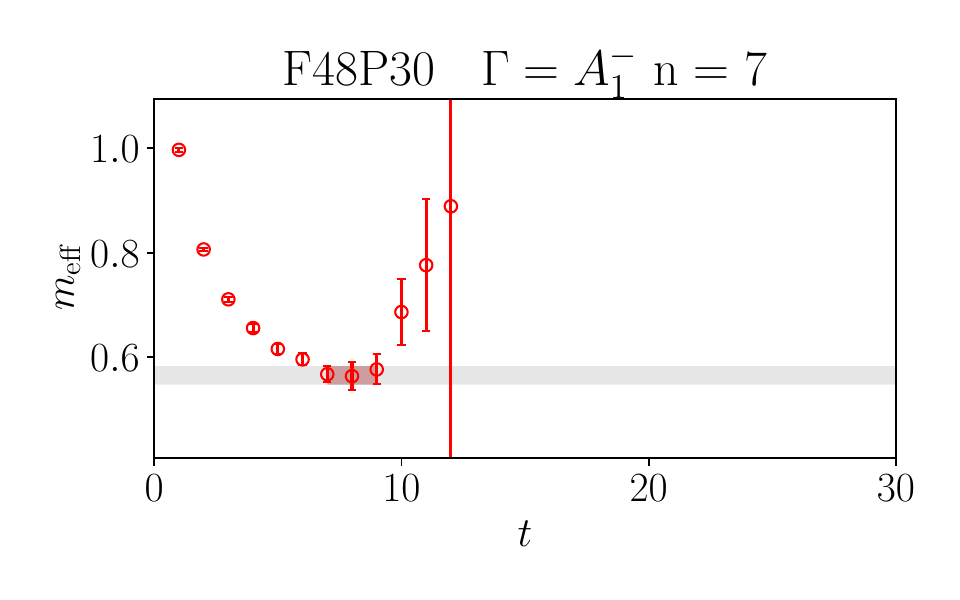}
\\
\includegraphics[width=0.32\columnwidth]{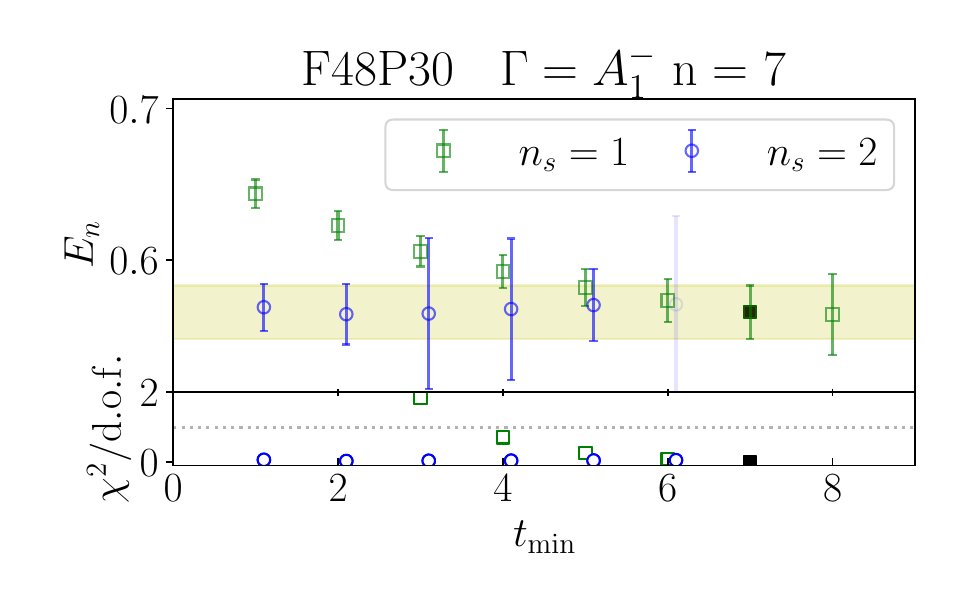}
\caption{Fitting for the $I=1$ $\pi\pi\pi$ channel for ensemble F48P30.}
\label{fig:pipipi-I=1-fit-F48P30}
\end{figure}

\begin{figure}[htbp]
\centering
\includegraphics[width=0.32\columnwidth]{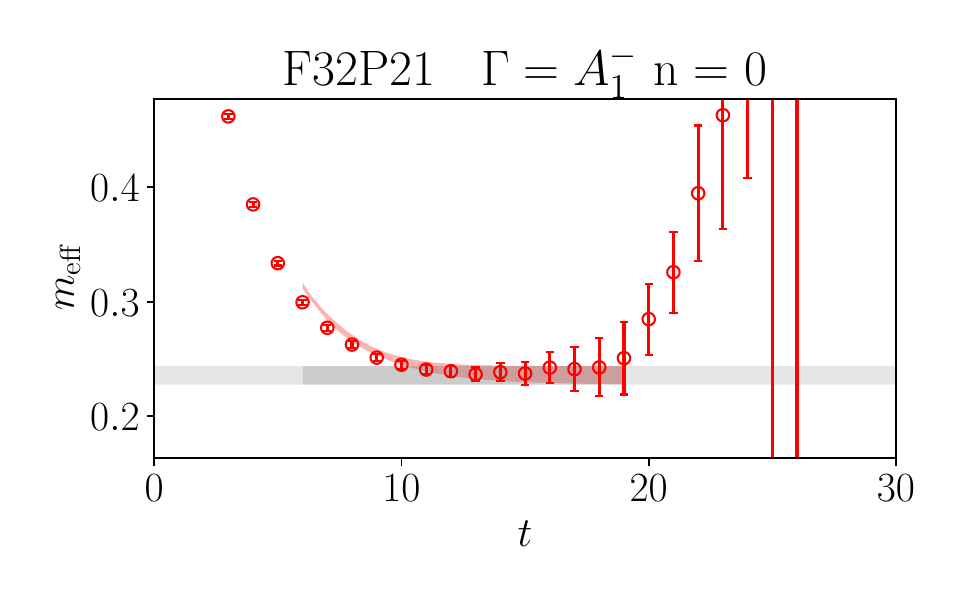}
\includegraphics[width=0.32\columnwidth]{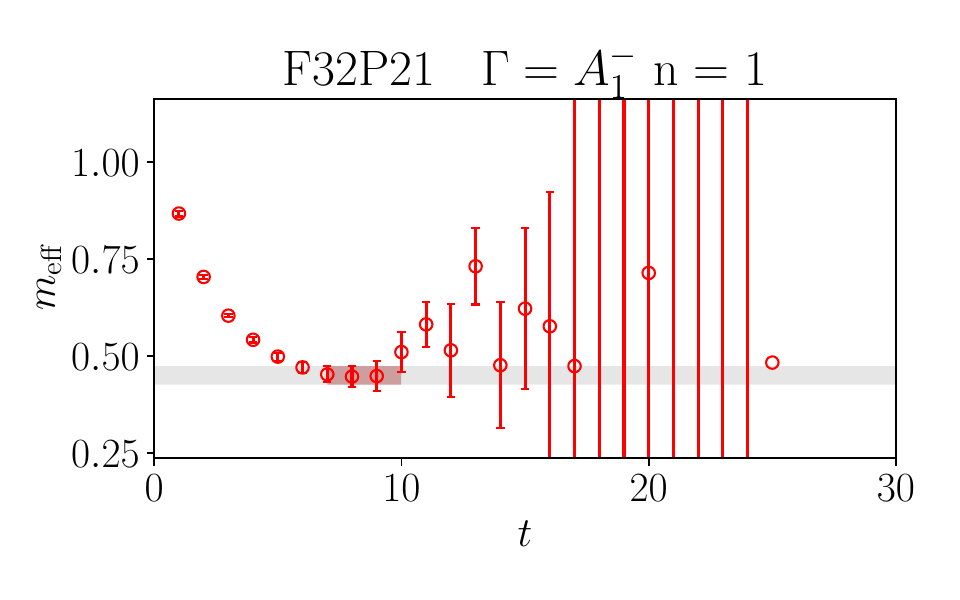}
\includegraphics[width=0.32\columnwidth]{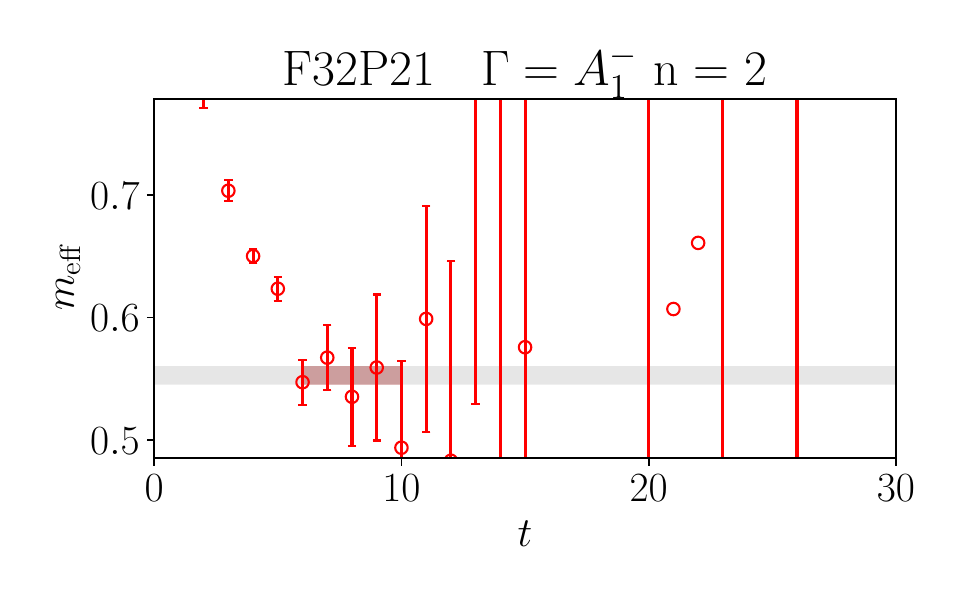}
\\
\includegraphics[width=0.32\columnwidth]{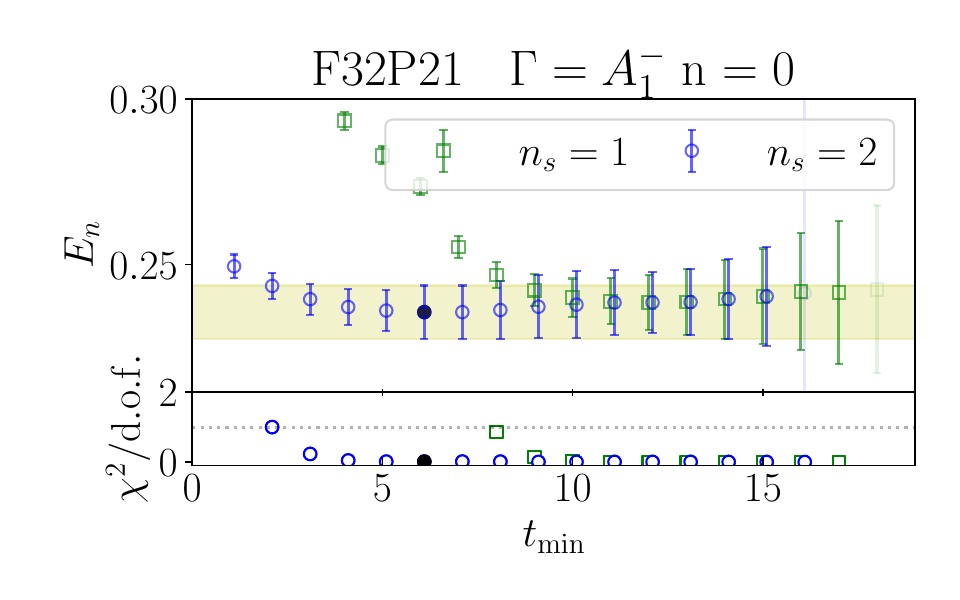}
\includegraphics[width=0.32\columnwidth]{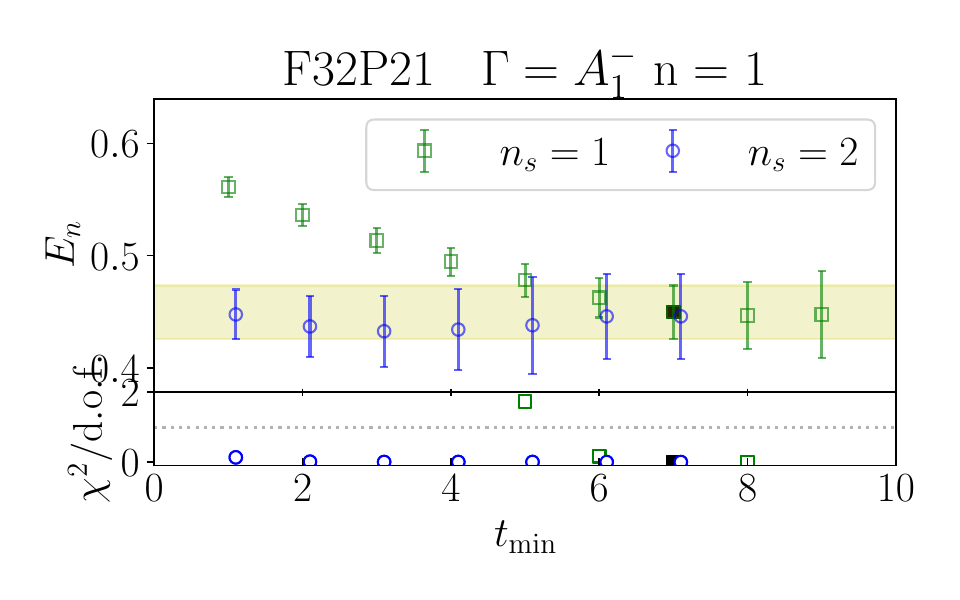}
\includegraphics[width=0.32\columnwidth]{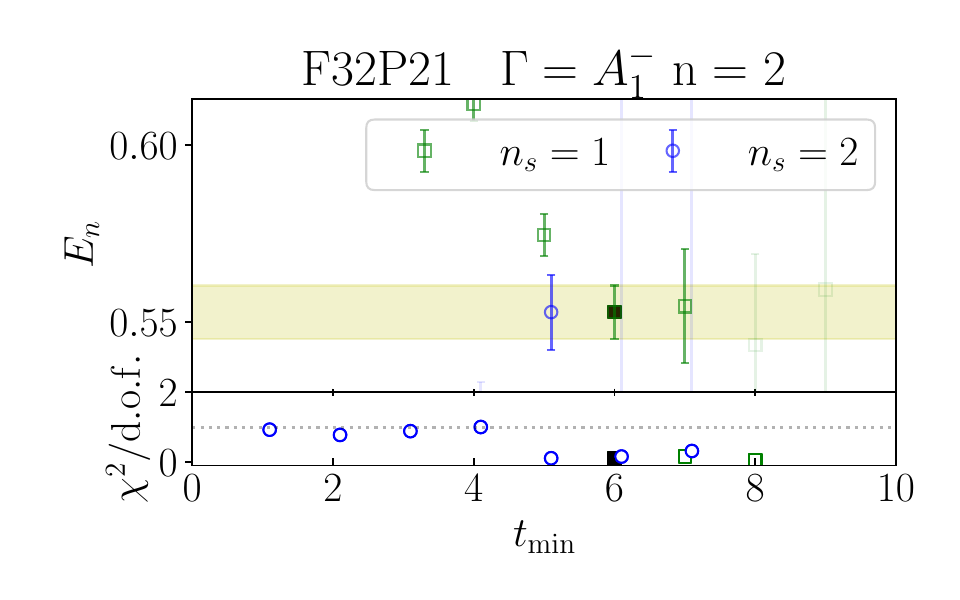}
\\
\includegraphics[width=0.32\columnwidth]{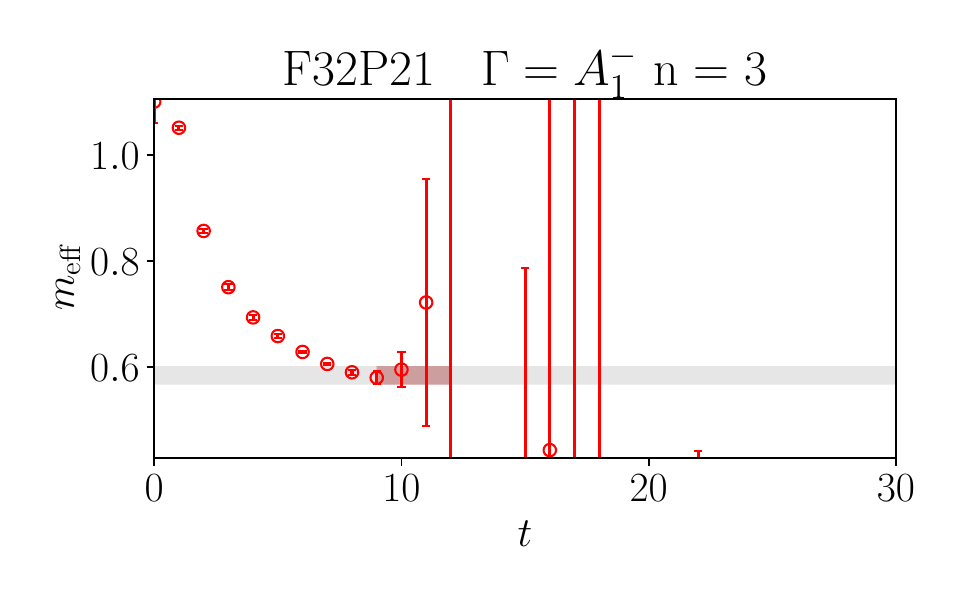}
\includegraphics[width=0.32\columnwidth]{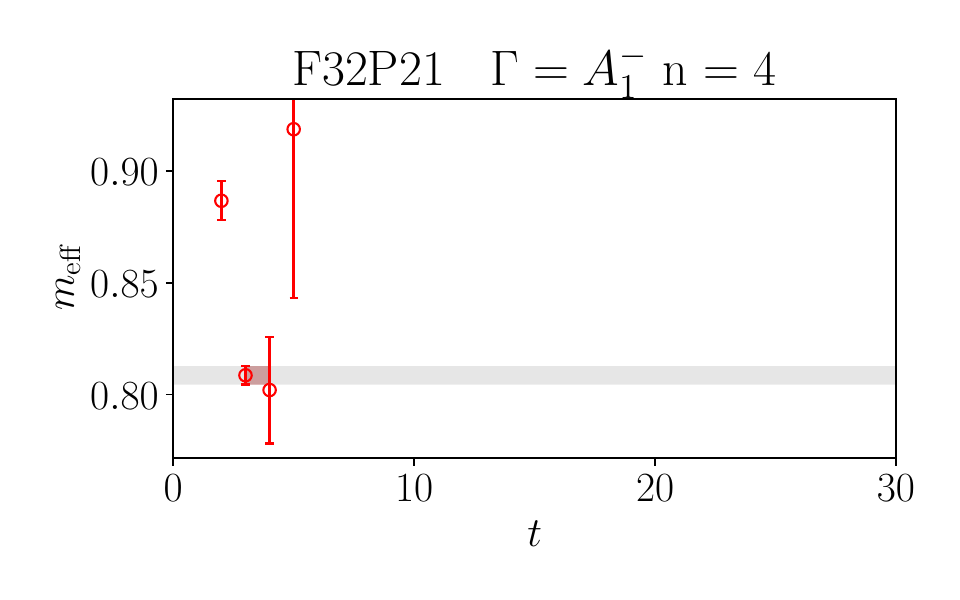}
\\
\includegraphics[width=0.32\columnwidth]{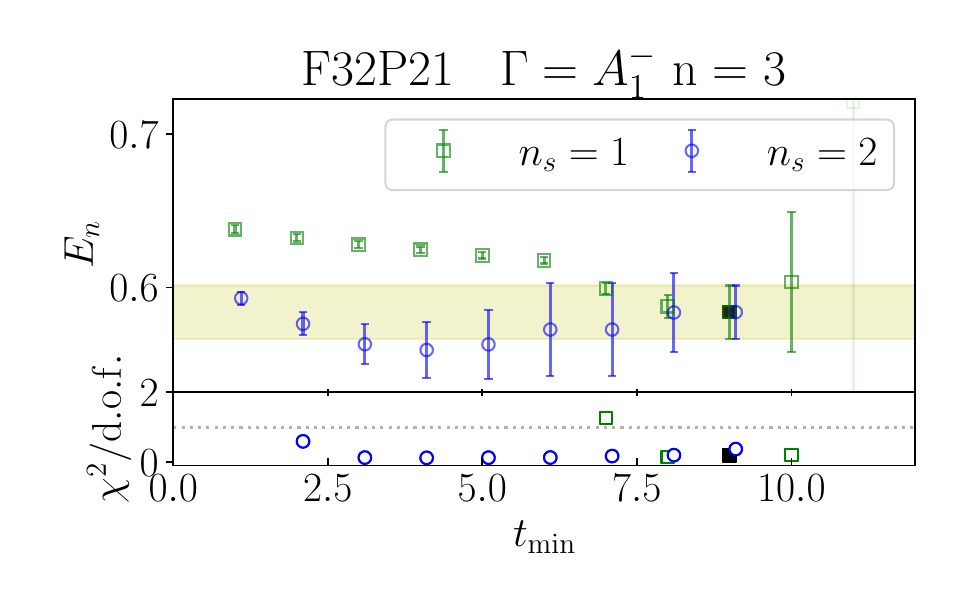}
\includegraphics[width=0.32\columnwidth]{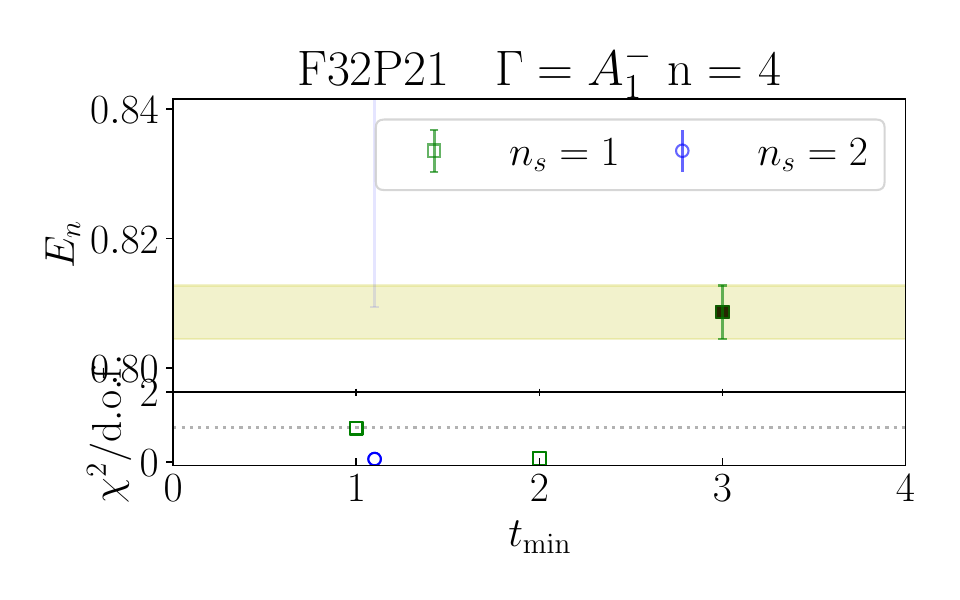}
\caption{Fitting for the $I=1$ $\pi\pi\pi$ channel for ensemble F32P21.}
\label{fig:pipipi-I=1-fit-F32P21}
\end{figure}

\begin{figure}[htbp]
\centering
\includegraphics[width=0.32\columnwidth]{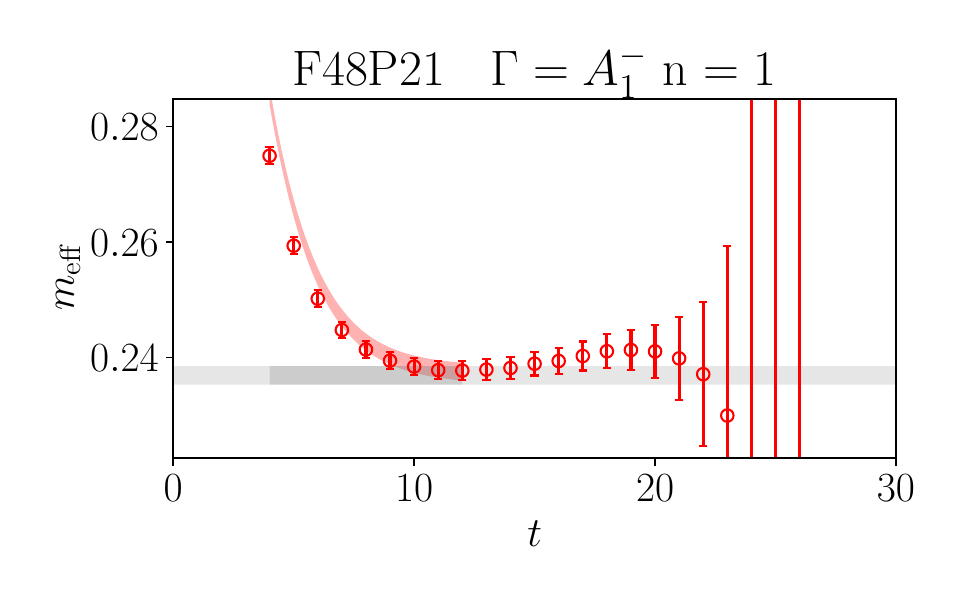}
\includegraphics[width=0.32\columnwidth]{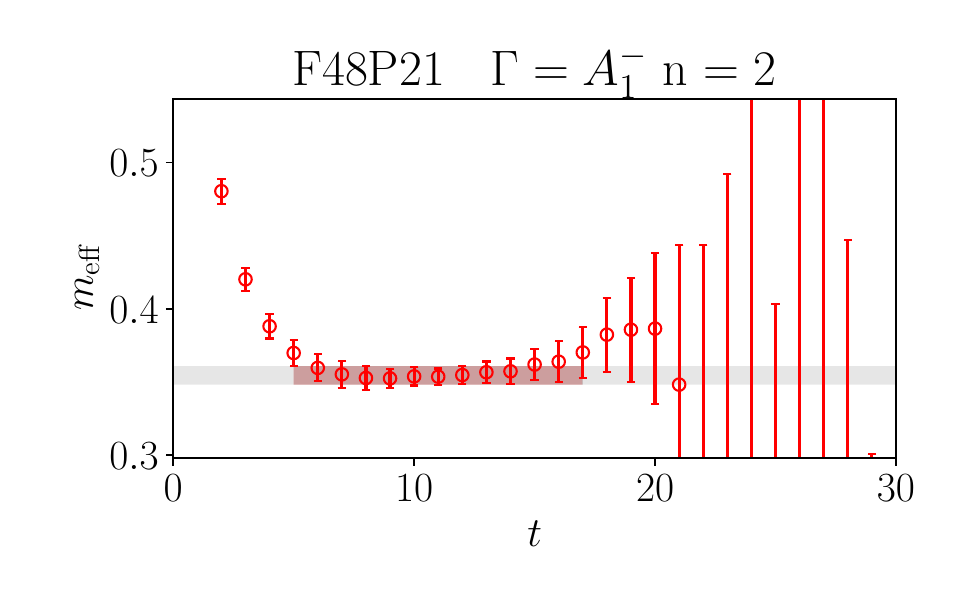}
\includegraphics[width=0.32\columnwidth]{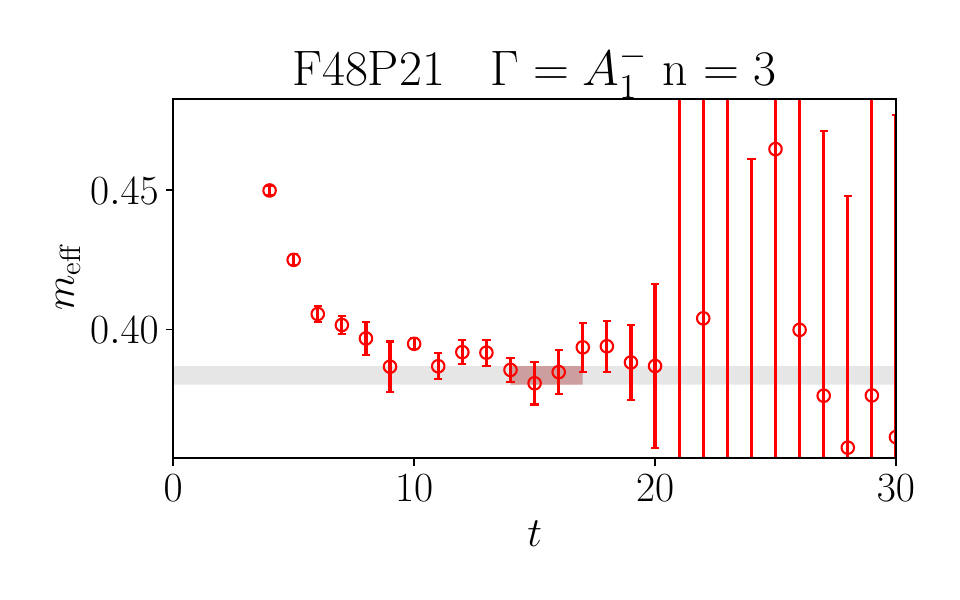}
\\
\includegraphics[width=0.32\columnwidth]{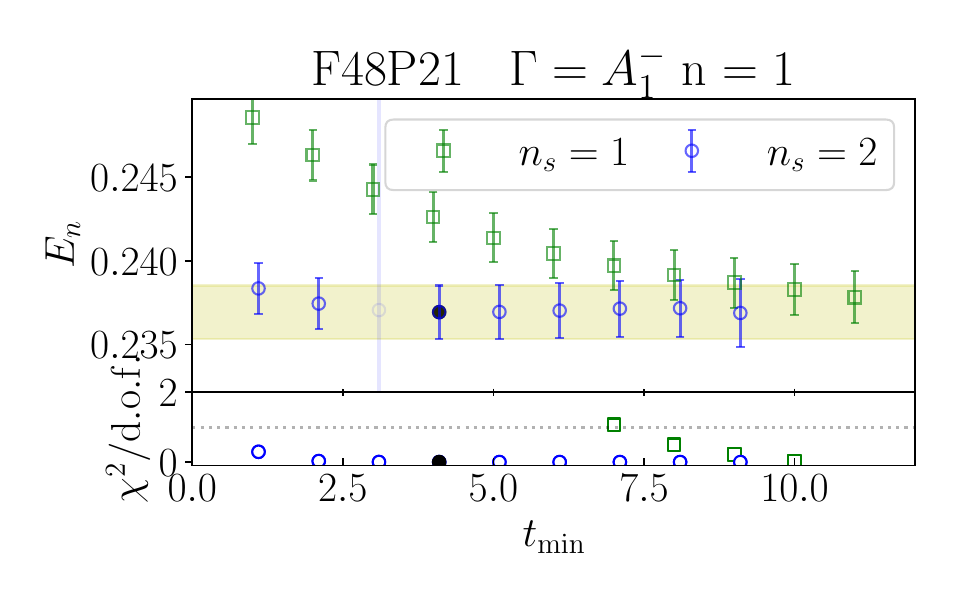}
\includegraphics[width=0.32\columnwidth]{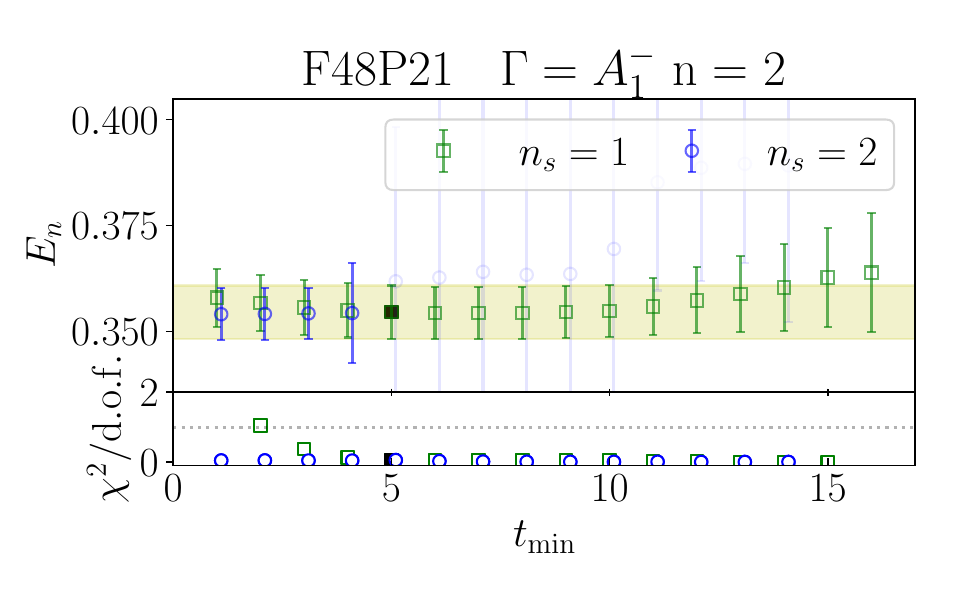}
\includegraphics[width=0.32\columnwidth]{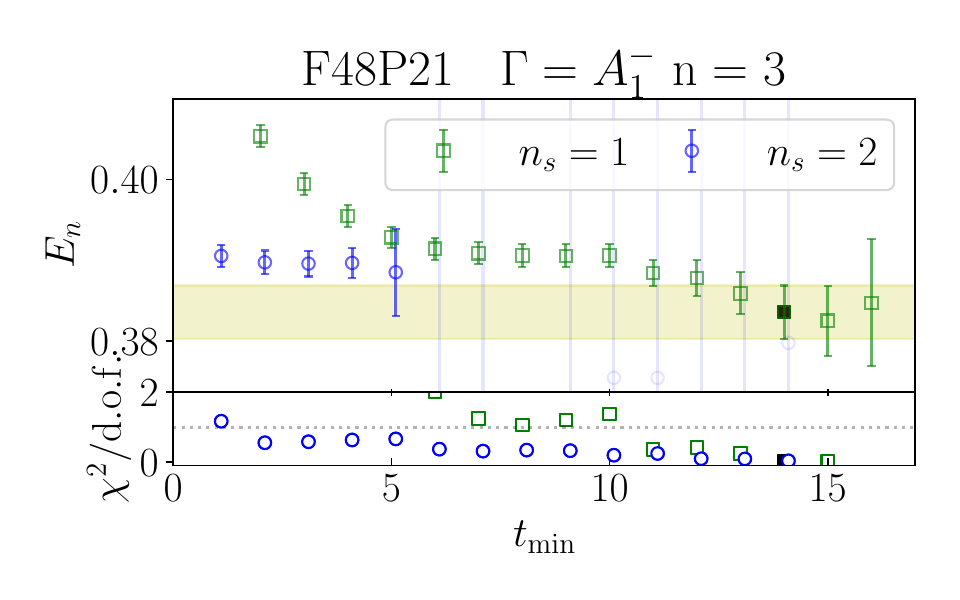}
\\
\includegraphics[width=0.32\columnwidth]{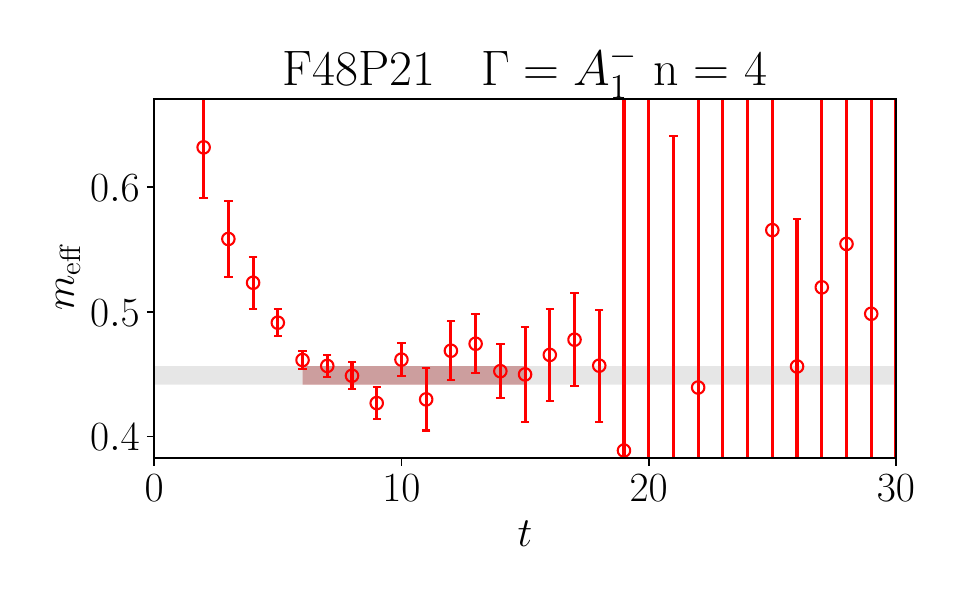}
\includegraphics[width=0.32\columnwidth]{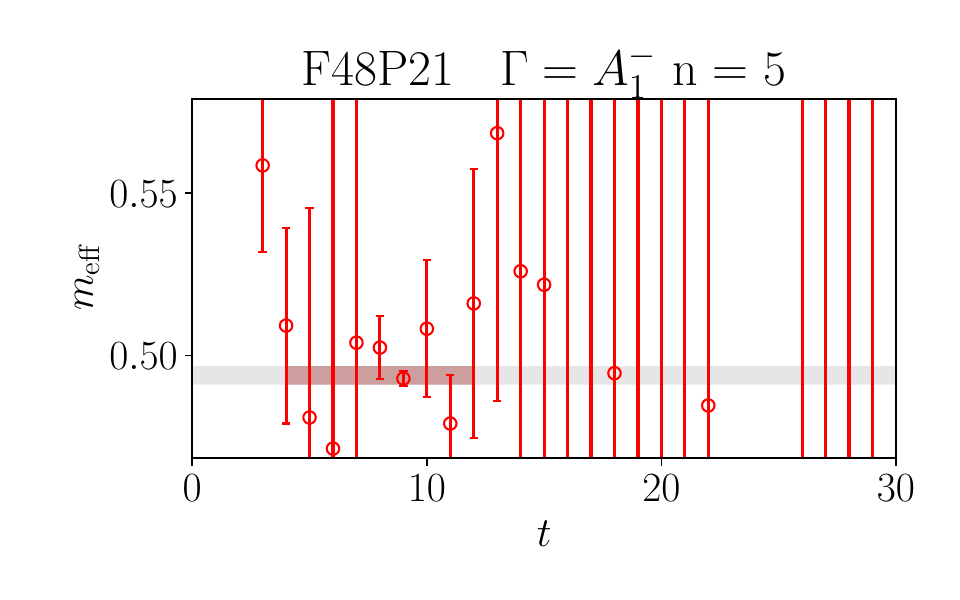}
\includegraphics[width=0.32\columnwidth]{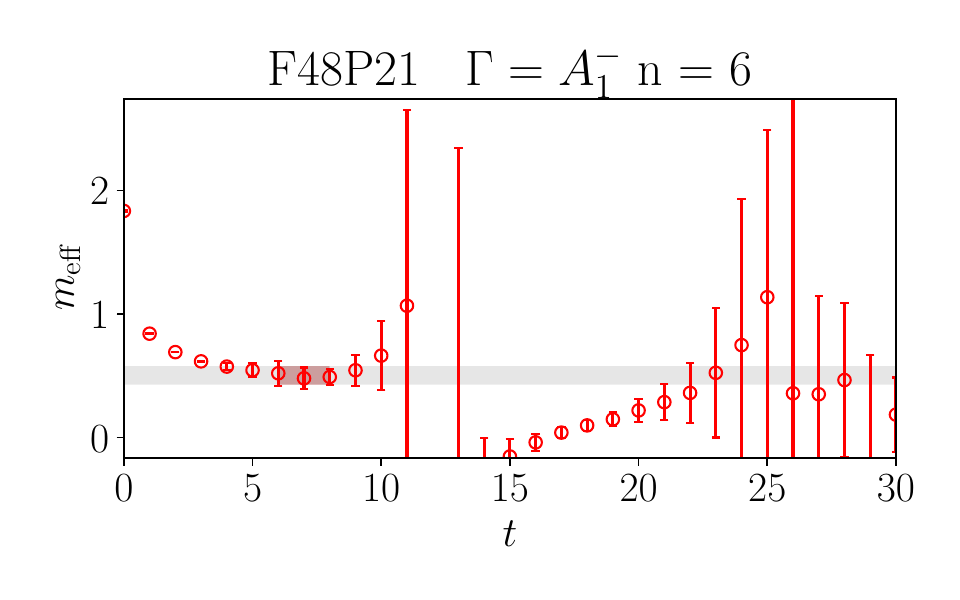}
\\
\includegraphics[width=0.32\columnwidth]{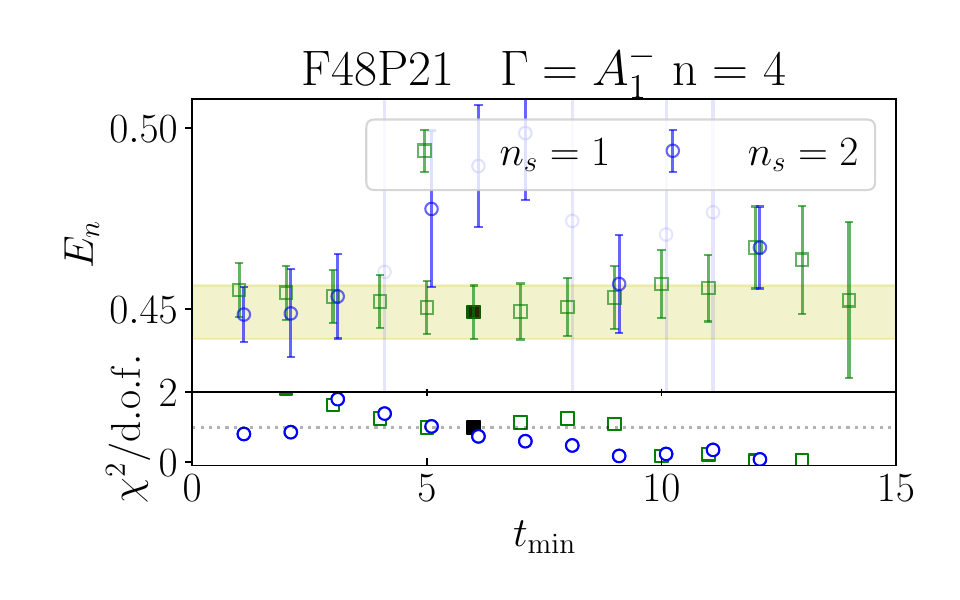}
\includegraphics[width=0.32\columnwidth]{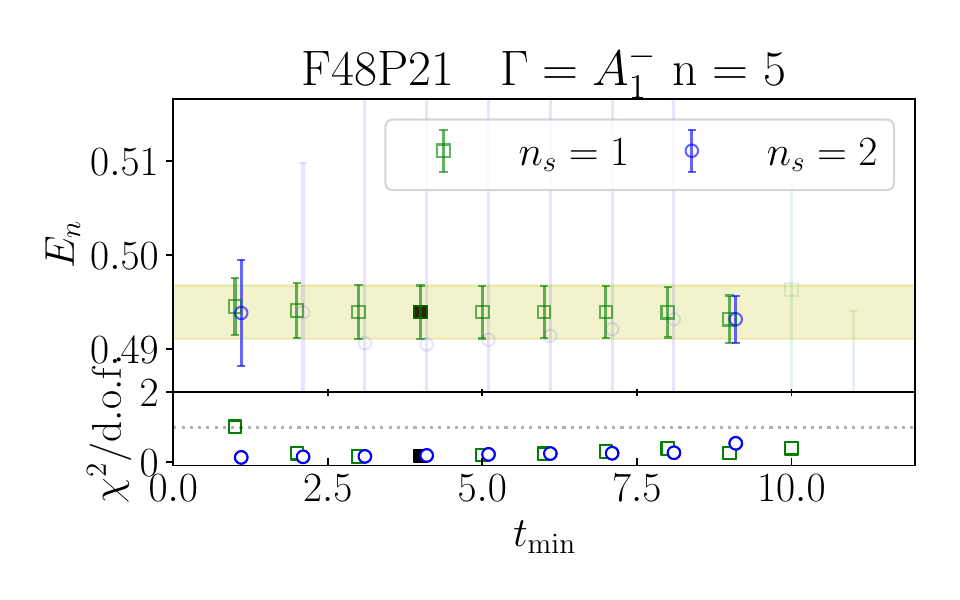}
\includegraphics[width=0.32\columnwidth]{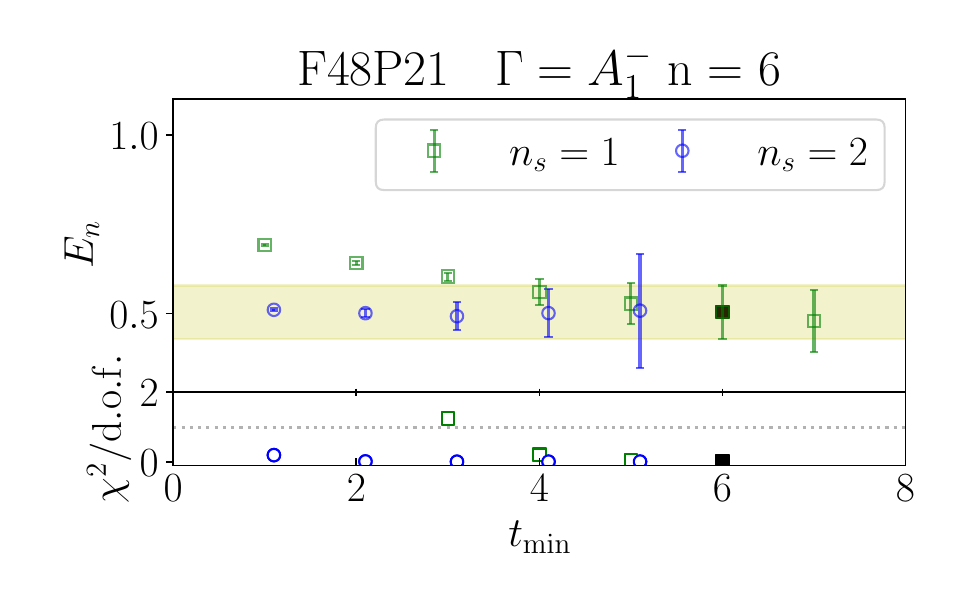}
\caption{Fitting for the $I=1$ $\pi\pi\pi$ channel for ensemble F48P21.}
\label{fig:pipipi-I=1-fit-F48P21}
\end{figure}

The fits and their dependence on the fitting range are shown in Figs.~\ref{fig:pipi-I=0-fit-F32P30}, \ref{fig:pipi-I=0-fit-F48P30}, \ref{fig:pipi-I=0-fit-F32P21}, \ref{fig:pipi-I=0-fit-F48P21}, \ref{fig:pipi-I=1-fit-F32P30}, \ref{fig:pipi-I=1-fit-F48P30}, \ref{fig:pipi-I=1-fit-F32P21}, \ref{fig:pipi-I=1-fit-F48P21}, \ref{fig:pipi-I=2-fit-F32P30}, \ref{fig:pipi-I=2-fit-F48P30}, \ref{fig:pipi-I=2-fit-F32P21}, \ref{fig:pipi-I=2-fit-F48P21}, \ref{fig:pipipi-I=1-fit-F32P30}, \ref{fig:pipipi-I=1-fit-F48P30}, \ref{fig:pipipi-I=1-fit-F32P21}, and \ref{fig:pipipi-I=1-fit-F48P21}. Statistical uncertainties are estimated via the Jackknife. For each eigenvalue from the GEVP, two plots are provided: the upper plot shows the effective mass plots with the red band representing the fit and the gray band indicating the extracted energy; the lower plot shows the extracted energy as a function of the starting point of the one- or two-state fit, with the filled circle or box marking the chosen starting value.

\clearpage
\subsection{Two- and three-body input}

Particular choices of physical inputs and cutoffs define our model space, we will explore with different fits to our LQCD data. Specifically, we consider
\begin{itemize}
    \item [(a)] all possible combinations of terms in \cref{eq:supp:c}; 
    \item [(b)] all two-body energy eigenvalues supplemented by either light, heavy or both three-body energy eigenvalues;
    \item [(c)] $3\times 3\times 3$ combinations of cutoffs ($\sigma_{\rm MP},|l_{\rm max}|,|k_{\rm max}|$);
    \item[(d)] mIAM4($l_1^r,l_2^r,l_3^r,l_4^r$), mIAM3($l_1^r,l_2^r,l_3^r,l_4^{r,{\rm FLAG}}$) and
mIAM2($l_1^r,l_2^r,l_3^{r,{\rm FLAG}},l_4^{r,{\rm FLAG}}$) correspondingly to the FLAG values~\cite{FlavourLatticeAveragingGroupFLAG:2024oxs} of the CHPT low-energy constants $\l_i^r$, similar to Ref.~\cite{Mai:2019pqr};
\end{itemize}
Overall, we have tested $\sim 2000$ different scenarios. Best fit parameters of the combined fits to two- and three-body finite-volume spectra are provided in \cref{tab:fit-results-all}. These include fits to either heavy, light, or both ensembles using the modified Inverse Amplitude method (mIAM)~\cite{Hanhart:2008mx} for the two-body part and three-body force of the general form
\begin{align}
    c^{\alpha\beta}=c_c^{\alpha\beta}+\frac{c_p^{\alpha\beta}}{s-m_0^2}\,,
    \quad
    {\alpha=\beta}\in \{\sigma\pi,\rho\pi\}\,,
    \label{eq:supp:c}
\end{align}
corresponding to the pertinent isobar-spectator channel. Fits in \cref{tab:fit-results-all} are obtained for fixed cutoffs $MP/M_\pi^2=1$, $|k_{\rm max}|=\sqrt{3\cdot5^2}\frac{2\pi}{aL}$, $|l_{\rm max}|=\sqrt{3}\frac{2\pi}{aL}$, while other relevant choices are provided in the attached files. Overall, as explicitly checked, the systematics associated with the choice of cutoff are sub-leading to statistical or other systematic uncertainties discussed in the main text. The coefficients defined above fix both the angular momentum (JLS) and helicity basis (HB, index $i$, $j$). The projected three-body force is defined as 
\begin{align}
    \left[C_{\rm HB}\right]_{(\bm{p}^\prime,j)(\bm{p},i)}=&
    \frac{1}{4\pi} 
    \mathfrak{D}^{0*}_{0,-\lambda(j)}(\phi_{-\bm{p'}},\theta_{-\bm{p'}},0)
    C_{ji}(s,p',p)
    \mathfrak{D}^{0}_{0,-\lambda(i)}(\phi_{-\bm{p}},\theta_{-\bm{p}},0)
    \\
    &\text{with}\quad
    C_{ji}(s,p',p)=U_{jL'} [C_{\rm JLS}(s,p',p)]_{L'L}U_{Li}
    \quad\text{for}\quad
    U_{Lj}=\begin{pmatrix}
        -1 & 0 & 0 \\
        0 & 1 & 0 \\
        0 & 0 & 1 \\
    \end{pmatrix}_{Lj}\\
    &\text{with}\quad
    C_{\rm JLS}=
    \begin{pmatrix}
       (p')^1 c_{\pi\rho,\pi\rho}(p)^1&0&0\\
        0&(p')^0c_{\pi\sigma,\pi\sigma}(p)^0&0\\
        0&0&0\\
    \end{pmatrix}
\end{align}

\begin{table}[b]
\centering
\tiny
\caption{Fit results for cutoff $\sigma_{\rm MP}=1M_\pi^2$, $|l_{max}|=\sqrt{3}\frac{2\pi}{aL}$, $|k_{max}|=\sqrt{3\cdot5^2} \frac{2\pi}{aL}$. Fitted spectra include heavy, light, and all ensembles in the top, middle, and bottom segments of the table, respectively. Everywhere, two- and three-body levels are fitted simultaneously, including cross-correlations. The best fits used for pole position extraction are $\{5,23,194,203\}$, the overall best fit ($203$) is marked by bold font.}
\addtolength{\tabcolsep}{2pt}
\label{tab:fit-results-all}
\begin{tabular}{l|llll|rrrrr|rrrrr|cc}
\toprule
\# & \multicolumn{4}{c|}{$\{l_1^r,l_2^r,l_3^r,l_4^r\}\times1000$} & $c^\sigma_c/{M_\pi^2}$ & $c^\sigma_p$ & $c^\rho_c/{M_\pi^2}$ & $c^\rho_p$ & $m_0/{M_\pi}$ & $c^\sigma_c/{M_\pi^2}$ & $c^\sigma_p$ & $c^\rho_c/{M_\pi^2}$ & $c^\rho_p$ & $m_0/{M_\pi}$ & 
$\chi^2/{\rm d.o.f.}$ & $N_{\rm data}$ \\
\midrule
$2$ & $-4.56$ & $4.33$ & $11.69$ & $4.49$ & -- & -- & -- & -- & -- & -- & $10.71$ & -- & -- & $4.07$ & $1.22$ & $33$ \\
$5$ & $-4.57$ & $4.32$ & $11.68$ & -- & -- & -- & -- & -- & -- & -- & $10.71$ & -- & -- & $4.07$ & $1.18$ & $33$ \\
$8$ & $-4.78$ & $3.88$ & -- & -- & -- & -- & -- & -- & -- & -- & $2.23$ & -- & -- & $4.02$ & $2.32$ & $33$ \\
$11$ & $-5.05$ & $4.24$ & $10.78$ & $-15.24$ & -- & -- & -- & -- & -- & -- & -- & -- & $26.36$ & $4.14$ & $1.44$ & $33$ \\
$14$ & $-4.56$ & $4.35$ & $11.37$ & -- & -- & -- & -- & -- & -- & -- & -- & -- & $22.14$ & $4.11$ & $1.28$ & $33$ \\
$17$ & $-4.77$ & $3.92$ & -- & -- & -- & -- & -- & -- & -- & -- & -- & -- & $30.76$ & $4.21$ & $2.4$ & $33$ \\
$20$ & $-4.62$ & $4.21$ & $8.52$ & $4.58$ & -- & -- & -- & -- & -- & $19.57$ & $117.21$ & -- & -- & $4.4$ & $1.16$ & $33$ \\
$23$ & $-4.62$ & $4.21$ & $8.51$ & -- & -- & -- & -- & -- & -- & $19.6$ & $117.77$ & -- & -- & $4.4$ & $1.12$ & $33$ \\
$29$ & $-5.37$ & $4.17$ & $8.18$ & $-22.34$ & -- & -- & -- & -- & -- & -- & -- & $9.19$ & $38.14$ & $4.21$ & $1.72$ & $33$ \\
$32$ & $-4.57$ & $4.32$ & $11.34$ & -- & -- & -- & -- & -- & -- & -- & -- & $-0.01$ & $9.78$ & $3.98$ & $1.31$ & $33$ \\
$35$ & $-4.78$ & $3.9$ & -- & -- & -- & -- & -- & -- & -- & -- & -- & $4.7$ & $16.16$ & $4.$ & $2.46$ & $33$ \\
$38$ & $-4.84$ & $4.4$ & $10.85$ & $-12.99$ & -- & -- & -- & -- & -- & $0.$ & -- & -- & -- & -- & $2.26$ & $33$ \\
$41$ & $-4.31$ & $4.89$ & $14.03$ & -- & -- & -- & -- & -- & -- & $-7.75$ & -- & -- & -- & -- & $2.58$ & $33$ \\
$44$ & $-4.47$ & $4.57$ & -- & -- & -- & -- & -- & -- & -- & $15.27$ & -- & -- & -- & -- & $3.63$ & $33$ \\
$56$ & $-4.31$ & $4.88$ & $10.76$ & $2.8$ & -- & -- & -- & -- & -- & $0.32$ & -- & $-0.83$ & -- & -- & $2.81$ & $33$ \\
$59$ & $-4.55$ & $4.36$ & $12.47$ & -- & -- & -- & -- & -- & -- & $-4.1$ & -- & $-8.05$ & -- & -- & $1.42$ & $33$ \\
$62$ & $-4.66$ & $4.13$ & -- & -- & -- & -- & -- & -- & -- & $1.1$ & -- & $-8.52$ & -- & -- & $2.39$ & $33$ \\
\midrule
$65$ & $-4.65$ & $4.15$ & $6.64$ & $4.3$ & -- & $14.67$ & -- & -- & $4.95$ & -- & -- & -- & -- & -- & $1.03$ & $27$ \\
$68$ & $-4.65$ & $4.15$ & $6.62$ & -- & -- & $14.79$ & -- & -- & $4.95$ & -- & -- & -- & -- & -- & $0.99$ & $27$ \\
$71$ & $-4.76$ & $3.92$ & -- & -- & -- & $12.9$ & -- & -- & $4.92$ & -- & -- & -- & -- & -- & $1.14$ & $27$ \\
$74$ & $-4.63$ & $4.17$ & $6.59$ & $4.84$ & -- & -- & -- & $61720.$ & $70.78$ & -- & -- & -- & -- & -- & $1.06$ & $27$ \\
$77$ & $-4.64$ & $4.17$ & $6.55$ & -- & -- & -- & -- & $59071.6$ & $69.$ & -- & -- & -- & -- & -- & $1.02$ & $27$ \\
$80$ & $-4.76$ & $3.94$ & -- & -- & -- & -- & -- & $60050.6$ & $69.47$ & -- & -- & -- & -- & -- & $1.16$ & $27$ \\
$83$ & $-4.65$ & $4.13$ & $6.17$ & $4.77$ & $23.02$ & $761.8$ & -- & -- & $6.96$ & -- & -- & -- & -- & -- & $1.06$ & $27$ \\
$86$ & $-4.66$ & $4.13$ & $6.15$ & -- & $21.42$ & $668.27$ & -- & -- & $6.82$ & -- & -- & -- & -- & -- & $1.01$ & $27$ \\
$89$ & $-4.77$ & $3.92$ & -- & -- & $93.54$ & $9302.87$ & -- & -- & $10.75$ & -- & -- & -- & -- & -- & $1.13$ & $27$ \\
$92$ & $-4.63$ & $4.17$ & $6.59$ & $4.84$ & -- & -- & $-12.35$ & $763.47$ & $142.71$ & -- & -- & -- & -- & -- & $1.12$ & $27$ \\
$95$ & $-4.64$ & $4.17$ & $6.55$ & -- & -- & -- & $-12.47$ & $657.45$ & $874.9$ & -- & -- & -- & -- & -- & $1.07$ & $27$ \\
$98$ & $-4.76$ & $3.94$ & -- & -- & -- & -- & $-12.51$ & $30.47$ & $364.77$ & -- & -- & -- & -- & -- & $1.22$ & $27$ \\
$101$ & $-4.63$ & $4.19$ & $6.9$ & $4.3$ & $-2.92$ & -- & -- & -- & -- & -- & -- & -- & -- & -- & $1.03$ & $27$ \\
$104$ & $-4.63$ & $4.19$ & $6.89$ & -- & $-2.94$ & -- & -- & -- & -- & -- & -- & -- & -- & -- & $0.99$ & $27$ \\
$107$ & $-4.75$ & $3.95$ & -- & -- & $-2.64$ & -- & -- & -- & -- & -- & -- & -- & -- & -- & $1.15$ & $27$ \\
$119$ & $-4.64$ & $4.16$ & $6.76$ & $4.32$ & $-1.86$ & -- & $-10.92$ & -- & -- & -- & -- & -- & -- & -- & $1.05$ & $27$ \\
$122$ & $-4.65$ & $4.16$ & $6.75$ & -- & $-1.89$ & -- & $-10.93$ & -- & -- & -- & -- & -- & -- & -- & $1.$ & $27$ \\
$125$ & $-4.76$ & $3.93$ & -- & -- & $-1.26$ & -- & $-11.53$ & -- & -- & -- & -- & -- & -- & -- & $1.16$ & $27$ \\
\midrule
$191$ & $-4.47$ & $4.72$ & $12.31$ & $-2.02$ & -- & $103.16$ & -- & -- & $6.68$ & -- & $35.74$ & -- & -- & $4.54$ & $2.36$ & $36$ \\
$194$ & $-4.56$ & $4.34$ & $11.79$ & -- & -- & $55.34$ & -- & -- & $5.98$ & -- & $10.62$ & -- & -- & $4.07$ & $1.11$ & $36$ \\
$197$ & $-4.77$ & $3.89$ & -- & -- & -- & $41.22$ & -- & -- & $5.81$ & -- & $0.87$ & -- & -- & $3.95$ & $2.25$ & $36$ \\
$200$ & $-4.62$ & $4.21$ & $8.51$ & $4.61$ & $15.34$ & $416.83$ & -- & -- & $6.47$ & $19.45$ & $115.28$ & -- & -- & $4.4$ & $1.12$ & $36$ \\
$\bf 203$ & $\bf -4.62$ & $\bf 4.21$ & $\bf 8.5$ & -- & $\bf 15.24$ & $\bf 415.65$ & -- & -- & $\bf 6.47$ & $\bf 19.46$ & $\bf 115.5$ & -- & -- & $\bf 4.4$ & $\bf 1.08$ & $\bf 36$ \\
$206$ & $-4.76$ & $3.93$ & -- & -- & $129.73$ & $18028.$ & -- & -- & $12.45$ & $272.48$ & $15731.$ & -- & -- & $8.47$ & $1.31$ & $36$ \\
$209$ & $-4.47$ & $4.49$ & $11.48$ & $5.91$ & -- & -- & -- & $104.33$ & $6.24$ & -- & -- & -- & $34.24$ & $4.24$ & $1.35$ & $36$ \\
$212$ & $-4.53$ & $4.41$ & $11.43$ & -- & -- & -- & -- & $123.52$ & $5.97$ & -- & -- & -- & $17.19$ & $4.06$ & $1.26$ & $36$ \\
$215$ & $-4.74$ & $3.97$ & -- & -- & -- & -- & -- & $125.48$ & $5.68$ & -- & -- & -- & $0.02$ & $3.86$ & $2.92$ & $36$ \\
$218$ & $-4.63$ & $4.39$ & $10.56$ & $-2.85$ & -- & -- & $3.55$ & $317.27$ & $6.13$ & -- & -- & $9.28$ & $16.38$ & $4.17$ & $1.59$ & $36$ \\
$221$ & $-4.48$ & $4.5$ & $11.09$ & -- & -- & -- & $6.3$ & $427.06$ & $6.75$ & -- & -- & $69.13$ & $401.66$ & $4.59$ & $1.37$ & $36$ \\
$224$ & $-4.71$ & $4.02$ & -- & -- & -- & -- & $2.07$ & $915.38$ & $5.89$ & -- & -- & $11.12$ & $31.23$ & $4.01$ & $2.66$ & $36$ \\
\bottomrule
\end{tabular}
\addtolength{\tabcolsep}{-2pt}
\end{table}

\subsection{Model average procedure}
To find the average across the different results of the lattice data analyses, we employ the procedure developed in Ref.~\cite{EuropeanTwistedMass:2014osg}. Namely, starting from $N$ computations with mean values $x_k$ and uncertainties $\sigma_{x,k}$ ($k=1,\cdots,N$) their average $x$ and uncertainty $\sigma_x$ are given by
\begin{gather}
    \label{eq:averaging}
    x = \sum_{k=1}^N \omega_k ~ x_k ~ , ~ \qquad
    \sigma_x^2 = \sigma_{x,\mathrm{stat}}^2 + \sigma_{x,\mathrm{syst}}^2 ~ , ~ \qquad
    \sigma_{x,\mathrm{stat}}^2=\sum_{k=1}^N \omega_k~\sigma_{x,k}^2 , ~ \qquad
    \sigma_{x,\mathrm{syst}}^2=\sum_{k=1}^N \omega_k ~ (x_k - x)^2~ , ~
\end{gather}
where $\omega_k$ represents the weight associated with the $k$-th determination. The weights $\omega_k$ are determined according to the Akaike Information Criterion (AIC)\,\cite{Akaike}, namely 
\begin{equation}
    \label{eq:AIC}
    \omega_k = A e^{- (\chi_k^2 + 2 N_\mathrm{parms} - N_\mathrm{data}) / 2}~ , ~
\end{equation}
where $\chi_k^2$ is the $\chi^2$ values of the $k$-th fit, $N_\mathrm{parms}$ is the number of free parameters, $N_\mathrm{data}$ the number of data points and $A$ is a normalization constant ensuring that $\sum_{k=1}^N w_k=1$.
The previous formula can be generalized in the case of multiple variables. For instance, to compute the covariance $\sigma_{xy}$ between $x$ and another variable $y$, we use
\begin{gather}
    \label{eq:averaging1}
    \sigma_{xy}^2 = \sigma_{xy,\mathrm{stat}}^2 + \sigma_{xy,\mathrm{syst}}^2 ~ , ~ \qquad
    \sigma_{xy,\mathrm{stat}}^2=\sum_{k=1}^N \omega_k~\sigma_{xy,k}^2 , ~ \qquad
    \sigma_{xy,\mathrm{syst}}^2=\sum_{k=1}^N \omega_k ~ (x_k - x)(y_k - y)~ , ~
\end{gather}


\end{document}